\documentclass[12pt,preprint]{aastex}

\shorttitle{X-Atlas}
\shortauthors{Westbrook et al.}

\begin{document}
\title{X-Atlas: An Online Archive of Chandra's Stellar High Energy
Transmission Gratings Observations}

\author{Owen W. Westbrook\altaffilmark{1}, Nancy Remage Evans, Scott
J. Wolk, Vinay L. Kashyap, Joy S. Nichols, Peter
J. Mendygral\altaffilmark{2}, Jonathan D. Slavin, Bradley Spitzbart} 
\affil{Harvard-Smithsonian Center for Astrophysics, 60 Garden Street,
Cambridge, MA 02138}
\email{nevans@cfa.harvard.edu}

\and

\author{Wayne L. Waldron}
\affil{Eureka Scientific, Inc., 2452 Delmer Street Suite 100, Oakland,
CA 94602-3017}
\altaffiltext{1}{present address: Massachusetts Institute of
Technology, Cambridge, MA}
\altaffiltext{2}{present address: University of Minnesota, Twin
Cities, MN}

\begin{abstract}
\begin{center}
\today
\end{center}
The high-resolution X-ray spectroscopy made
possible by the 1999 deployment of the {\it Chandra X-ray Observatory}
has revolutionized our understanding of stellar X-ray emission.  Many
puzzles remain, though, particularly regarding the mechanisms of
X-ray emission from OB stars.  Although numerous individual stars have
been observed in high-resolution, realizing the full
scientific potential of these observations will necessitate studying
the high-resolution $Chandra$ dataset as a whole.  To facilitate the
rapid comparison and characterization of stellar spectra, we have
compiled a uniformly processed database of
all stars observed with the $Chandra$ High Energy Transmission Grating
(HETG).  This database, known as X-Atlas, is accessible through a web
interface with searching, data retrieval, and interactive plotting
capabilities.  For each target, X-Atlas also features predictions of
the low-resolution ACIS spectra convolved from the HETG
data for comparison with stellar sources in archival ACIS images.
Preliminary analyses of the hardness ratios, quantiles, and
spectral fits derived from the predicted ACIS spectra reveal
systematic differences between the high-mass and low-mass stars in the
atlas and offer evidence for at least two distinct classes of
high-mass stars.  A high degree of X-ray variability is also seen in
both high and low-mass stars, including Capella, long thought to
exhibit minimal variability.  X-Atlas contains over 130 observations of
approximately 25 high-mass stars and 40 low-mass stars and will be
updated as additional stellar HETG observations become public.  The
atlas has recently expanded to non-stellar point sources,
and Low Energy Transmission Grating (LETG) observations are currently
being added as well.
\end{abstract}


\section{INTRODUCTION\label{s:intro}}

$Chandra$ has made high-resolution X-ray spectra of bright 
objects readily available.  Over 130 observations of stars have been 
made in high resolution and analyzed, resulting in both insight 
and further questions about X-ray production, magnetic fields, stellar
winds, and line shapes.  However,  the extracted gratings spectra
are not currently expected to be part of the level 3 data products
slated to be released by the {\it Chandra X-ray Center} in 2008.  We have therefore compiled a database, known as X-Atlas,\footnote{
{\tt http://cxc.harvard.edu/XATLAS/}} of all stars observed with the
$Chandra$ high and medium energy gratings (HETG) and extracted the
spectra in a uniform manner.

Additionally, although over 60 stars have been observed in high
spectral resolution, thousands of stellar sources have been imaged
with the $Chandra$ ACIS CCDs, which have sufficient energy resolution
to produce an extractable low-resolution spectrum for each source.
To enable comparisons of the analyses made at low-resolution with
results from the high-resolution spectra, we have used the HETG
grating observations in X-Atlas to predict the low-resolution ACIS
spectra by convolving the HETG spectra with the response functions of
the ACIS detectors.  These predicted low-resolution spectra, as well
as hardness ratios and quantiles calculated from them, are a valuable
tool for interpreting the physical properties of the large sample of
stars observed only at low spectral resolution.  For further
comparison, we have also extracted the low-resolution zeroth-order
spectrum from the HETG observations.  

As a demonstration of the capabilities of the database, we
present some early science results from X-Atlas.  We use the
high-resolution spectra to develop line ratio metrics for quick
characterization of the target.  Analyses of the
predicted hardness ratios and quantiles of the stars in the atlas
indicate that high-mass and low-mass stars exhibit distinct
coronal temperature regimes and suggest that at least two classes of
high-mass OB stars exist.  Variability analyses performed on each star
reveal that a high proportion of the stars in X-Atlas exhibit
variability during their $Chandra$ observation period.  Even stars
generally considered minimally-variable, such as Capella, may undergo
count rate fluctuations between observations while remaining stable
over the time scale of a typical observation.

Both the high-resolution and predicted low-resolution spectra are made
available in FITS and ASCII text formats through a dedicated web 
interface.  This web interface allows users to search the database and create
customized spectral plots of up to three sources, selecting either
individual lines or spectral regions for intercomparison.  We note
that a similar spectral browsing utility has been created by
\citet{gon07} for the plotting of Reflection Grating Spectrometer
(RGS) observations from {\it XMM-Newton}, though without the data retrieval
and low-resolution predictions offered by X-Atlas.  In addition to offering interactive plotting capabilities, we also present on the
X-Atlas web page the spectra in montages as functions of spectral type
and object class for high-mass OB stars, Wolf-Rayet stars, and low-mass stars.
Hardness ratios and quantiles also available for plotting and retrieval on
the web page.

The X-Atlas spectral database will be updated as additional stellar
HETG observations are carried out and archived. Further, we have
applied the reduction and analysis methods used for normal
stars to HETG observations of other point sources, including neutron stars,
galactic novae, active galatic nuclei, and other exotic objects.
All LETG observations of both stellar and non-stellar
point sources will also eventually be included in the database.

We describe the high-resolution spectral atlas in \S\ref{s:hires},
and the low-resolution predictions in \S\ref{s:lores}.  We discuss
the characteristics of the sources in X-Atlas in \S\ref{s:discuss},
and summarize in \S\ref{s:summary}.

\section{HIGH-RESOLUTION ATLAS\label{s:hires}}

The initial motivation for creating X-Atlas was to make available for
comparison all of the {\it Chandra} high-resolution gratings
observations of OB stars.  X-Atlas has since been expanded to include
low-mass stars as well.  The database currently features over 130
observations\footnote{as of July 2007} of roughly 25 high-mass and 40
low-mass stars (Section 2.5), with additional observations being added as
they are made public.

\subsection{HETG Observations and Data Reduction}

The HETG consists of two grating arms: the High Energy Grating (HEG),
optimized for a spectral range of approximately 1.5 to 15 \AA, and the
Medium Energy Grating (MEG), sensitive up to 30 \AA\ \citep{wei02}.  The
dispersed spectra produced by these grating arms cross at their respective
zeroth orders, creating the elongated 'x' for which X-Atlas is named.

We have reprocessed each observation in the X-Atlas database uniformly,
eliminating the need for weighing different analysis techniques and
assumptions from observation to observation and enabling quick
comparisons of X-ray sources in the atlas.  The {\it Chandra} team has
maintained a robust software package, {\it CIAO}, for the reduction and analysis of $Chandra$
data \citep{fru06}.  Using the most recent version (3.4) of {\it
CIAO}, we have created a Perl pipeline script of the appropriate {\it
CIAO} tools and `threads'\footnote{{\tt
http://asc.harvard.edu/ciao/threads/}} needed to reduce and analyze
each observation and have added our own IDL routines and S-Lang
scripts for further analysis.  The calibration files used for this
processing are those available with the $Chandra$ Calibration Database
(CALDB) 3.3.0.  After retrieving the necessary files from the {\it Chandra} Data Archive
(CDA), this pipeline carries out a number of standard X-ray reduction
steps, removing various instrumental distortions, locating the
$0^{th}$-order centroid and the dispersed spectra of each gratings arm,
classifying the dispersed photons by spectral order, and coadding the
positive and negative grating spectral orders of the HEG and MEG
spectra before proceeding to processing and analyses developed
specifically for X-Atlas.

Data preparation for a given observation begins with the level 1 event
file, a list of the arrival times and energies of incoming photons
that is available from the Chandra data archive.  Most {\it Chandra}
archival data has undergone Standard Data Processing (SDP) during the
most recent reprocessing run, which began in February 2006, but for
those observations that have not yet been reprocessed, additional
corrections are necessary.

Some portion of the X-ray events detected by {\it Chandra} are
caused by cosmic ray interactions in the CCD, and if not removed,
these `afterglow' events can be interpreted as faint sources.  
Possible cosmic ray events are flagged during SDP, and are filtered
out in the level 2 event file.  Note that the particular method used
in SDP versions prior to 7.4.0 has been shown to reject 3-5\% of valid
source photons\footnote{http://cxc.harvard.edu/ciao/threads/acisdetectafterglow/}.
This issue has been corrected in more recent processing versions, but
for archival data that has not yet been reprocessed appropriately, we
reset the afterglow flags on the level 1 event file and re-identify
afterglows and bad pixels with the updated reduction methods.

In addition to afterglow detection, SDP also randomizes the
coordinates of events within the pixel where they are detected to
avoid aliasing effects and remove the 'gridded' appearance of the
data.\footnote{http://cxc.harvard.edu/ciao/threads/acispixrand/}  For
X-Atlas reductions, we removed this randomization.  Observations were
also corrected for the contamination of the optical/UV filter on
ACIS, a time-dependent drift in the detector gain values, and the
charge transfer inefficiency (CTI) of the detectors as charge is moved
from pixel to pixel during readout.

The dispersed photons are extracted after the data are prepped as
described above.  The {\it CIAO} wrapper script {\it tgdetect} was
used to locate the centroid of the $0^{th}$-order image.  While {\it
tgdetect} is very accurate for the majority of HETG observations, if
the intended target is bright enough to cause the $0^{th}$-order image
to pile up, {\it tgdetect} may fail to properly detect the
centroid.\footnote{http://space.mit.edu/ASC/docs/memo\_fzero\_1.3.ps}
However, some of the most severe pileup effects, such as pulse saturation,
where the energy of a piled event is greater than the instrumental
threshold, are generally absent from stellar HETG observations in the
atlas.  In cases where {\it tgdetect} does not accurately identify the
position of the zeroth order, the contributed {\it CIAO} tool {\it
findzero} will usually obtain a correct value.  In addition, {\it
CXC's} current reprocessing effort of the entire archive correctly
determines the zeroth-order position for piled-up sources.  In the interest of
applying uniform processing to each observation, we did not implement
more sophisticated centroid detection schemes.  It is expected that
any in-depth scientific analysis of the stars in the atlas will
require separate reprocessing of the appropriate observations.

After centroid detection, various {\it CIAO} tools determine the
locations of the grating arms and assign events from each arm to the
appropriate spectral orders.  At this point, a level 2 event file is
created by filtering out bad grades previously marked as
afterglows and hot pixels, applying corrections for the effects of
excess streaking on chip 8, and selecting only those events falling
within the good time intervals determined by SDP.

The grating spectra are then extracted from the destreaked level 2
event file.  In addition, we create grating Redistribution Matrix
Files (gRMFs) to map between detector pulse height and energy space and
Ancillary Response Files (gARFs) to measure the  instrumental area and
quantum efficiency of the HETG/ACIS-S configuration.  The positive and
negative spectral orders of each arm are then coadded and stored in
six pulse height amplitude (PHA type I) spectral files, each of which
corresponds to either the HEG or MEG arm of HETG and to one of the
coadded $\pm1^{st}$, $\pm2^{nd}$, or $\pm3^{rd}$ orders.  With the
high-resolution spectrum extracted, data reduction continues with
processing routines developed specifically for X-Atlas, starting with the
creation of the predicted low-resolution ACIS spectra, described in
section \S\ref{s:lores}.

This data reduction procedure, including the convolution of the
predicted low-resolution ACIS spectra, was developed for stellar
observations but is applicable to any X-ray point source.
The current release of X-Atlas includes all point sources observed
with HETG, including cataclysmic variables, X-ray binaries, and active
galactic nuclei.  The same methods are being adapted for use with observations
of point sources using the ACIS-S/LETG, HRC-S/LETG, and
HRC-I/LETG configurations as well.  Because of the difficulties
involved in the spectroscopy of extended sources, only point sources
will be included in X-Atlas.  There are no standard {\it CIAO} tools
for the analysis of grating spectra from extended sources.  The
emission lines in such spectra will take on the shape of the
zeroth-order spatial structure convolved with the instrumental Line
Spread Function (LSF).  Grating analysis of such sources therefore
requires spatial/spectral analysis because there is not a unique
mapping of wavelength to diffraction angle for extended sources.

\subsection{Spectral Line Metrics}

Generally, it is difficult to summarize the information contained
in a high-resolution spectrum in a few variables that capture the
relevant details about the behavior of the source.  This is because
a coronal source spectrum is dominated by a large number of lines
from many elements at different ionization stages, in addition to
having an underlying continuum that is determined by the metallicity
of the plasma.  We have identified a set of line and continuum
fluxes that together encompass the major parameters that describe
coronal plasmas: temperature, composition, and density.  We measure
the counts in typically strong lines, such as the H-like and He-like
lines of O, Ne, Mg, Si, and S, as well as the prominent Fe\,XVII lines
at 15\AA\ and 17\AA.  The counts due to the continuum are also measured
over different wavelength ranges ($<12.1$\AA, $>12.1$\AA) to characterize
the general shape of the spectra.  The various ratios formed by these
fluxes serve as proxy for a complete description of the spectrum, but
in lower dimensional space.  These metrics are listed in Table~\ref{tblmetrics},
and are provided in addition to the nominal hardness ratios and quantile
widths that are calculated after converting the grating spectra to the
lower resolution spectra corresponding to the imaging CCDs.

\subsection{Coadding Multiple Observations of the Same Target}

For observations of a single target that have been split over multiple
observing periods, additional processing is necessary to combine the
various exposures into a single dataset.  We accomplish this
by adding the PHA spectra from each observation together after
ensuring that the source or instrument did not experience major
variability between observations.

To determine whether a set of observations should be coadded, we
developed a number of simple tests.  First, we check to see if the
actual source appeared to be variable from one observation to the
next by examining the count rates in both the $0^{th}$-order image
spectrum and the HEG and MEG grating spectra.  The differences in the
total count rates between observations should not be significantly
greater than the combined error in the count rates.  If either the
$0^{th}$-order or gratings spectra count rates have shifted more than
two-sigma between observations, then either the source or the
instrumental effective area has likely changed between observations.

To roughly determine the degree to which changes in the effective area
could have contributed to a shift in the observed count rates from one
observation to the next, we compare the average fractional changes in
the effective area to the previously calculated fractional changes in
the count rate.  As long as the fractional shifts in effective area
are statistically small compared to the fluctuations in the count
rates, changes in the HETG/ACIS-S instrument between each observation
can be deemed insignificant.
 
After confirming that the source did not vary between observations,
the actual coadding process consists of summing together the
$\pm$$1^{st}$ order PHA datasets from each observation in their
original binning.  The wavelength bins for each observation should
coincide; if not, the observations must be rebinned to the same
wavelength grid.  Methods for testing the accuracy of the coaddition are
described in section 3.3.  There are currently four X-Atlas targets
for which the coaddition of multiple observations has been performed
(Table \ref{tblcoadd}).

\subsection{Extracting Spectra from Crowded Star Fields}

A significant proportion of the stars featured in the atlas are
proximate to other X-ray bright sources, presenting both a scientific
bonus and a technical challenge.  Since relatively few stars have been
observed at high-resolution, HETG observations of relatively crowded
star fields such as the Trapezium or Cyg OB2 present valuable
opportunities to enhance the science return of a single observation.
Actually performing multiple spectral extractions often proves
difficult, particularly if the spectral arms or zeroth-order image of a source
fall on the dispersed spectra of another.  In such cases, all sources
must be extracted simultaneously, starting with the tool {\it
tg\_create\_mask}.\footnote{http://cxc.harvard.edu/ciao/ahelp/tg\_create\_mask.html}
 Otherwise, artifacts may be present in the extracted spectra.  In
general, we recommend that X-Atlas users extract any spectra of
interest themselves before undertaking detailed spectral analysis,
especially for stars situated in crowded fields.

For the initial release of X-Atlas, we have extracted only the primary
target of the observation, usually the brightest star close to the ACIS-S
aimpoint.  We are beginning to extract secondary and tertiary sources
from observations of clusters such as the Trapezium or Cyg OB2, using
the {\it SIMBAD} database to identify bright sources near the
aimpoint.  We plan to limit the extractions to sources of more than
300 $0^{th}$-order counts located no more than 4' off-axis.  Beyond
4', the ACIS point-spread function broadens dramatically and the 90\%
encircled energy radius rises well above 1'' \citep{cxc05}.

\subsection{Observation Summary}

Although the X-Atlas archive will eventually contain all point
sources observed with HETG or LETG, this article is intended to only
cover stellar sources currently in the atlas.  At the time of
publication, X-Atlas consisted of all archived stellar HETG
observations, including the more than 130 observations shown in Table
\ref{tbl1}, with the exception of observations that suffered
instrumental malfunctions.

\subsection{X-Atlas Website Interface}

X-Atlas was conceived of as a public tool for research, and a primary
goal of the project was to make the data and analyses described here
readily available through a user-friendly web interface.  We sought to
present the data in a form intelligible to people unfamiliar with X-ray
spectra while still offering useful capabilities to the expert X-ray
astronomer.

The X-Atlas website is fully documented, with the main page containing
links to descriptions of the data processing procedures and threads
for navigating the website.  Links to various {\it Chandra X-ray
Center} pages, other stellar catalogs, and X-ray emission line lists
are also provided.  For an overview of the atlas's contents, a listing
of all observations, ordered by spectral class, can be pulled up, and
the proposal abstract and any publications associated with a
particular observation can be viewed.

\begin{figure}
\includegraphics[width=3.1in]{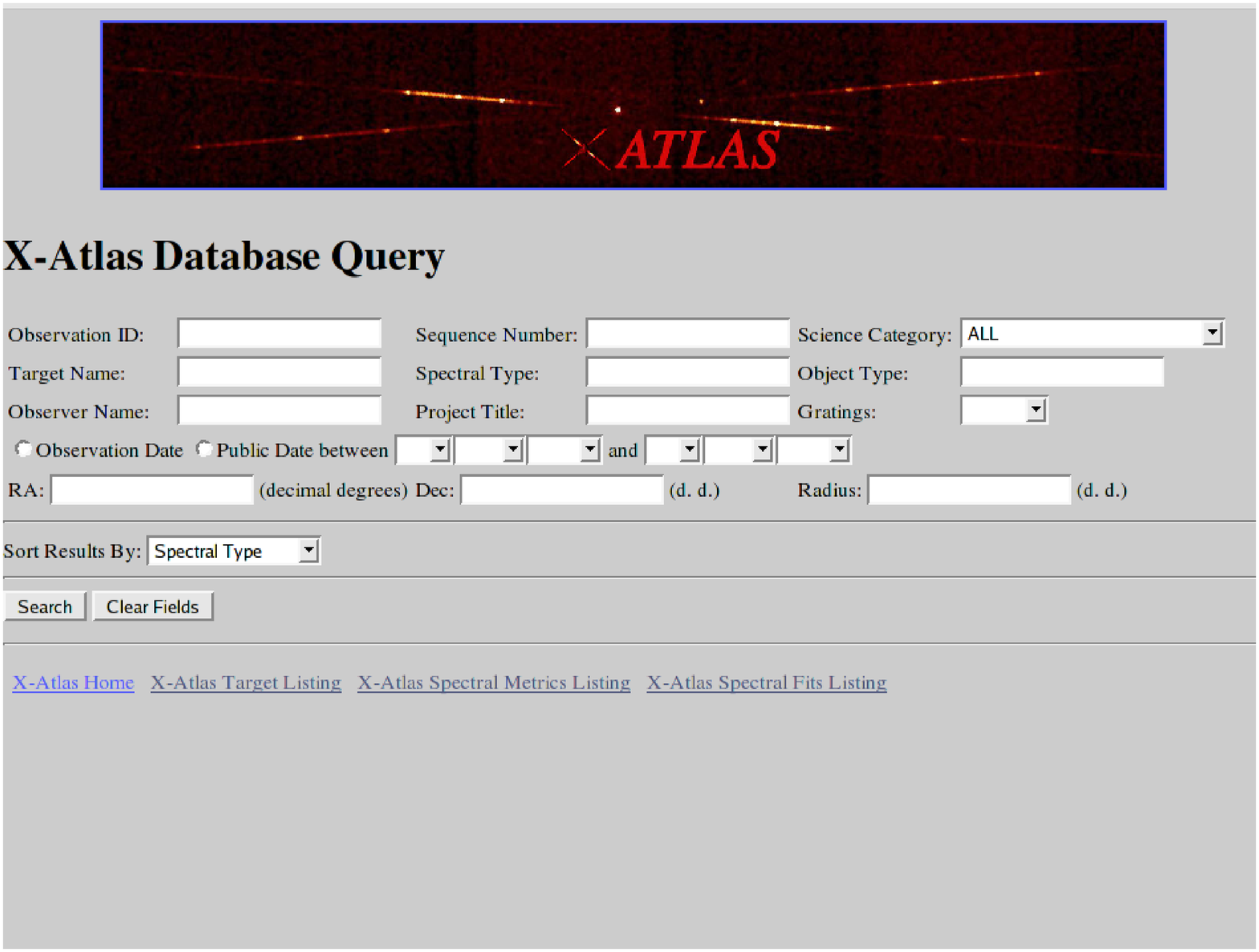}
\includegraphics[width=3.1in]{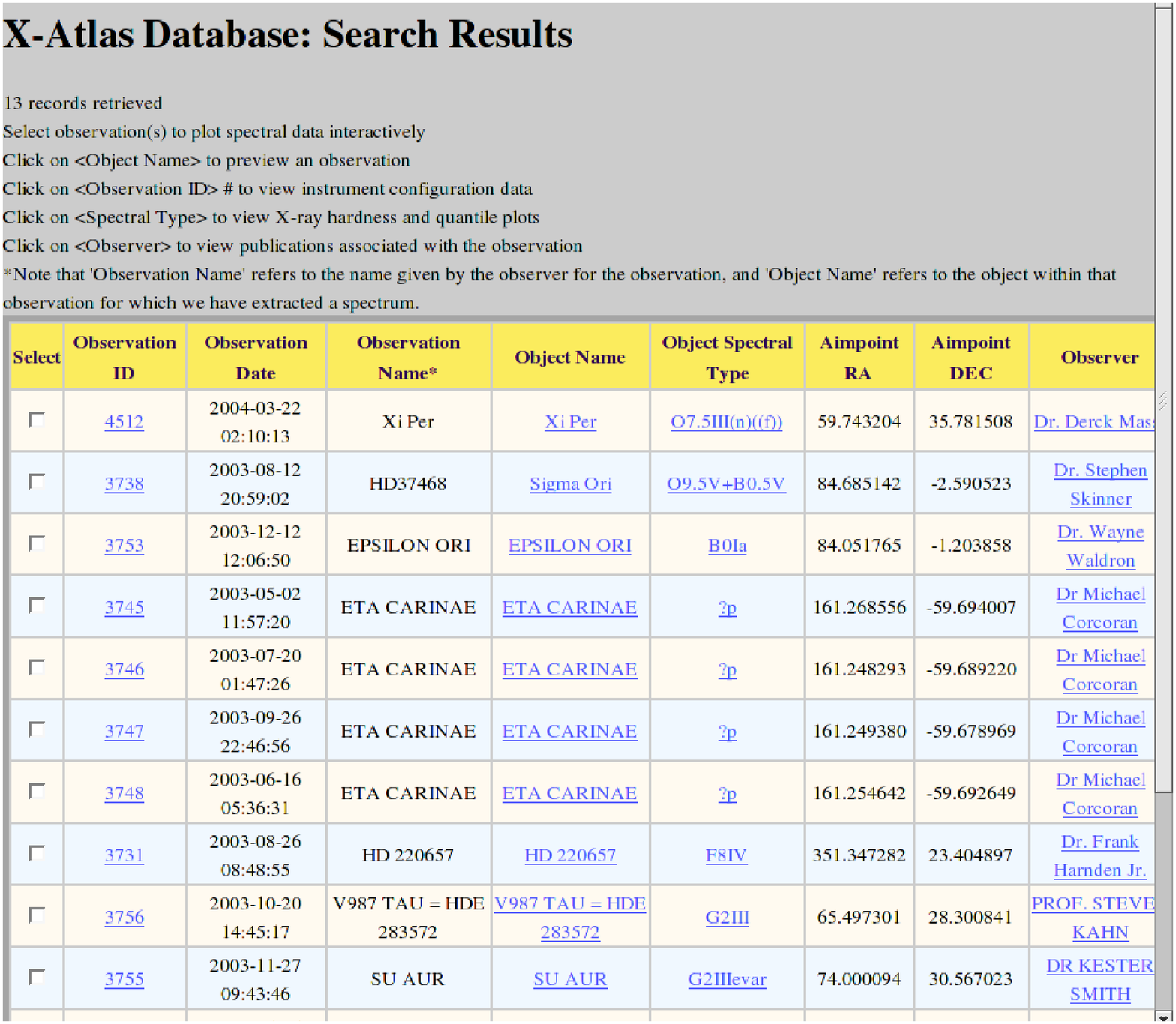}
\newline
\includegraphics[width=3.1in]{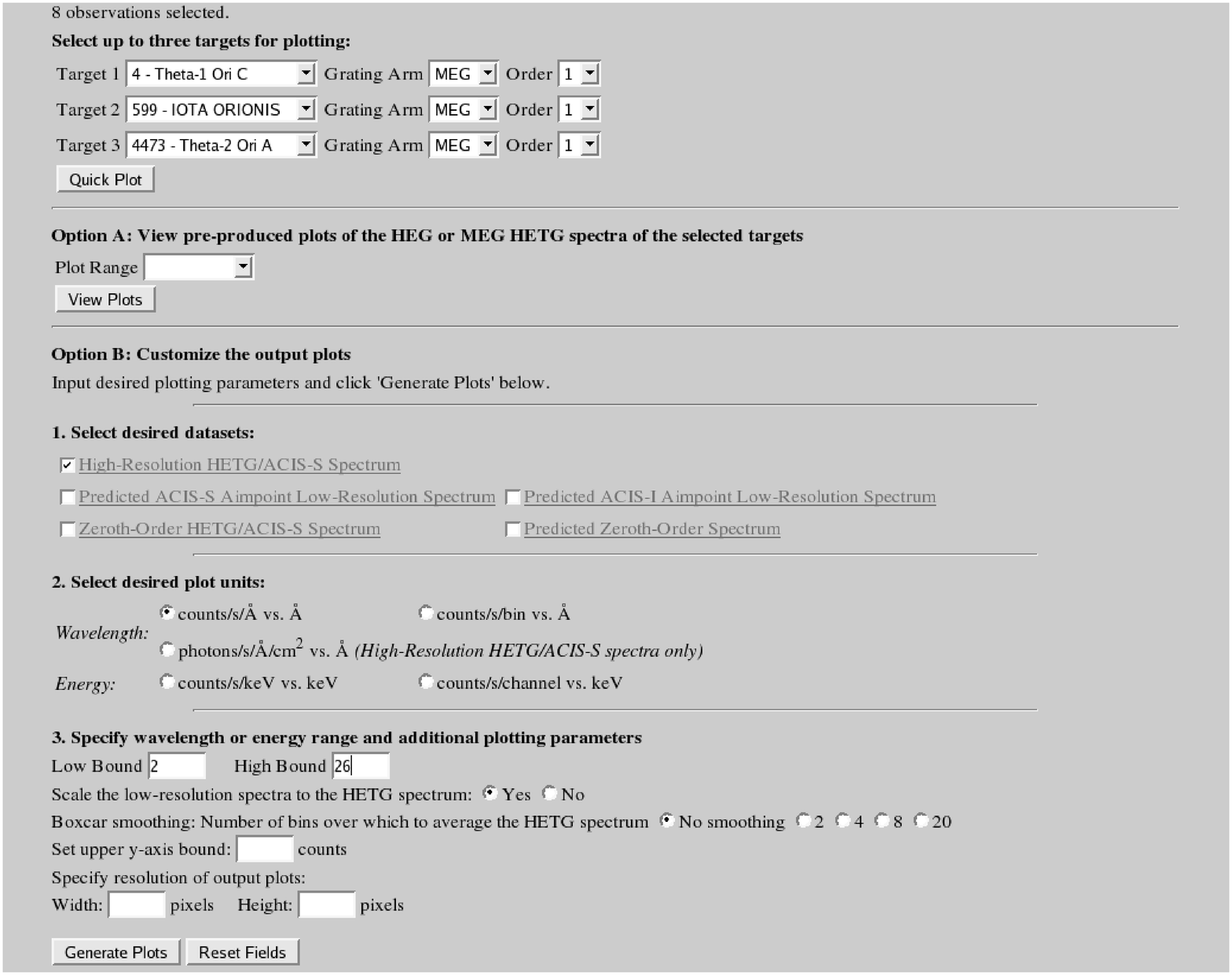}
\includegraphics[width=3.1in]{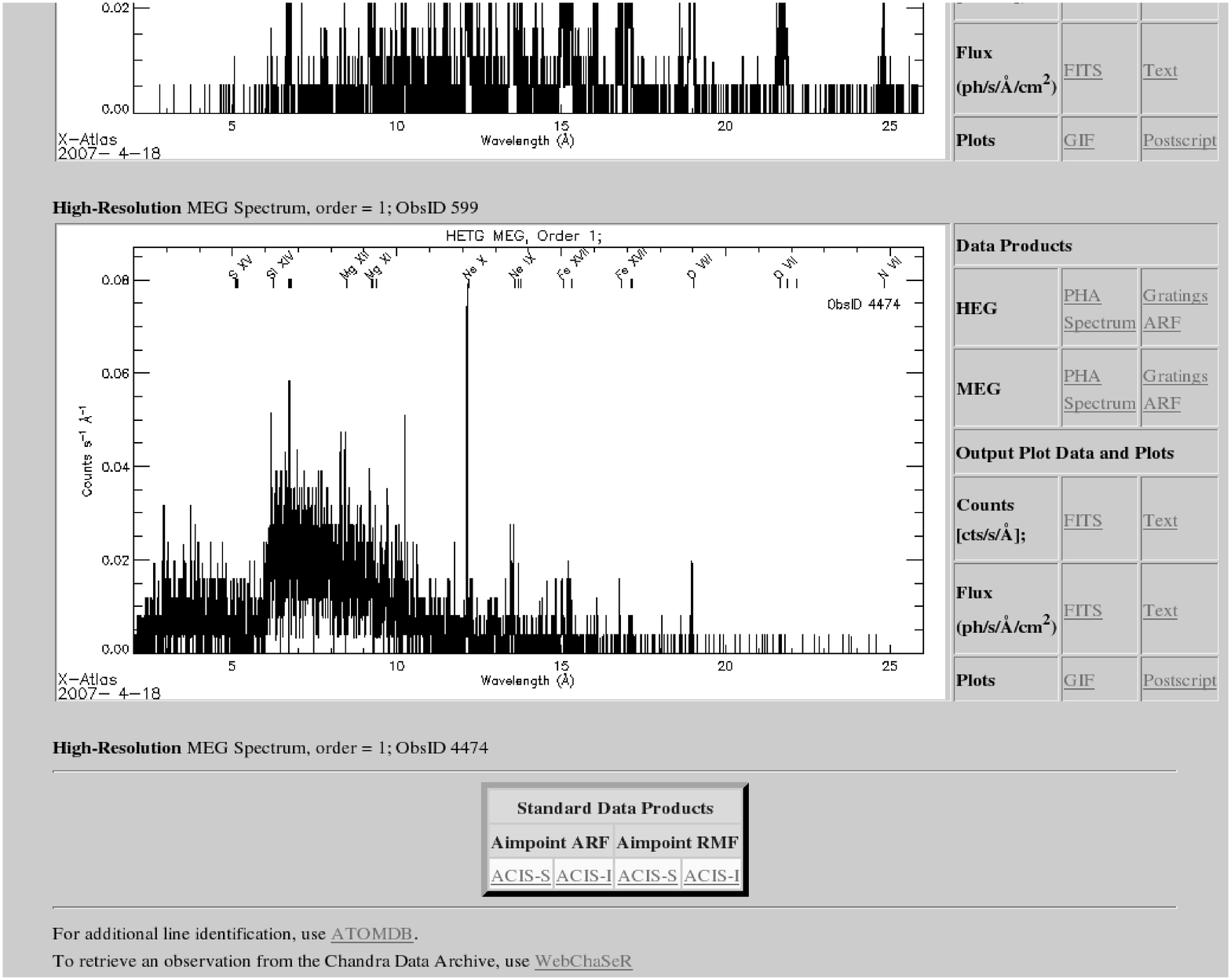}
\caption{\scriptsize The X-Atlas web interface. The database can be
searched on a variety of observation and target parameters ({\it top
left}) and the basic astrophysical parameters, spectra, X-ray color
data, and variability of sources in the observations can be previewed
({\it top right}).  After selecting observations of interest,
customized spectral plots of up to three sources of interest can be created ({\it bottom left}) and both the output plots and
spectral data for those sources can be retrieved, including the high-resolution HEG and
MEG spectra and the predicted low-resolution ACIS-S and ACIS-I spectra.
 \label{figxweb}}
\end{figure}

The central purpose of the website is to provide an interface for
searching, plotting, and retrieving the spectral data products
contained in the atlas (Figure \ref{figxweb}).  The search interface
is modeled after the query pages of the $Chandra$ Data Archive.
The database may be queried by observation or target parameters such
as observation ID, target name, spectral type, object type, science
category, and observation date, or through a cone search of the target
RA and declination. Results may be sorted by spectral type,
observation ID, target name, observation date, RA, or declination.

For those observations retrieved in the search, the instrumental
configuration of the observation, hardness and quantile data, known
physical parameters of the target, spectral fits of the $0^{th}$-order
and predicted low-resolution spectra, light curves and variability
analyses of the observation, the $0^{th}$-order image of the extracted
target, and preview plots of the HEG and MEG spectra, are readily
viewable.  The spectral data products for the observation are also
made available for download.  Observations returned by the search may
be marked for interactive, customized spectral plotting and data retrieval.

Three sample uses of the X-Atlas website are outlined below.  Note
that the procedures for generating the data products in the atlas are
discussed below (see \S\ref{s:lores} and \S\ref{s:discuss} ).  Although these tasks are by no means a
comprehensive list of the atlas's capabilities, we believe that three
particularly useful applications of the website will be to view target
preview information for individual stars, create spectral plots for
the comparison of multiple targets, and retrieve and compare spectral
metrics such as hardness ratios and quantiles for a large sample of the
observations in the database.

\subsubsection{Retrieving Target Information}

For the first task, we will retrieve target preview information,
including spectral plots, hardness ratios, quantiles, spectral fits,
and lightcurves of a star of interest, using FK Com as an example.  We
start at the X-Atlas search page, entering 'FK Com' as the target
name, and run the search.  We may also search the atlas without any
search filters to retrieve all observations in the database and
select FK Com from the list of observations returned.

In either case, after executing the search, we will find an entry for
FK Com, Obsid 614, among the observations retrieved.  Clicking on the
Observation ID will call up $Chandra's$ instrumental configuration during the
observation.  Clicking instead on the target name (FK Com) will bring
up a preview of the target extracted from of the observation.  Under
some basic target information is a snapshot of the
$0^{th}$-order image along with the $0^{th}$-order count rate and a rough
estimation of the degree of pileup for a gauge of the reliability of
the $0^{th}$-order data.  Below is a listing of predicted hardness
ratios and quantiles (see Section 4.1) based on the predicted ACIS-S
and ACIS-I spectra of FK Com (see Section 3), followed by plots of the spectral
fits of the $0^{th}$-order and predicted ACIS-S and
ACIS-I spectra (see Section 3.4).  The relevant fit parameters are
listed as well, including N$_{H}$, two values of kT, metal abundances,
total flux, the flux emission ratio of the two kT components, and $\chi^{2}$ values of
the fit.  Below the spectral fits is a light curve constructed from
the HETG HEG and MEG data and binned at intervals of 1200 and 4000 s.
Variability results from GLvary (described in section 4.2) accompany
the plot.  Following the light curve are preview plots of the $1^{st}$
order HEG and MEG spectra in units of counts/s/\AA\ vs. \AA.  Those
wishing to create their own plots of the spectra can simply click on
the 'Customize Spectral Plots' button provided on the page.  Finally,
at the bottom of the preview page, various 
spectral data products are offered for download, including PHA files
for the HEG and MEG spectra, FITS and text files of the reduced HEG
and MEG spectra, the predicted low-resolution ACIS and $0^{th}$-order spectra, and the
actual $0^{th}$-order spectrum, and the accompanying gARFs and aimpoint ARFs and RMFs.

\subsubsection{Customizing Spectral Plots}

A second, more powerful use of the atlas is to create comparison plots
of up to three targets at once.  For the following example, we will compare
the MEG spectra and the Fe XVII emission lines for three T Tauri stars:
SU Aur, Speedy Mic, and TV Crt.  To do so,
we return to the main X-Atlas search page, enter 'T Tauri' as
the object type and 'Normal stars and WD' as the science category, and
execute the search.  This particular search will return over ten
observations of T Tauri stars, ordered by spectral type by default.
As we scroll through the list, we can click on the checkboxes in the
leftmost column for all observations of SU Aur (ObsID 3755), Speedy
Mic (ObsIDs 2536 and 3491), and TV Crt (ObsID 3728).  After selecting
the appropriate observations, we click on 'Select Observation(s) for
Plotting' to call up the plotting interface.

The plotting interface allows for the creation of customized spectral
plots for up to three targets at once.  Three plotting options
are offered.  The first, marked as 'Option A,' is to view pre-produced
plots of the high-resolution HEG or MEG spectra over one of a number
of pre-determined spectral ranges in units of counts and wavelength
(counts/s/\AA\ vs. \AA).  Alternatively, the output plots may be customized through a variety of
plotting options.  The most flexible option, 'Option B,' allows the
plotting of multiple datasets of differing spectral resolution and the
choice of plot units and spectral range of the output.  The available
datasets include the high-resolution HETG/ACIS-S HEG and MEG spectra,
the predicted low-resolution ACIS-S or ACIS-I Aimpoint data, and both
the extracted and predicted low-resolution HETG/ACIS-S $0^{th}$-order
spectra.  These data may be plotted in units of counts/s/\AA\ or
counts/s/bin vs. \AA, in counts/s/keV or counts/s/bin vs. keV, or, for the HETG
spectra only, in flux units of ph/s/\AA/cm$^{2}$ vs. \AA. Optional parameters
may be set to scale the low-resolution spectra, boxcar smooth the HETG
spectra, define upper bounds for the y-axis, or control the pixel
resolution of the output plots.  These plots are generated with
identifications of common spectral lines provided.

To plot the MEG spectra of SU Aur, Speedy Mic, and TV Crt from 5 \AA\ to 25 \AA, we
begin at 'Option B' on the webpage.  First, up to three
desired targets are selected, and for each, the grating arm
(HEG or MEG) and order (1, 2, or 3) of interest.  For this example, we
will select the $1^{st}$ order MEG spectra of SU Aur (ObsID 3755),
Speedy Mic (ObsID 2536), and TV Crt (ObsID 3728).  The high-resolution
HETG/ACIS-S spectrum is already selected as the default dataset for step 2 and
counts/s/\AA\ vs. \AA\ is already selected as the default plot units
for step 3.  In step 4, we enter '5' as the low bound and '25' as the
high bound and click on the button to 'Generate Plots.'  To instead
view the predicted ACIS-S and ACIS-I spectra and plot in energy space,
we deselect the high-resolution HETG spectrum and select the predicted
ACIS spectra instead, choosing counts/s/keV vs. keV as the plot units.
The plot range may be either specified or left blank to use the
default values. Any of the low-resolution datasets may also be scaled
to the HETG spectrum for viewing on the same plot.

\subsubsection{Emission Line Plots}

Besides viewing pre-produced plots or specifying the dataset, plot
units, and plot range, a final plotting option is offered for the more
specific task of making comparisons of emission line profiles of the
HEG or MEG spectra.  Up to six common X-ray emission lines can be
chosen from the drop-down boxes provided to create line profile montages in units of
wavelength (counts/s/\AA\ vs. \AA or counts/s/bin vs. \AA), energy
(counts/s/keV vs. keV), or Doppler velocity (counts/s/\AA\ vs. km/s).
For instance, to view the Fe XVII emission lines for SU Aur, Speedy
Mic, and TV Crt in wavelength units (counts/s/\AA\ vs. \AA), we
leave all of the fields under 'Option B' blank and instead proceed to
'Option C.'  Using any five of the six drop-down boxes, we
select the Fe XVII lines at 15.01, 15.26, 16.78, 17.05, and 17.10
\AA.  We then make sure that the default plot units, counts/s/\AA\
vs. \AA, are selected.  We may specify the distance from line center
of each line profile plot and choose between viewing all lines of a
given star on the same scale or plotting each line completely
independently.  When we are satisfied with the settings, we may click
'Generate Plot.'

The interactively customized plots take only a few seconds to create.
The plots are then displayed on the screen along with options to
download the plots in GIF, JPEF, and/or postscript form.  However, we intend the
plotting interface to be used mainly as a browsing tool rather than an
end in itself.  Therefore, the datasets used to generate the plots are
provided for download in FITS files and ASCII text tables, and the HEG
and MEG PHA spectral files for the target are offered as well, along
with the appropriate grating ARFs and the ACIS-S and ACIS-I aimpoint
ARFs and RMFs.

\subsubsection{Retrieving X-ray Properties of the X-Atlas Dataset}

Besides the capabilities for viewing the observation data and analysis products for
individual observations or creating customized spectral plots, the
X-Atlas website also provides the means to make quick comparisons of
the X-ray properties of the entire X-Atlas dataset.  For any subset of
observations in the database, hardness ratios and energy quantiles
derived from the predicted ACIS-S and ACIS-I spectra and spectral
metrics calculated from the HETG data may be retrieved by clicking on
the respective links for 'X-Atlas Hardness Ratios and Quantiles' and
'X-Atlas Spectral Metrics' on the X-Atlas home page, the observation
search page, or any of the target preview pages.  These links will
call up tables of these metrics for every X-Atlas observation that can
be filtered by the values of each metric and sorted by a variety of
target and  observation parameters as well as by any of
the individual metrics themselves.  A similar page is provided for the spectral
fits of the predicted ACIS spectra.

\subsection{Sample High-Resolution Spectra: Spectral Montages}

X-Atlas was conceived as a means of enabling comparisons of stellar
spectra across spectral type, hardness, or other characteristics
of interest.  For this purpose, we have constructed montages of the
HETG spectra of both high-mass and low-mass stars to accompany the
dynamic plotting capabilities of the website.  Only MEG spectra
are displayed below, but both HEG and MEG montages will be available on
the website, along with montages of the convolved low-resolution
spectra shown and described in Section 3.  The spectra in the montages
are displayed in counts s$^{-1}$ \AA$^{-1}$ vs. \AA\ and are normalized for the
purpose of comparison.

\begin{figure}
\includegraphics[width=6.2in]{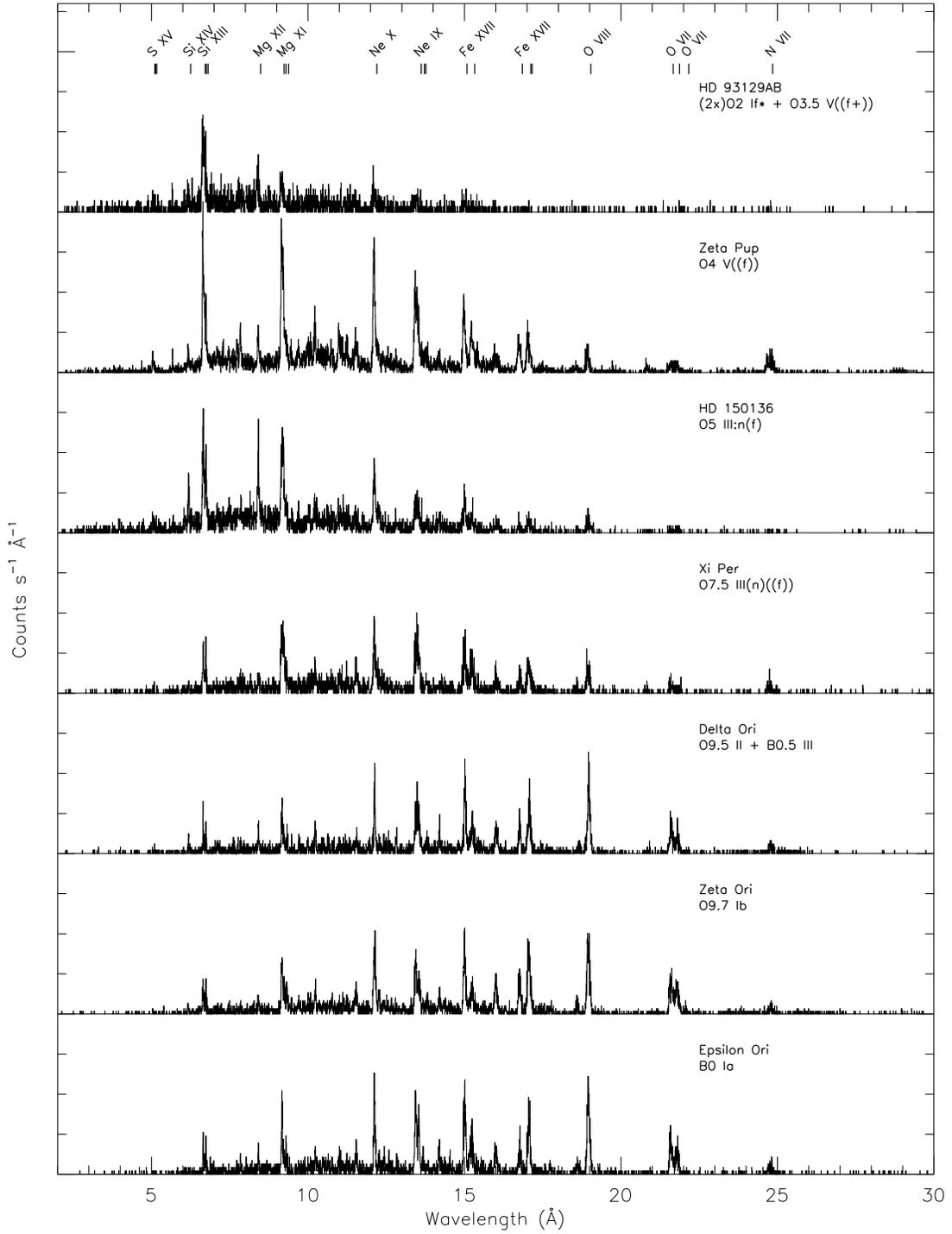}
\caption{\scriptsize Montage of the high-resolution MEG spectra ([counts/s/\AA]) of
normal OB supergiants and giants, ordered by spectral class.
The locations of some strong lines are marked.
The spectra from different sources are offset for clarity.  \label{figobsgh}}
\end{figure}

\begin{figure}
\includegraphics[width=6.2in]{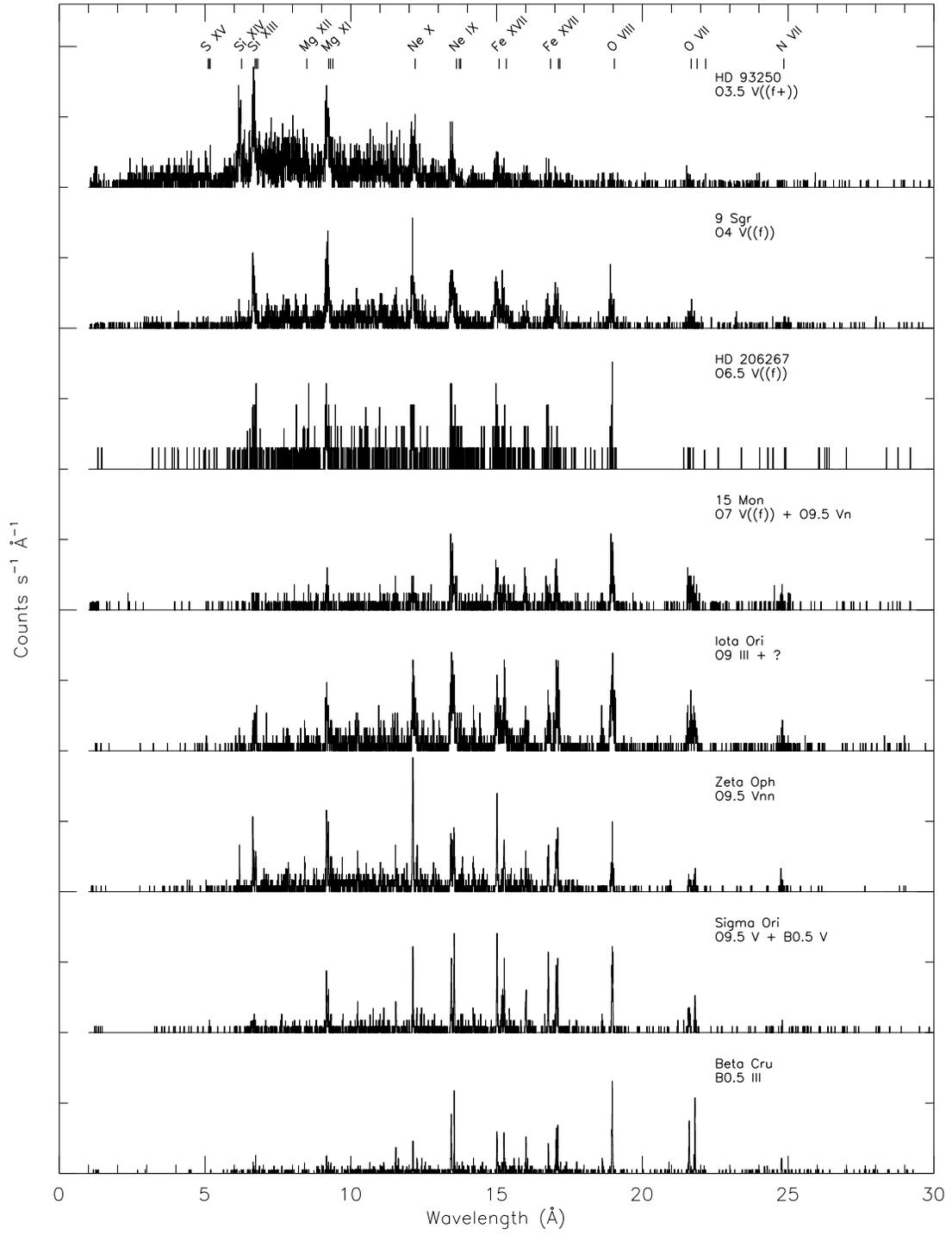}
\caption{\scriptsize As in Figure~\ref{figobsgh}, for
normal OB giants, sub-giants and dwarfs.\label{figobmsh}}
\end{figure}

\begin{figure}
\includegraphics[width=6.2in]{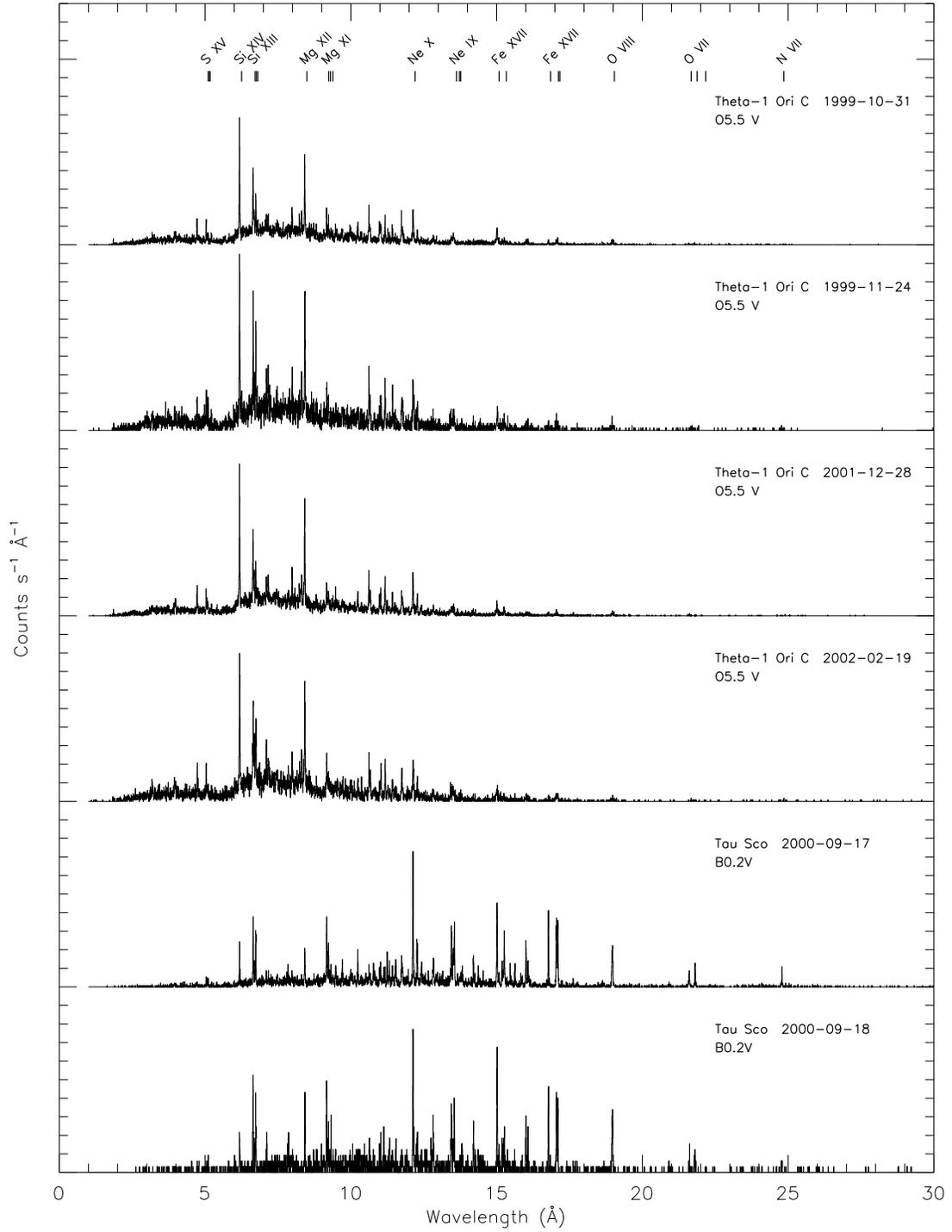}
\caption{\scriptsize As in Figure~\ref{figobsgh}, for
four observations $\theta^{1}$ Ori C and two observations
of $\tau$ Sco, two high-mass stars thought to possess magnetic fields
responsible for generating X-rays. \label{figobmagh}} 
\end{figure}

\begin{figure}
\includegraphics[width=6.2in]{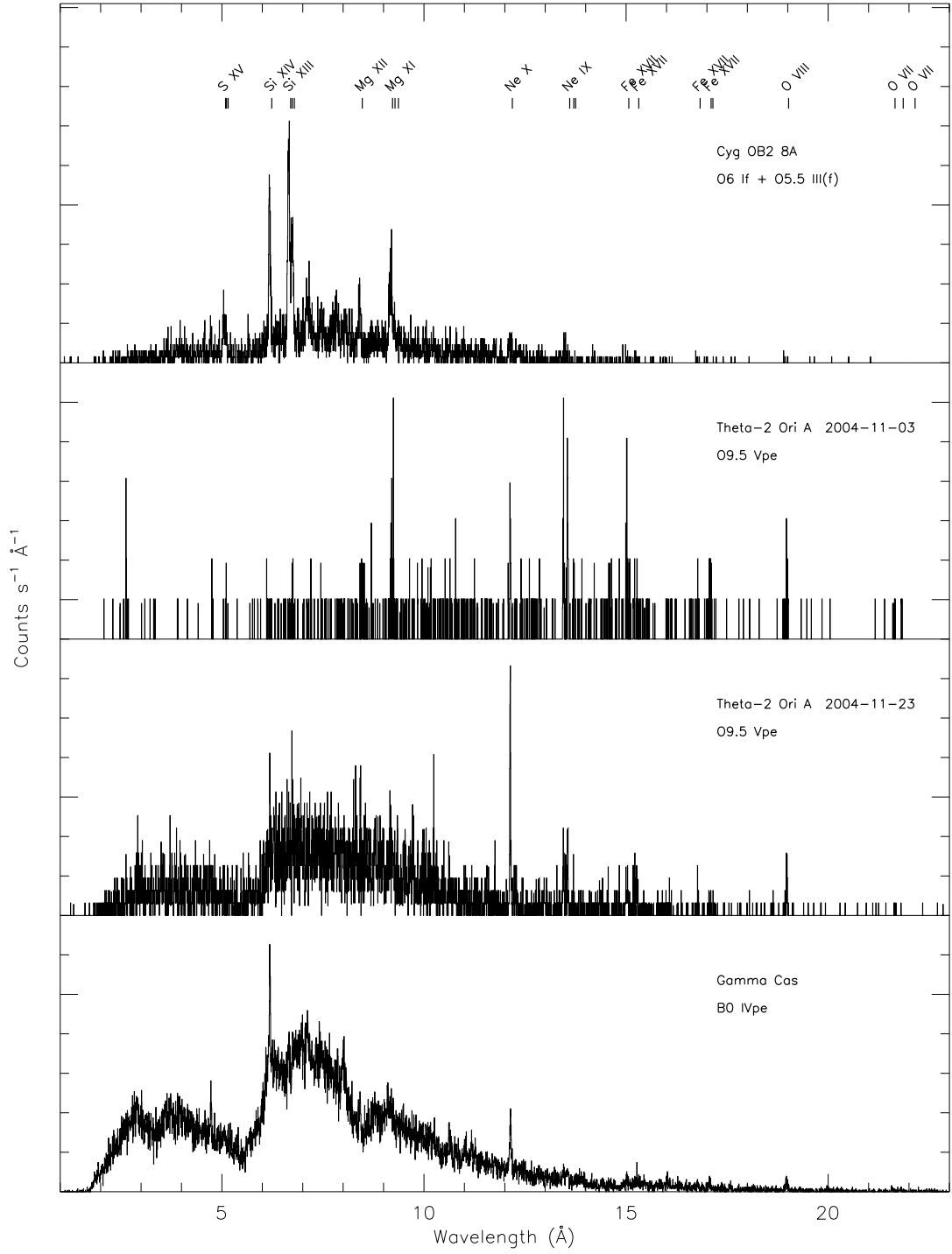}
\caption{\scriptsize As in Figure~\ref{figobsgh}, for
three peculiar OB stars, $\gamma$ Cas, Cyg OB2 8A, and $\theta^{2}$
Ori A, approximately ordered by X-ray hardness, descending from
hardest to softest. \label{figobpech}}
\end{figure}

\begin{figure}
\includegraphics[width=6.2in]{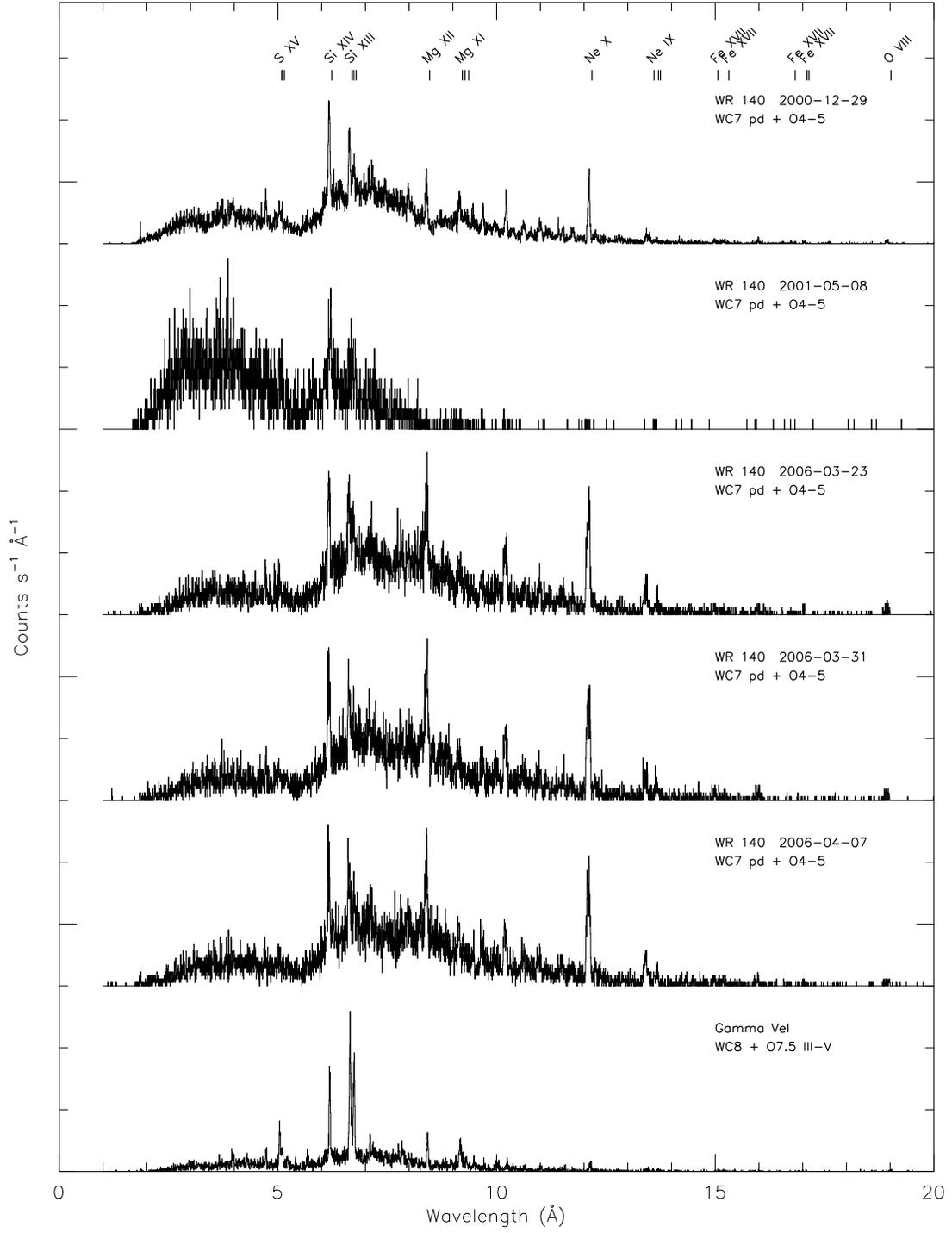}
\caption{\scriptsize As in Figure~\ref{figobsgh}, for
two Wolf-Rayet stars observed thus far with HETG. \label{figwrh}}
\end{figure}

\begin{figure}
\includegraphics[width=6.2in]{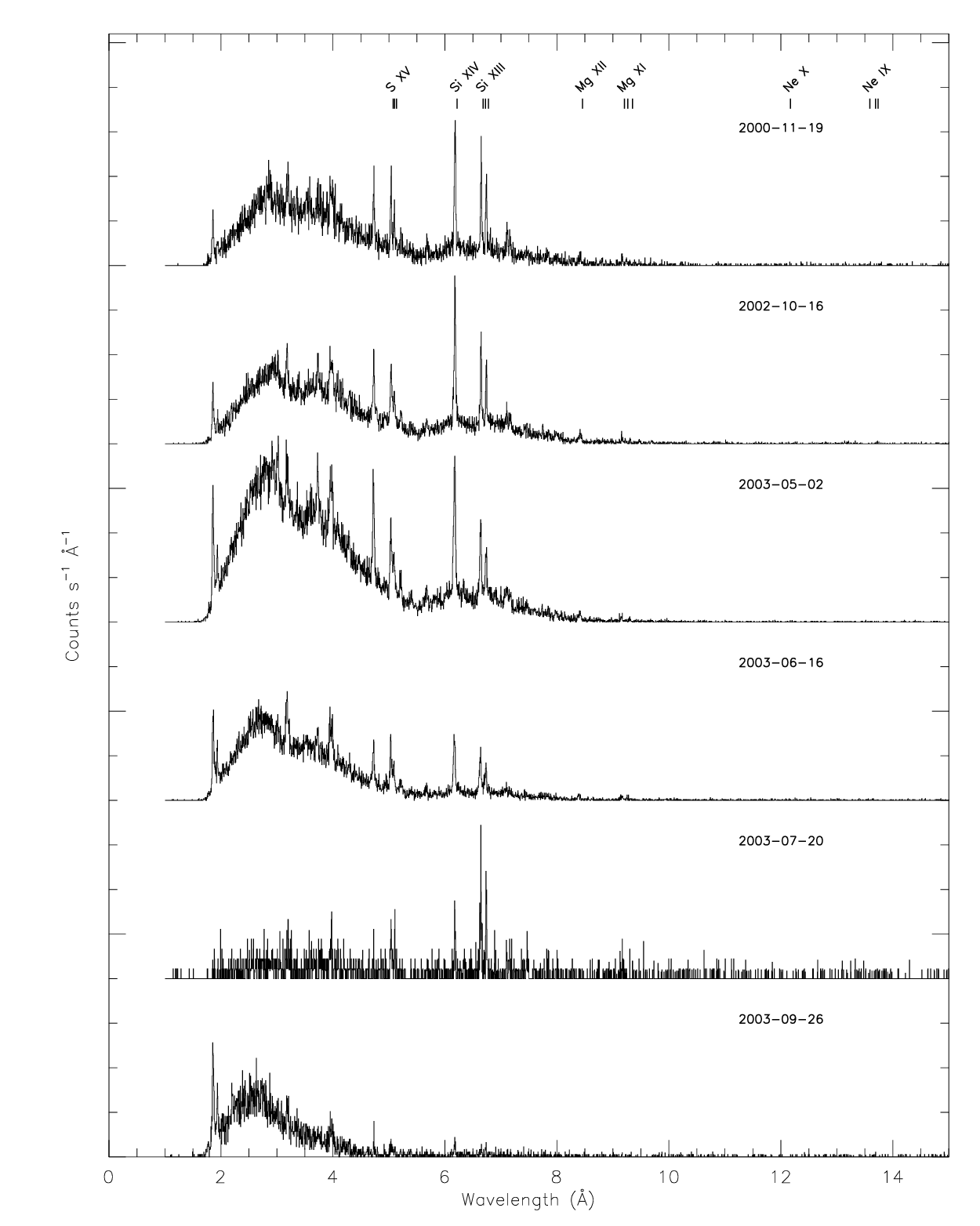}
\caption{\scriptsize As in Figure~\ref{figobsgh}, for
six HETG observations of $\eta$ Carinae. $\eta$ Carinae underwent an
X-ray eclipse in mid-2003: the 2003-05-02 observation was made close to
X-ray maximum, and the 2003-07-02 observation was taken near X-ray
minimum \citep{cor05}. \label{figetah}}
\end{figure}

Normal OB stars are divided into one montage of supergiants and giants
(Figure \ref{figobsgh}) and another of giants, sub-giants, and dwarfs
(Figure \ref{figobmsh}), both ordered by spectral type.  More exotic
high-mass stars are displayed as well.  A separate montage is offered
of two stars, $\theta^{1}$ Ori C and $\tau$ Sco, in which a magnetic
field clearly plays a role in X-ray emission (Gagne, et al. 2005; 
Donati, et al. 2007; Figure \ref{figobmagh}).
Several peculiar OB stars with unconfirmed X-ray emission mechanisms
are shown as well in Figure \ref{figobpech}.  Two Wolf-Rayet stars are
shown in Figure \ref{figwrh}, including $\gamma$ Vel and five
observations of WR 140.  A series of spectra of $\eta$ Car passing
through periastron was obtained by \citet{cor05} and is displayed in
Figure \ref{figetah}.

The discovery of X-ray emission from high-mass stars in the Carina
nebula \citep{sew79,sew82} and in the Cyg OB2 association
\citep{har79} with the {\sl Einstein} satellite in the Carina nebula
\citep{sew79,sew82} and in Cyg OB2 \citep{har79} came as a surprise.
Since then, studies of the energy distributions of massive stars,
their spectra, and their line profiles have continued to enlighten and
puzzle us.  Displaying normal OB stars in montages as a function of optical spectral type
reveals a clear shift of mean X-ray flux to lower energies at lower
photospheric temperatures for both the supergiants and giants (Figure \ref{figobsgh}) and giants,
sub-giants, and dwarfs (Figure \ref{figobmsh}).  These findings were also
noted by \citet{walb06}.  As he discusses, the constant ratio between
bolometric and X-ray luminosities implies such a spectral progression
since bolometric luminosity and temperature are related.  Stellar
winds correlated with luminosity are likely to be the connection with
X-ray luminosity.

\begin{figure}
\includegraphics[width=6.2in]{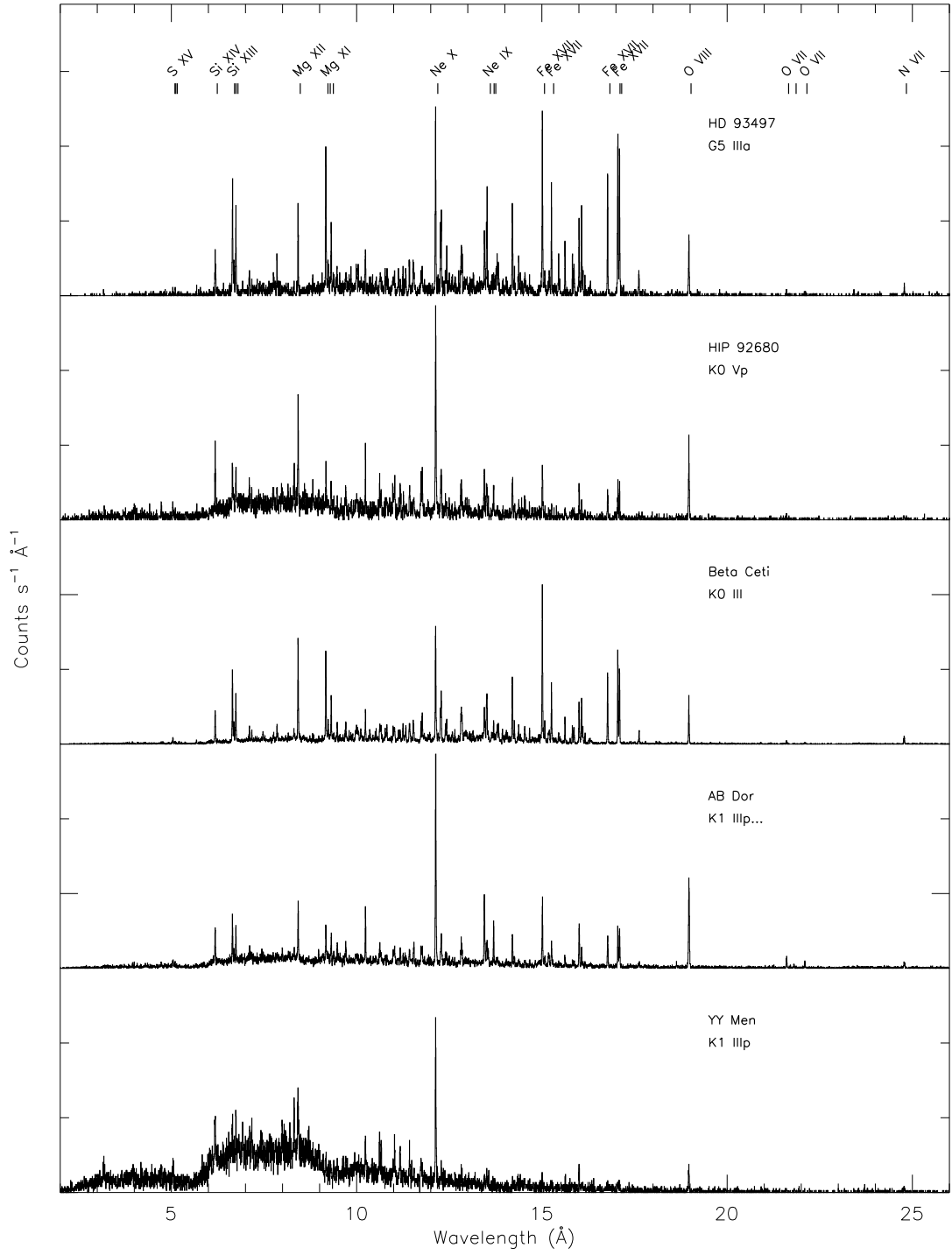}
\caption{\scriptsize As in Figure~\ref{figobsgh}, for
normal and rapidly rotating low-mass stars in the
atlas. \label{figcoolnormh}}
\end{figure}

\begin{figure}
\includegraphics[width=6.2in]{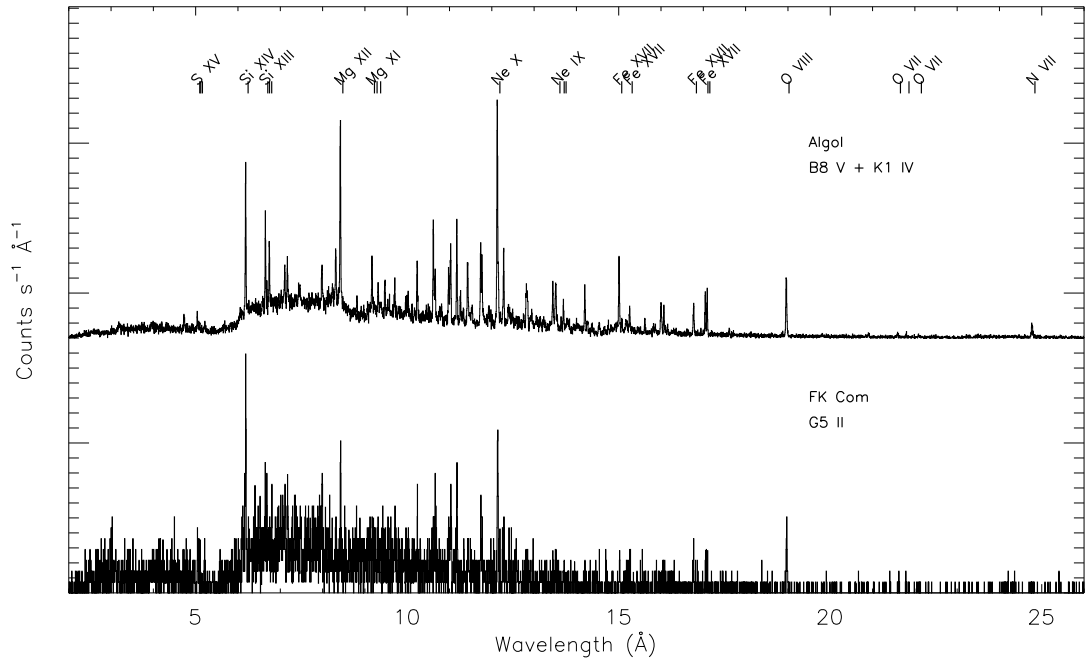}
\caption{\scriptsize As in Figure~\ref{figobsgh}, for
two unique stellar systems, Algol ({\it top}) and FK Com ({\it
bottom}). Algol is a three-star system with a semi-detached B8 V/K1 IV
eclipsing binary in orbit with an A V star.  The X-ray emission is
believed to originate from the K1 star \citep{whi80}.  FK Com is an
active, variable G-type giant rotating near breakup velocity,
possibly a recently coalesced binary \citep{hue93}.\label{figcoolpech}}
\end{figure}

\begin{figure}
\includegraphics[width=6.2in]{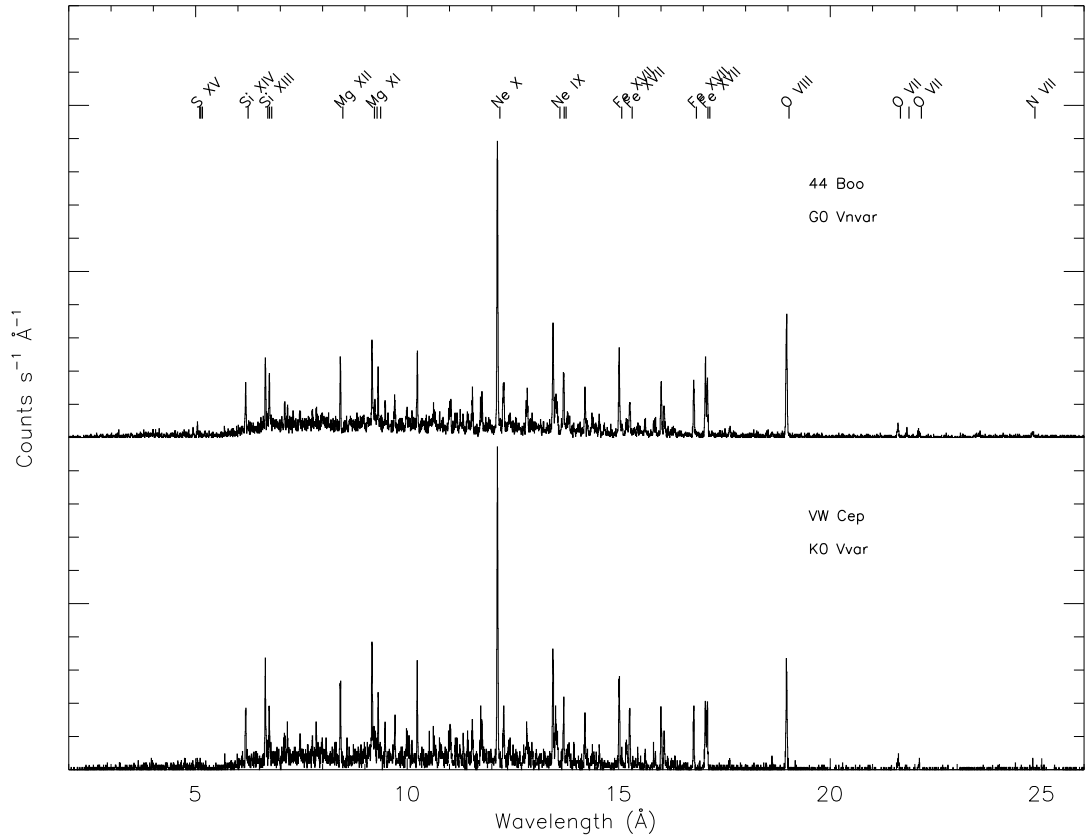}
\caption{\scriptsize As in Figure~\ref{figobsgh}, for
two W UMa--type stars, 44 Boo and VW Cep. \label{figwumah}}
\end{figure}

\begin{figure}
\includegraphics[width=6.2in]{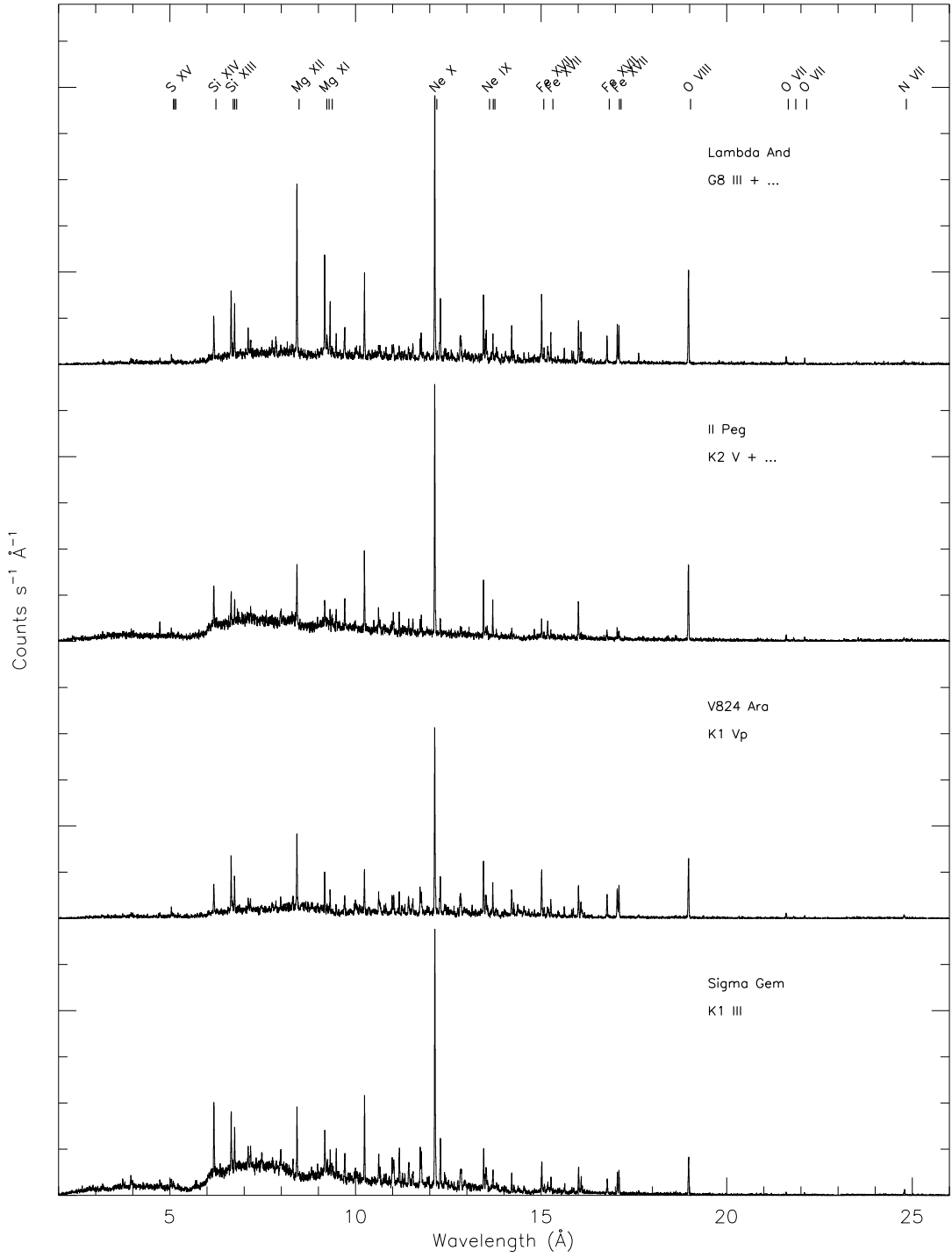}
\caption{\scriptsize As in Figure~\ref{figobsgh}, for
RS CVn stars in the atlas. \label{figrscvnh}}
\end{figure}

\begin{figure}
\includegraphics[width=6.2in]{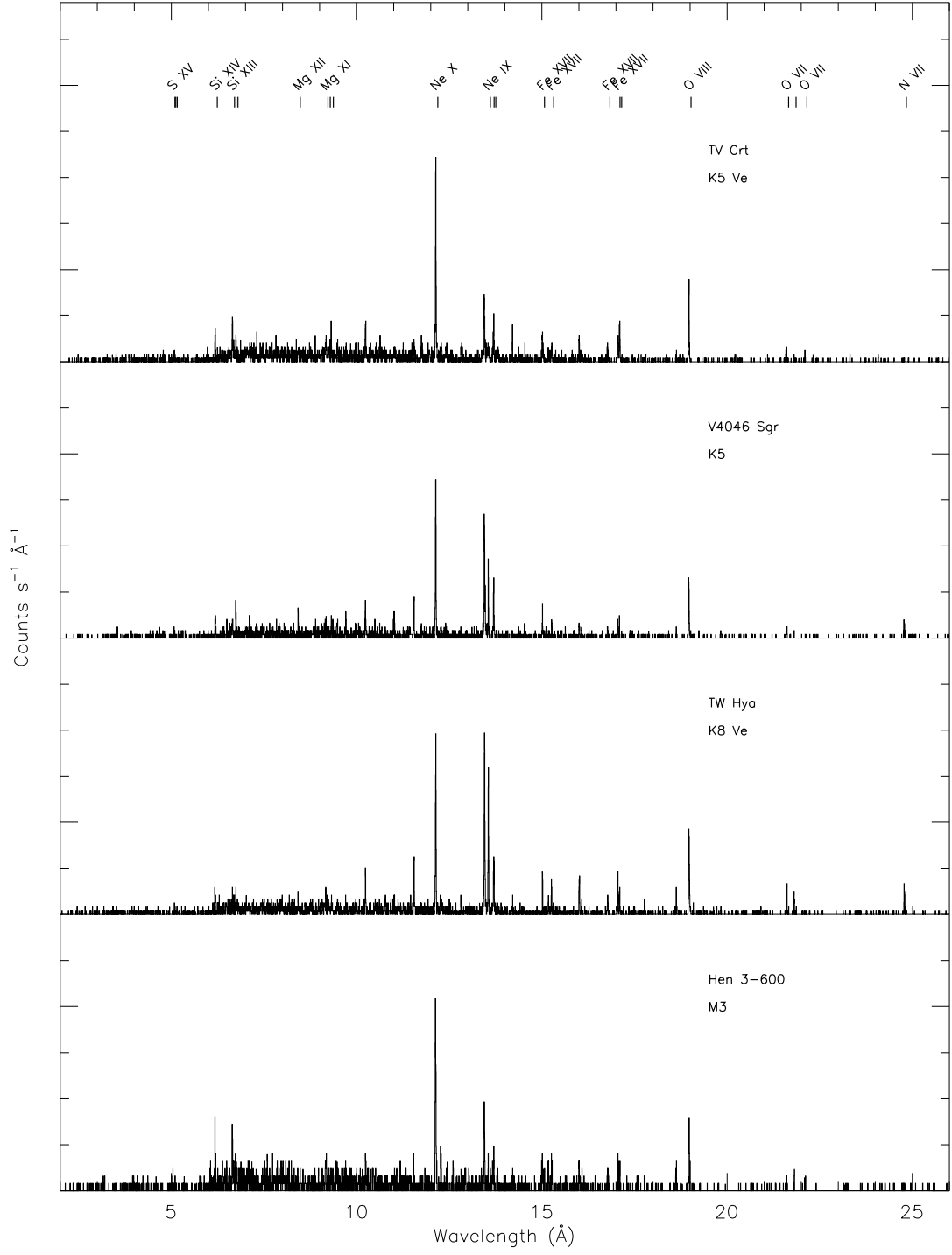}
\caption{\scriptsize As in Figure~\ref{figobsgh}, for
T Tauri-type stars in the atlas. \label{figttsh}}
\end{figure}

\begin{figure}
\includegraphics[width=6.2in]{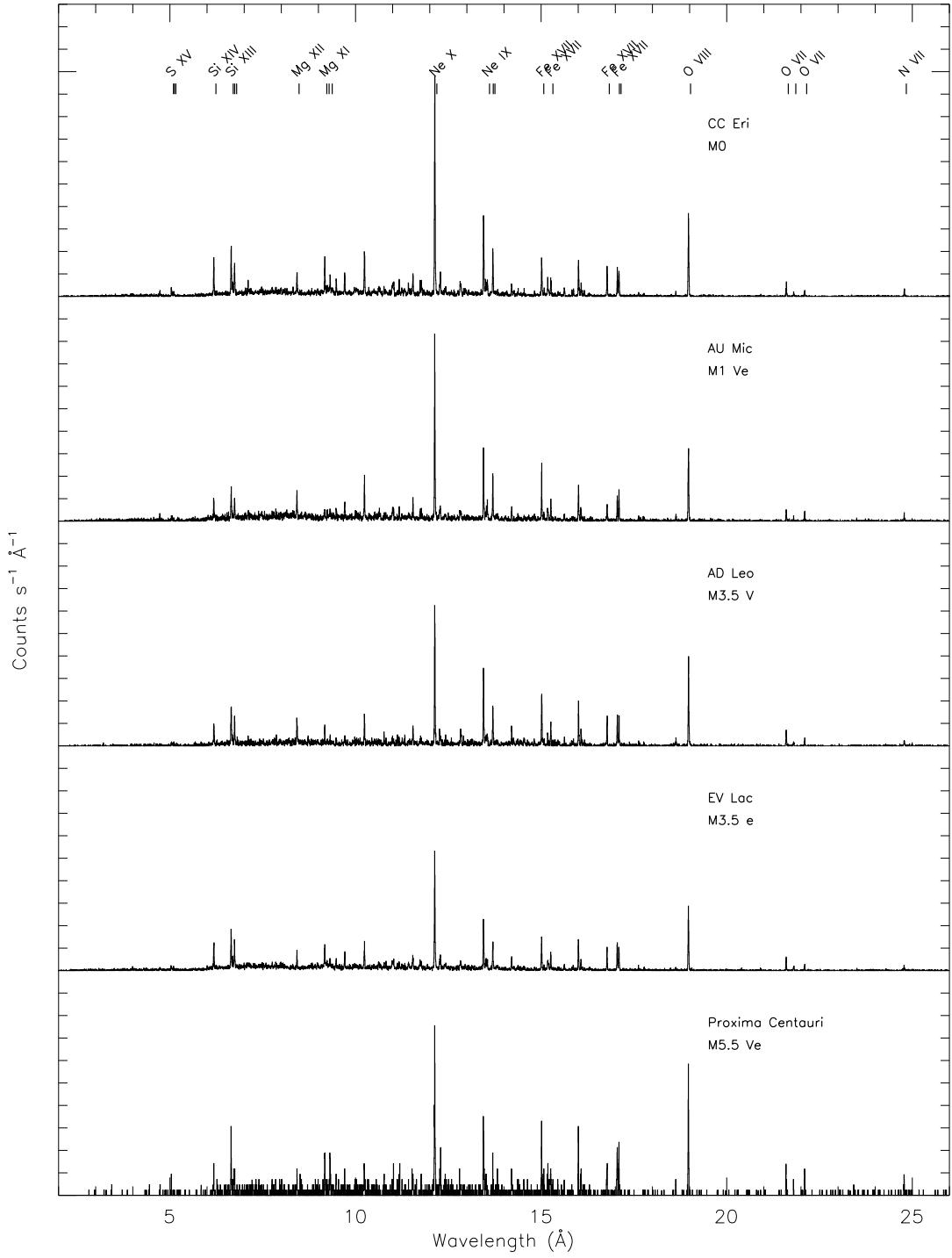}
\caption{\scriptsize As in Figure~\ref{figobsgh}, for
active and flaring low-mass stars. \label{figflareh}}
\end{figure}

Because of the diversity of astrophysical mechanisms at work in
low-mass stars, spectral montages of these stars are organized by
stellar type.  In particular, separate montages have been made of
normal or rapidly rotating low-mass stars (Figure
\ref{figcoolnormh}), W UMa-type stars (Figure \ref{figwumah}), RS CVn
stars (Figure \ref{figrscvnh}), pre-main sequence T Tauri-type stars
(Figure \ref{figttsh}), and active or flaring low-mass stars (Figure
\ref{figflareh}).  Two unique objects, Algol and FK Com, are also
shown (Figure \ref{figcoolpech}).

\subsection{Sample High-Resolution Spectra: Line Profiles}

\begin{figure}
\includegraphics[width=6.2in]{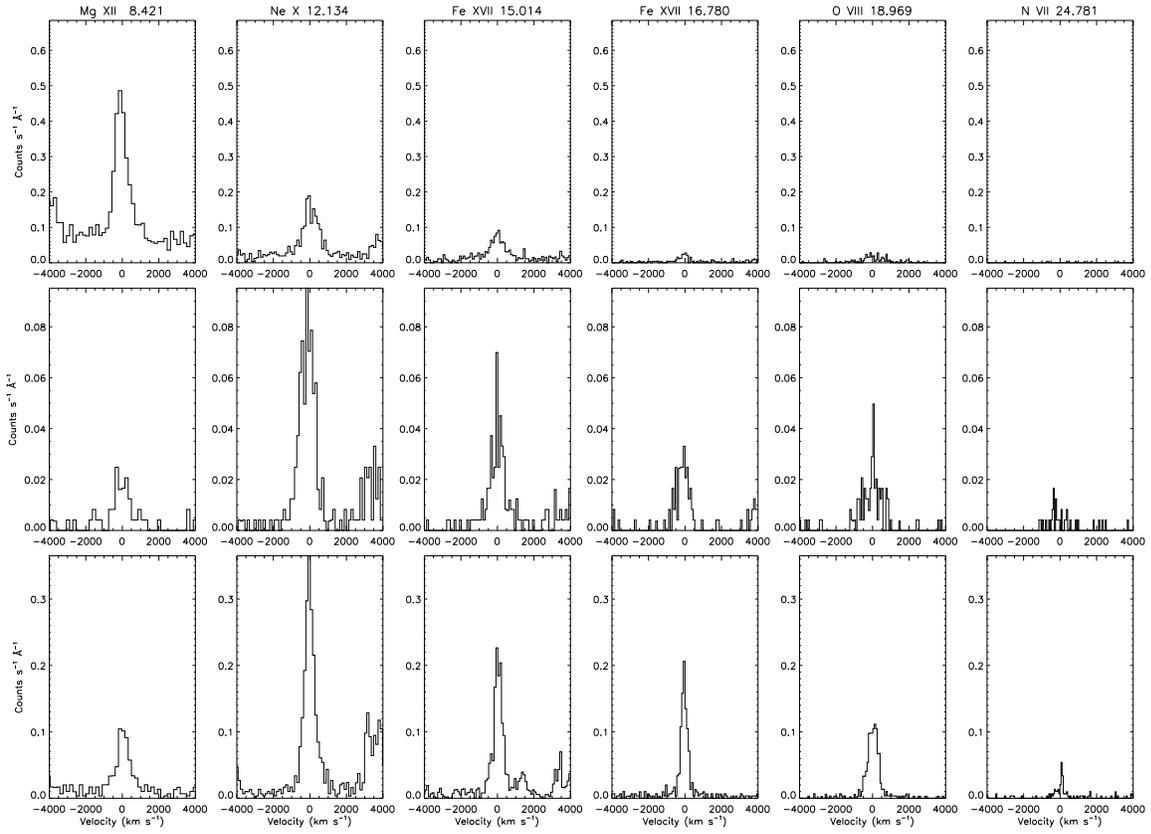}
\caption{\scriptsize Line profiles of selected lines in MEG spectra.
Columns from left to right: Mg\,XII~$\lambda$8.421,
Ne\,X~$\lambda$12.134, Fe\,XVII~$\lambda$15.014, Fe\,XVII~$\lambda$16.780,
O\,VIII~$\lambda$18.969, and N\,VII~$\lambda$24.781.
Descending from top row: $\theta^{1}$~Ori~C, $\zeta$~Oph, and $\tau$~Sco.
\label{figlinprof}.}
\end{figure}

\begin{figure}
\includegraphics[width=6.2in]{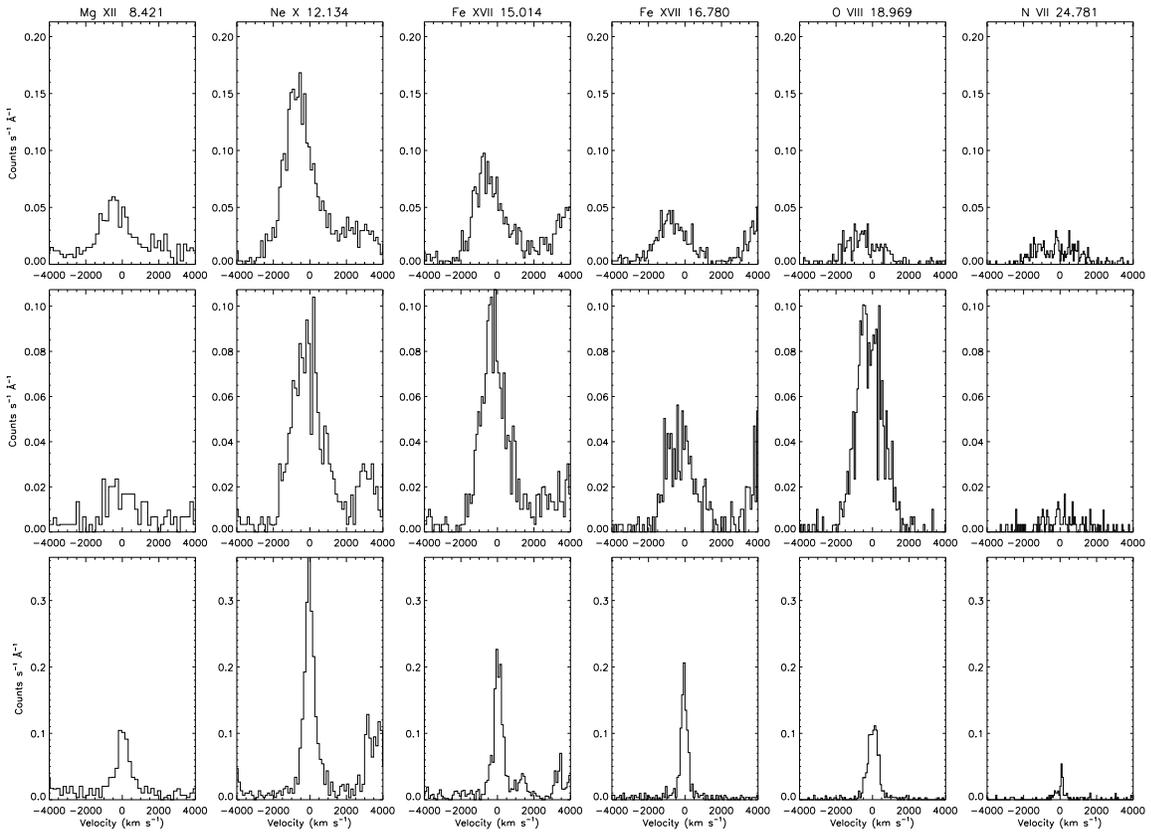}
\caption{\scriptsize As in Figure~\ref{figlinprof}, for the sources,
descending from top row: $\zeta$~Pup,
$\zeta$~Ori, $\iota$~Ori. \label{figlprof2}}
\end{figure}

Line profiles have been a key diagnostic and continuing puzzle for
both high-mass and low-mass stars.  In addition to spectral montages, X-Atlas also
offers comparison plots of selected targets and emission lines to
illustrate some of their variety (Figures \ref{figlinprof}, \ref{figlprof2}).
As described previously, the interactive plotting interface allows the
creation of customized line profile comparisons for up to three
observations at a time.

\section{LOW-RESOLUTION ATLAS\label{s:lores}}

The high-resolution spectra can be used to predict the low-resolution
ACIS-S or ACIS-I image spectra that a source would have produced had
the HETG not been in place.  This low-resolution prediction can then
be analyzed with spectral fitting tools such as {\it Sherpa} and
compared with actual low-resolution ACIS image spectra.  By doing so,
we potentially gain greater insight into the spectral properties
of sources in existing ACIS images for which no high-resolution
grating observations currently exist.

\subsection{Procedure}

Convolving high-resolution HETG spectra to ACIS low-resolution imaging
mode spectra involves correcting for the differences in the effective
area response with and without the gratings in place.  The effective
area is a product of the physical light collecting area of the
telescope, and the efficiencies of the mirror reflection, filters and
gratings, and the detector quantum efficiency.  The extracted
high-resolution HEG and MEG spectra ($C_{HEG}$, $C_{MEG}$) and the
relevant effective areas ($gA_{HEG}$, $gA_{MEG}$) are first used to
project how the target object's high-resolution spectrum would appear
at the $0^{th}$-order ($f^{\rm 0th}$), or the ACIS-S or ACIS-I
aimpoints ($f^{\rm aimS}$, $f^{\rm aimI}$) with no grating in place.
The resulting fluxed spectra are then multiplied by the appropriate
$0^{th}$-order or aimpoint ARFs and convolved with the corresponding
Response Matrix File (RMF) to predict the low-resolution spectra of
each detector configuration ($C^{\rm 0th}$, $C^{\rm aimS}$, $C^{\rm aimI}$). 

The extraction of the high-resolution HEG and MEG spectra, described
in section 2.1, produces coadded $\pm1^{st}$, $\pm2^{nd}$, and $\pm3^{rd}$ order HEG
and MEG source and background spectra in units of counts per
wavelength bin.  An adaptive binning scheme is then applied to the
source and background data
using a custom-built IDL routine from PINTofALE \citep{kas00}
such that every bin in the spectrum meets a minimum signal-to-noise ratio
while overall flux is conserved.\footnote{
{\tt http://hea-www.harvard.edu/PINTofALE/doc/PoA.html\#SMOOTHIE}
}
The appropriate signal-to-noise ratio for {\it smoothie} is
determined independently for the source and background data and is
generally set to be $\approx 3-4$.   After
the spectra are binned, the source and background errors on each
adapted wavelength bin are calculated as the square root of the counts
in that bin.  The smoothed background is then subtracted from the
smoothed source spectrum for both HEG and MEG, with all resulting
instances of negative count levels set to zero.  Errors on the
background-subtracted counts/bin spectrum are calculated by
propagating the errors from the separate unsubtracted source and
background spectra.

Before the high-resolution HEG and MEG spectra can be converted to
low-resolution, we first must determine how the HETG spectrum would
appear in high-resolution if observed with the effective area of ACIS-S or ACIS-I
rather than the full HETG/ACIS-S configuration.  The first step in
this correction is to divide the HETG counts/bin spectrum by the
effective area of the HETG configuration.  The effective area of the
gratings, in units of counts cm$^{-2}$ photons$^{-1}$, is contained in
a gratings ARF (gARF), one of which is produced for each grating arm
and spectral order during the extraction of the high-resolution
spectrum.  The HEG and MEG gARF effective areas are interpolated onto
the spectral range of the appropriate high-resolution data sets, and the
adaptively binned high-resolution counts/bin spectrum is converted to a flux spectrum
(photons cm$^{-2}$ bin$^{-1}$) by dividing by the interpolated gARF
effective area where the gARF is nonzero.  Where the gARF effective
area is zero, the high-resolution flux is set to zero.  The errors on the flux
bins are calculated simply by dividing the counts/bin errors by
the gARF effective area and by setting the flux error equal to an arbitrary
large value where the effective area is zero.

In order to predict the low-resolution spectrum of the ACIS-S or
ACIS-I Aimpoint, the high-resolution HEG and MEG spectra must first be
combined.  Although the MEG sensitivity extends above 30 \AA, the HEG
spectra do not quite reach 20 \AA.  Therefore, for the combined
spectrum, we define a new common HEG and MEG wavelength grid from 0
to 30 \AA \ at the spectral resolution of the HEG data (0.0025 \AA\ per
bin).  The MEG wavelength grid offers slightly lower resolution
than the HEG (0.005 \AA\ per bin), so the MEG flux spectrum is
rebinned onto the common wavelength grid at the higher HEG resolution
using another {\it PINTofALE} \citep{kas00} function.\footnote{
{\tt http://hea-www.harvard.edu/PINTofALE/doc/PoA.html\#REBINW}
} The MEG flux errors are rebinned in similar fashion.

Next, we calculate the combined HEG flux ($f_{HEG}$) and MEG flux
($f_{MEG}$) over the common wavelength grid.  For HEG and MEG fluxes
within the HEG wavelength range, the flux level in each wavelength bin
is computed by weighting the HEG and MEG fluxes by their respective
errors.  The combined HEG+MEG flux,
\begin{equation}
f_{tot} =\ \frac{w_{HEG}f_{HEG}+w_{MEG}f_{MEG}}{w_{HEG}+w_{MEG}}\ ,
\end{equation}
where
\begin{equation}
w_{HEG} = \frac{1}{\sigma_{HEG}} ~,~ w_{MEG} = \frac{1}{\sigma_{MEG}} \,,
\end{equation}
where $\sigma_X$ are the bin-wise propagated errors in each spectrum.
For data beyond the HEG wavelength range, $w_{HEG}$ was set to 0.

To predict the high-resolution spectrum of either the ACIS-S
aimpoint, the ACIS-I aimpoint, or the HETG/ACIS-S $0^{th}$-order, the combined HEG
and MEG spectrum is then multiplied by the corresponding effective area
of the aimpoint of interest, information contained in standardized
ARFs available from the $Chandra$ Data Archive,
\begin{eqnarray}
f^{\rm 0th} &=& A^{\rm 0th} f_{tot} \nonumber \\
f^{\rm aimS} &=& A^{\rm aimS} f_{tot} \nonumber \\
f^{\rm aimI} &=& A^{\rm aimI} f_{tot} \,.
\end{eqnarray}
Finally, the resulting prediction of the high-resolution counts/bin spectrum at the desired
aimpoint is convolved with the appropriate RMF, producing a
low-resolution prediction of the ACIS-S aimpoint, ACIS-I aimpoint, or
$0^{th}$-order spectrum:
\begin{eqnarray}
C^{\rm 0th} &=& Conv(RMF^{\rm 0th}, f^{\rm 0th}) \nonumber \\
C^{\rm aimS} &=& Conv(RMF^{\rm aimS}, f^{\rm aimS}) \nonumber \\
C^{\rm aimI} &=& Conv(RMF^{\rm aimI}, f^{\rm aimI}) \,.
\end{eqnarray}
The ACIS aimpoint ARFs and RMFs are updated every $Chandra$ cycle by
the $Chandra$ Data Archive.  For consistency, we use the
Cycle 06 calibration products to generate the low-resolution ACIS
predictions for each observation in X-Atlas.  

Since the HEG and MEG coaddition process is entirely linear, the
combined HEG and MEG flux levels will fall between those found in the
separate HEG and MEG spectra.  However, contrary to intuition, the
convolved low-resolution spectra do not follow this rule, i.e.,
it is possible that the low-resolution spectrum derived by
combining both HEG and MEG data could occassionally be either
greater or less than each of the component low-resolution
spectra.  This effect can occur unless one of the conditions
\begin{eqnarray}
(a) && f_{HEG} \leq f_{tot} \leq f_{MEG} \nonumber \\
(b) && f_{HEG} \geq f_{tot} \geq f_{MEG} \,
\end{eqnarray}
holds exclusively for all wavelength bins (i.e., the averaged flux is greater
than the MEG flux and less than the HEG flux for every bin -- or the
reverse) then the following conditions will not hold:
\begin{eqnarray}
(a) && C^{\rm X}_{HEG} \leq C^{\rm X}_{tot} \leq C^{\rm X}_{MEG} \nonumber \\
(b) && C^{\rm X}_{HEG} \geq C^{\rm X}_{tot} \geq C^{\rm X}_{MEG} \,.
\end{eqnarray}

\subsection{Test Results: Predicted Zeroth-Order Spectrum}

To assess the accuracy of the convolution to low-resolution, we should
ideally compare our predictions with actual ACIS image spectra of a
given target.  Unfortunately, ACIS imaging mode data are not always available for sources
observed with the HETG.  But every grating observation does contain, in
addition to the high-resolution dispersed spectrum, a $0^{th}$-order
image from which a low-resolution spectrum may be extracted.\footnote{
{\tt http://cxc.harvard.edu/ciao/threads/pieces/index.html}}
Using the $0^{th}$-order ARF and RMF in place of the ACIS aimpoint ARF and RMF
in the high resolution to low resolution conversion process described above (\S\ref{s:lores}),
we can predict the $0^{th}$-order spectrum and compare
it with the actual extracted $0^{th}$-order spectrum to determine
whether the results of the convolution produce an accurate
low-resolution spectrum.

This approach is unfortunately of limited use when the
$0^{th}$-order image has suffered significant pileup.  Pileup occurs
when multiple photons are detected as a single event, causing both a
distortion the energy spectrum, as the apparent energy of a piled
event is actually the sum of two or more photons' energies, and a decrease in
the apparent count rate, as multiple photons are counted only once
\citep{cxc05}.  For evaluation of the low-resolution convolution
procedure, we therefore selected several observations with average
count rates below 0.05 cts/s within a fifteen pixel radius of the
$0^{th}$-order, indicating relatively low to mild pileup (Table
\ref{tblpa_zero}).  For these observations, we calculated the
$\chi^{2}$ values for the comparison of the observed and predicted
$0^{th}$-order spectra.  Errors used to calculate the $\chi^{2}$
values include an extra 10\% systematic error attributable to
effective area uncertainties \citep{dra06}.

\begin{figure}
\includegraphics[width=6.2in]{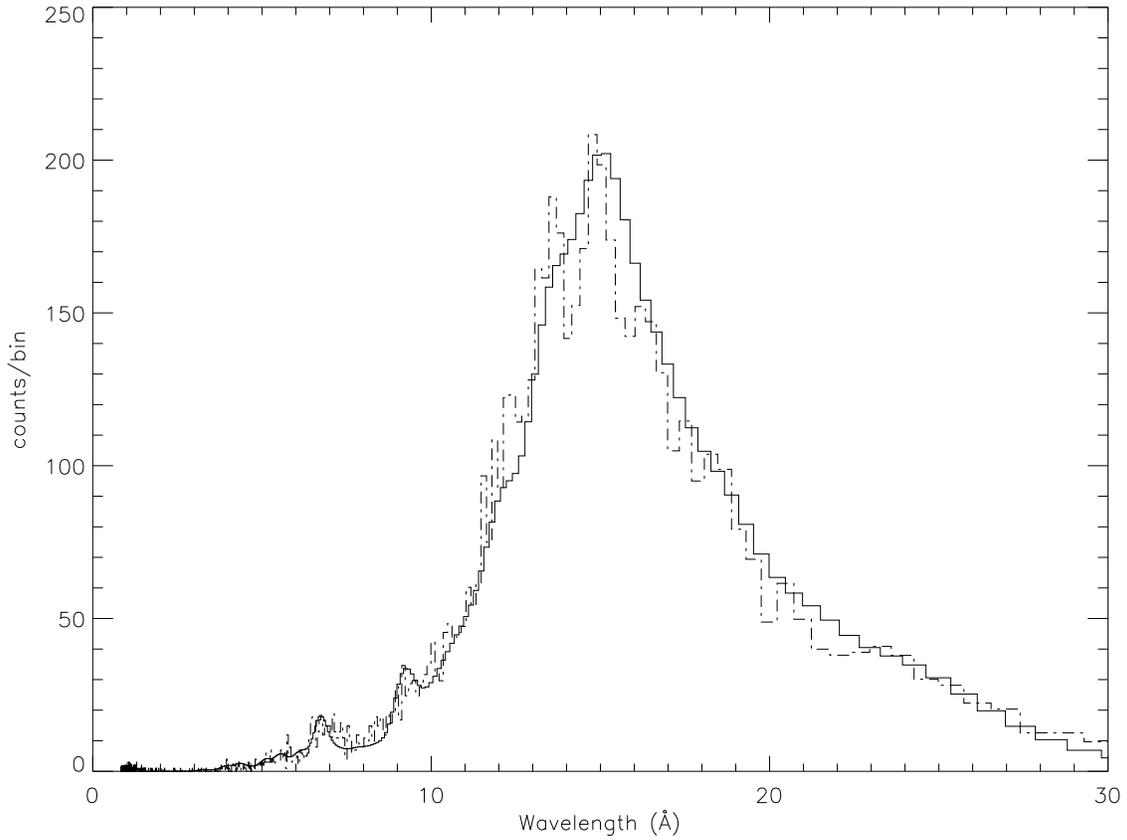}
\caption{\scriptsize Predicted (solid line) and actual (dashed line)
$0^{th}$-order spectra for $\xi$~Per (ObsID 4512).  Bin sizes vary
across the wavelength range, and are determined by the PHA binning
in the RMF.  For this observation, the predicted and observed spectra
agreed to a $\chi^{2}$ value of 381.7, with 340 non-zero bins.  A
10\% calibration error in the effective area was included in the
observed spectrum for the $\chi^{2}$ calculation. \label{figzeroth}}
\end{figure}

For these observations, the predicted low-resolution $0^{th}$-order
counts spectrum exhibits a close correspondence to the
actual $0^{th}$-order spectrum.  $\xi$ Per, from
observation ID 4512, is shown in Figure \ref{figzeroth}.  From the success of the $0^{th}$-order
convolution, we have confidence that our ACIS-S and ACIS-I aimpoint
predictions of low-resolution spectra are indeed accurate.

\subsection{Test Results: Coadded Observations}

For spectra that have been obtained over multiple observations (see
\S{2.3}), the principles and methods described in the previous
section for predicting the low-resolution spectrum can be applied to
confirm the accuracy with which the observations were coadded.
Specifically, comparing the low-resolution prediction of the
$0^{th}$-order spectrum of the coadded observation to the actual coadded
$0^{th}$-order spectrum can give a measure of the success of both the
coaddition and the convolution to low-resolution.

A low-resolution prediction of the $0^{th}$-order spectrum can be
generated from the HEG and MEG $1^{st}$ order spectra of the combined
observation using the same steps described in \S{3.1}.  The actual
extracted $0^{th}$-order spectrum of the combined observation is created
simply by adding together the $0^{th}$-order Pulse Invariant counts
spectra of the individual observations and rebinning if necessary.

\begin{figure}
\includegraphics[width=6.2in]{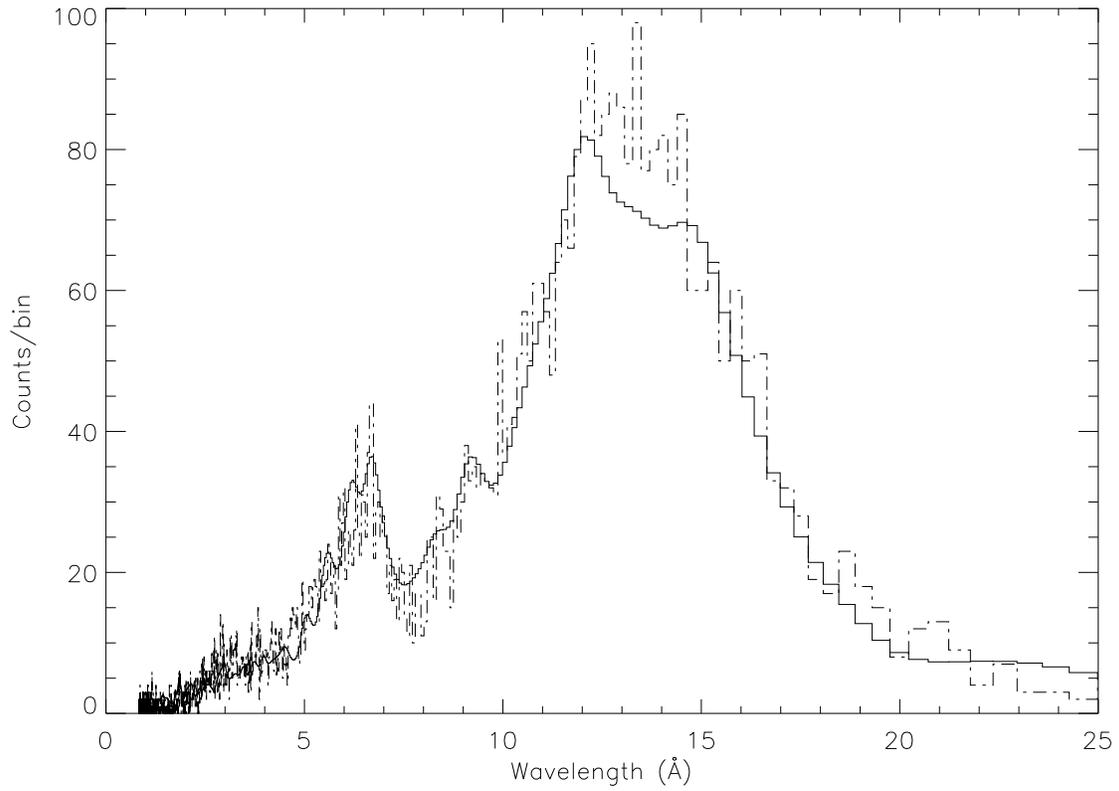}
\caption{\scriptsize Predicted (solid line) and actual (dashed line)
$0^{th}$-order spectra for HD 93250, produced by combining ObsIDs 5399, 5400,
7189, 7341, and 7342.  PHA binning is determined as in
Figure~\ref{figzeroth}. For this observation, the predicted and
observed spectra agreed to a $\chi^{2}$ value of 865.5, with 630
non-zero bins.  A 10\% calibration error in the effective area was
included in the observed spectrum for the $\chi^{2}$
calculation. \label{figcoadd}}
\end{figure}

The comparison of the extracted and predicted $0^{th}$-order spectra is
shown for the combined observation of HD 93250 in Figure
\ref{figcoadd}, composed of Observation IDs 5399, 5400, 7189, 7341,
and 7342.  $\chi^{2}$ values are listed in Table \ref{tblcoadd} and
include additional 10\% errors in the observed spectra to approximate
calibration errors in the effective area.

\subsection{Test Results: Spectral Fitting}

Spectral fitting offers a powerful means to extract stellar
information from low-resolution ACIS spectra, including absorption
($N_{H}$), temperature (kT), and metallicity.  We have performed
spectral fits on our predicted low-resolution ACIS spectra both to
extract such information and to further evaluate the convolution
methods that produced the predicted spectra.  Spectral fitting is art as well as science, and developing automated fitting routines for
bright sources such as are found in X-Atlas has proven difficult.  In
addition, we have been forced to use techniques developed for non-dispersed
spectra on spectra that have been first dispersed by HETG and then
un-dispersed with the methods described above.  However, by comparing
the results of our spectral fits with previously published values, we
have been able to verify the quality of our data and the potential for
extracting for accurate and useful scientific information from the
predicted ACIS spectra.

The predicted ACIS spectra cannot be fit directly, since the
statistical errors of the original observation are scrambled during
the convolution process, and the errors on the predicted counts in
the different channels are no longer independent.  In order to
obtain a realistic error on these spectra, we first adjust the
nominal exposure time such that a total of 1500 counts are predicted,
as in a well-exposed low-resolution ACIS observation.  We then add Poisson
deviates corresponding to the counts intensity in each bin to simulate
the statistical error of an actual observation.  (Note that this
process is similar to that adopted by the {\tt fakeit} commands
in {\sl Sherpa} and {\sl XSPEC}.)  We then use {\it Sherpa} to perform
two-temperature fits of the predicted ACIS-S spectra.  Where
$E(B-V)$ values are known (as is the case for a sample of high-mass stars),
we estimate the foreground H column density ($N_{\rm H}$),
\begin{equation}
N_{\rm H} = 5.9 \cdot 10^{21}\times E(B-V) ~~{\rm cm}^{-2}~~\,,
\end{equation}
and use this estimate to fix the column density during the spectral
fit.  For hot stars, we estimated $E(B-V)$ from the observed $(B-V)$
and the intrinsic $(B-V)_{0}$ for O and early B stars ($(B-V)_{0}$ =
-0.32 \citep{sk82}).
Where $E(B-V)$ values are not known, we choose $N_{\rm H}=10^{20}$~cm$^{-2}$
as the starting value.  We then employ {\it Sherpa} to fit the spectrum with
a 2-component {\tt mekal} (Mewe-Kaastra-Liedahl) thermal plasma model
\citep{mew85,mew86,kaa92,lie95}, using the {\tt cstat} statistic and the
Powell optimization method, extracting temperatures, metallicities, and
fluxes from each source.  In order to ensure that the fitting is robust,
we carry out the fitting process in multiple stages.  We first fit the
two temperatures, constrained to lie in disjoint regions ([0.01-1] and
[1-10] keV) for a fixed metallicity of 0.3 solar.  The temperatures
found in this way are used as starting values to refit the grouped
spectrum, this time using the $\chi^{2}$ model-variance statistic and the
Levenberg-Marquardt optimization method, and allowing for an overlap
in the temperature ranges ([0.01-2] and [1-10] keV).  The grouped
spectrum is then fit again, with $N_{\rm H}$ also being fit if
$E(B-V)$ is not known.  Next, the temperatures and $N_{\rm H}$ are
fixed at their best-fit values and the metallicity is fit.  In many
cases, especially when the $N_{\rm H}$ is known, this series of steps
yields an excellent fit.  
For those cases which could not be well fitted with four free                        
parameters ($N_{\rm H}$, two temperatures and metalicity)                             
as noted by temperatures up against range limits and/or  $\chi^2$/                   
d.o.f. $>$ 2, the                                                                    
data are refit in a modified manner.   The  temperatures 
(and $N_{\rm H}$, if necessary) are fit again, freezing                                           
the metallicity to the best-fit value, before finally fitting the                    
spectra with three free parameters ($N_{\rm H}$  and                                  
two temperatures).    We obtain error bars using the {\sl Sherpa}
command {\tt projection}.

A selection of the fit results is displayed in Table \ref{tblfit}.
The fit values compare favorably with published results
 for $\theta^{1}$ Ori C.  \citet{sch00}
report temperatures ranging from 5 to 61 $\times 10^{6}$ K, or
0.4 to 5.3 keV,  whereas our analyses finds
high-temperature components of about 10 to $>$15 keV and low-temperature
components of 1.1 to 1.4 keV.  For $\theta^{2}$ Ori A,
\citet{sch06} measure temperatures of above 25 $\times 10^{6}$ K (2.2
keV) in the low activity phase and temperatures from 3 to 100 $\times 10^{6}$
K (0.3 to 9 keV) in the high activity phase.  We find temperatures of
0.79 and 0.34 keV in the low phase and 3.52 and 1.14 keV for the high phase,
in reasonable agreement for a variable target.

For $\gamma$ Cas, \citet{smi04} report that the X-ray emission is
dominated by a heavily absorbed hot component at 12 keV, with cooler components near
0.1, 0.4, and 3 keV exhibiting much lower absorption.  We found only a
hot component at $>$15 keV with a relatively high absorption.  The
inability of our methods to differentiate the absorptions of the
different emission components mostly likely explains why the cooler
components were not detected despite comprising about 14\%
of the total emission \citep{smi04}.

Our methods produced highly consistent results over fifteen
observations of Capella.  Each observation
returned a primary temperature component between 0.56 and 0.61 keV,
with three observations also returning hotter components ranging from
4 to $>$10 keV.  \citet{can00} examined plasma diagnostics in the
high-resolution HETG spectra and found a broad range in temperatures from
log $T$ = 6.3 to 7.2 ($\sim$0.1 to 1.2 keV), while \citet{bri00} report a peak emission
measure near log $T$ = 6.8 (0.4 keV) from extreme ultraviolet
measurements.

While spectral fits of the Capella observations usually returned only
one temperature, all eight observations of the RS CVn star IM Peg returned two
emission components when fit.  The high temperature component ranged from 1.3
to 3.4 keV, while the low temperature component fell between 0.3 and
1.0 keV.  These values agree reasonably well with \citet{hue01}, who
found emission peaks at log $T$ = 7.2 ($\sim$1.2 keV) in the non-flaring state and
log $T$ = 7.6 ($\sim$2.5 keV) in the flaring state.

On the whole, our results are similar to those published by more
painstaking high-resolution analyses.  Our goal is not to supplant these more accurate
methods but to establish the credibility of our predicted
low-resolution ACIS data.  The variations between our results and those published emphasize the
oversimplification that is necessarily involved in reducing the
coronal temperature to two distinct values.

\subsection{Sample Predicted Low-Resolution Spectra: Spectral Montages}

\begin{figure}
\includegraphics[width=6.2in]{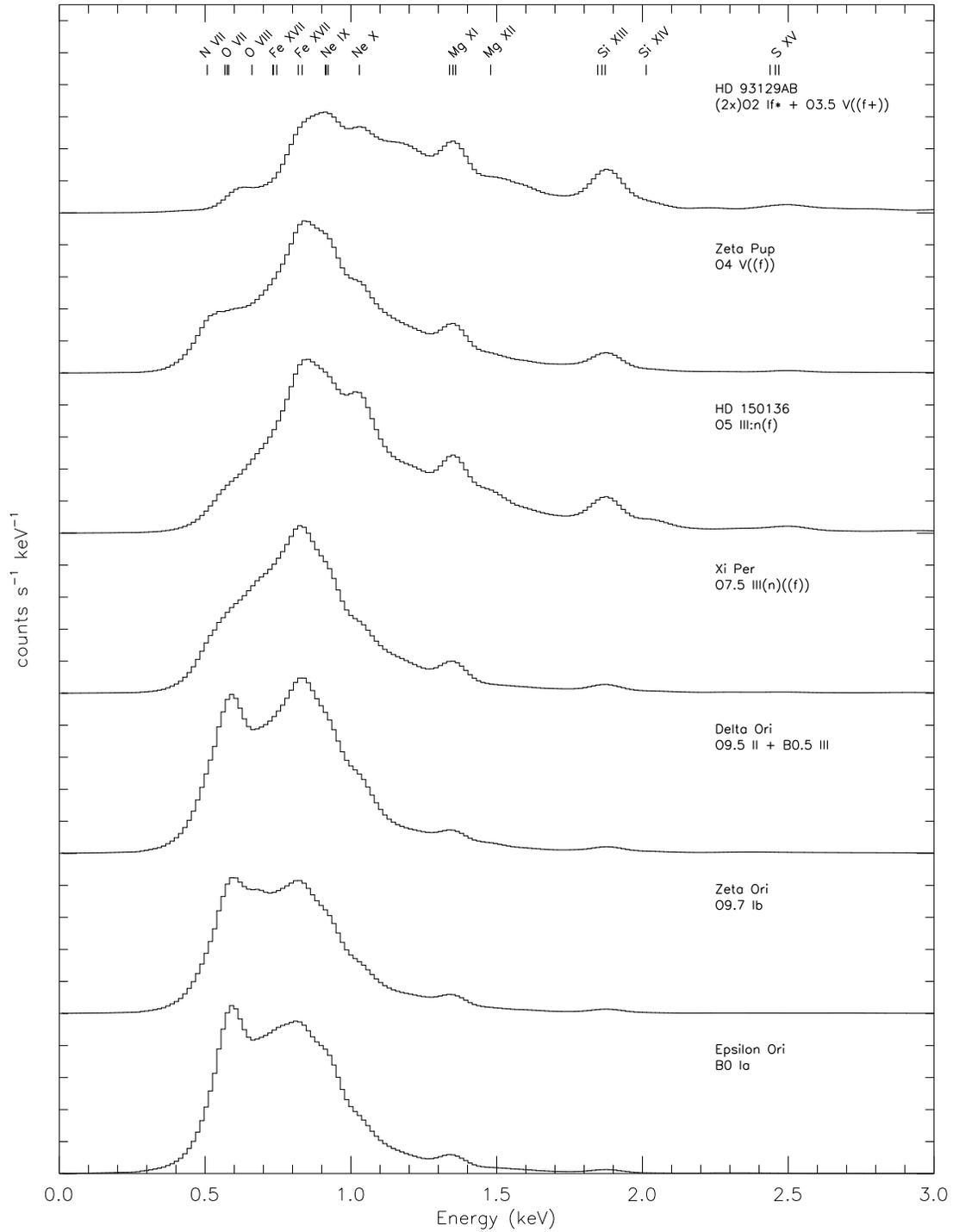}
\caption{\scriptsize The predicted low-resolution ACIS-S spectra
([counts/s/keV]) of OB
supergiants and giants, ordered by spectral class.  The spectrum of each
source is offset for clarity.
The diminishing
strength of S, Si, Mg, and Ne lines and the growth of O lines in later
spectral classes is still evident in low-resolution. \label{figobsgl}}
\end{figure}

\begin{figure}
\includegraphics[width=6.2in]{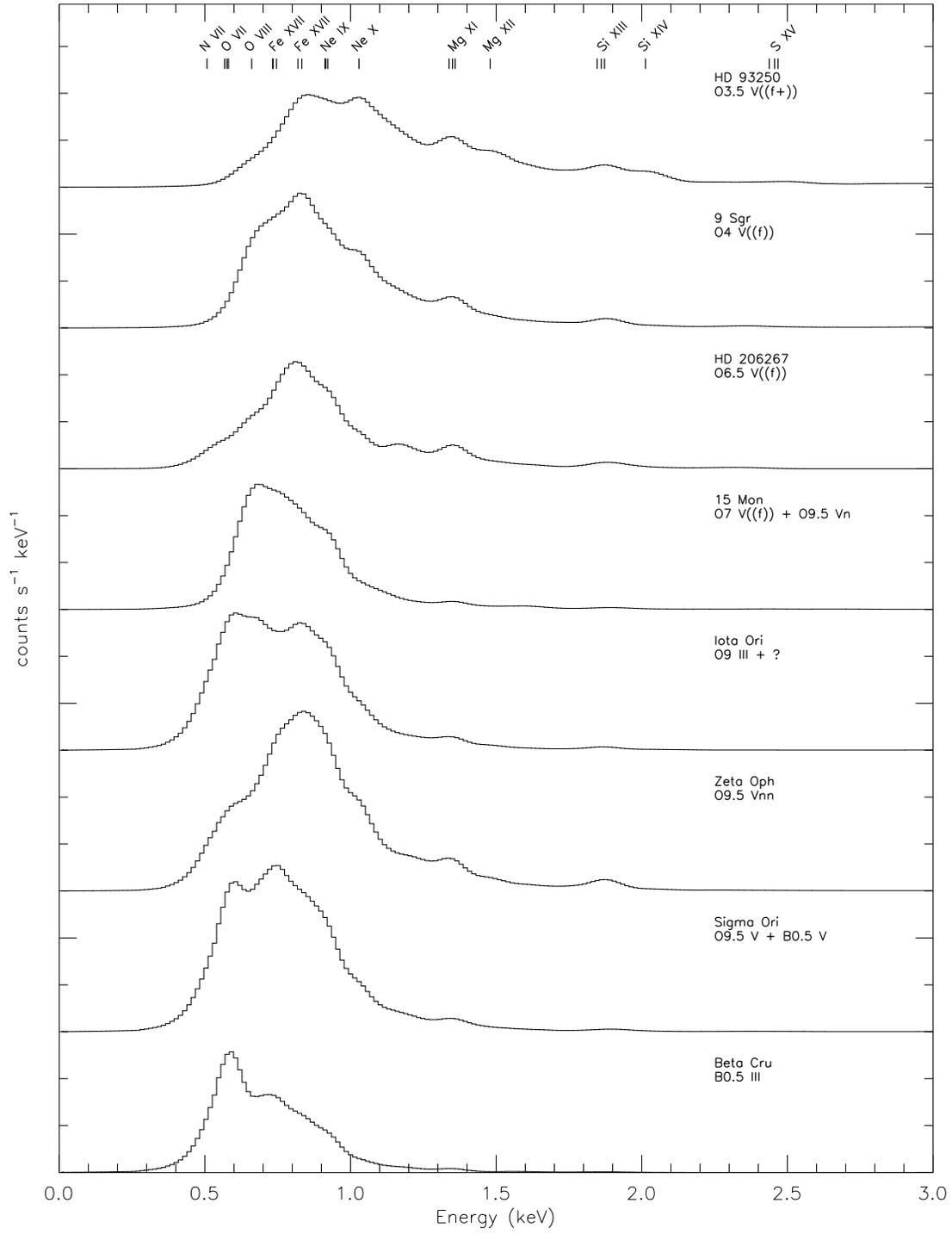}
\caption{\scriptsize As in Figure~\ref{figobsgl}, for
OB giants, sub-giants and dwarfs, ordered by spectral class. \label{figobmsl}}
\end{figure}

\begin{figure}
\includegraphics[width=6.2in]{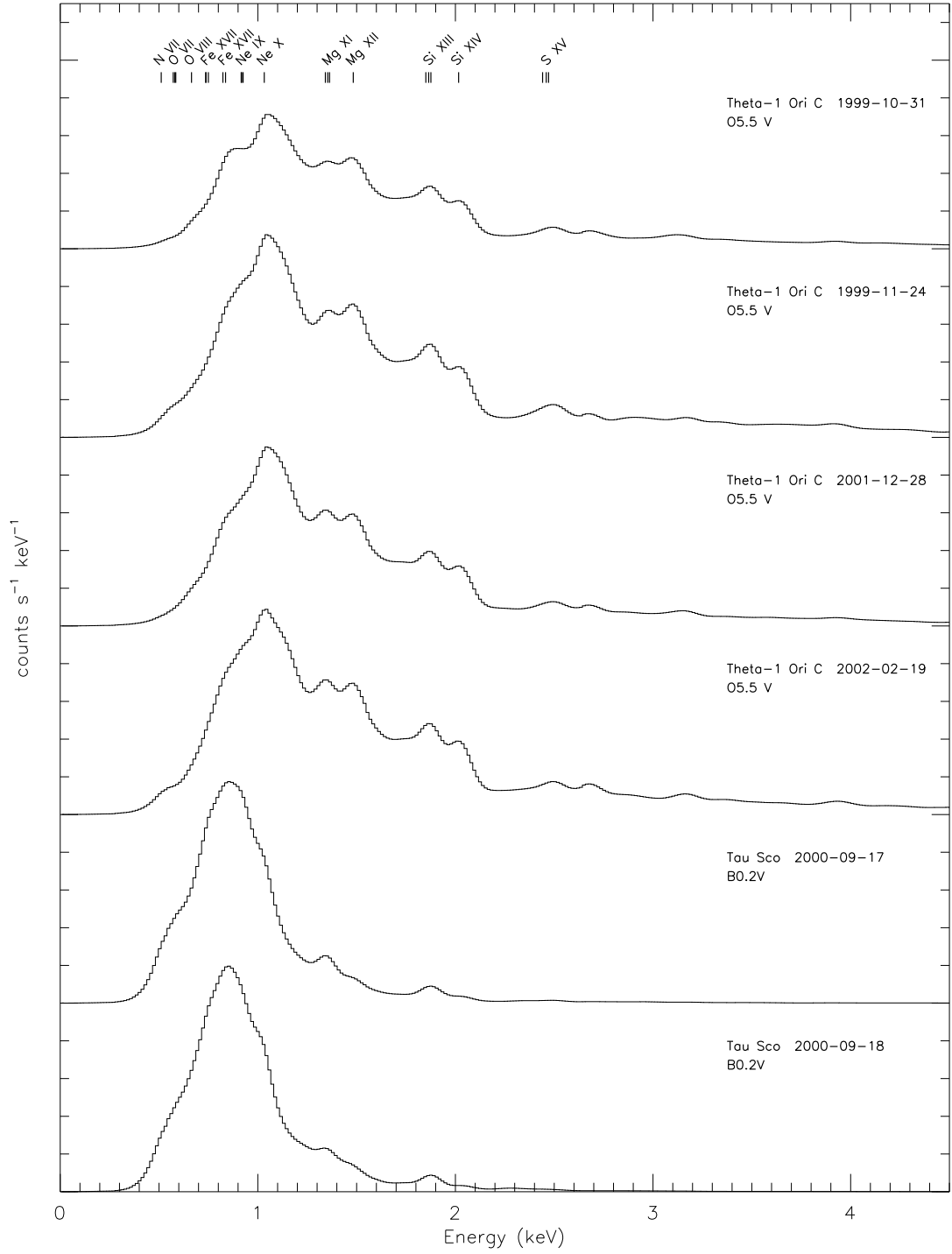}
\caption{\scriptsize As in Figure~\ref{figobsgl}, for
the two OB stars, $\theta^{1}$ Ori C and $\tau$ Sco, thought to have
strong magnetic fields responsible for their X-ray
emissions. \label{figobmagl}}
\end{figure}

\begin{figure}
\includegraphics[width=6.2in]{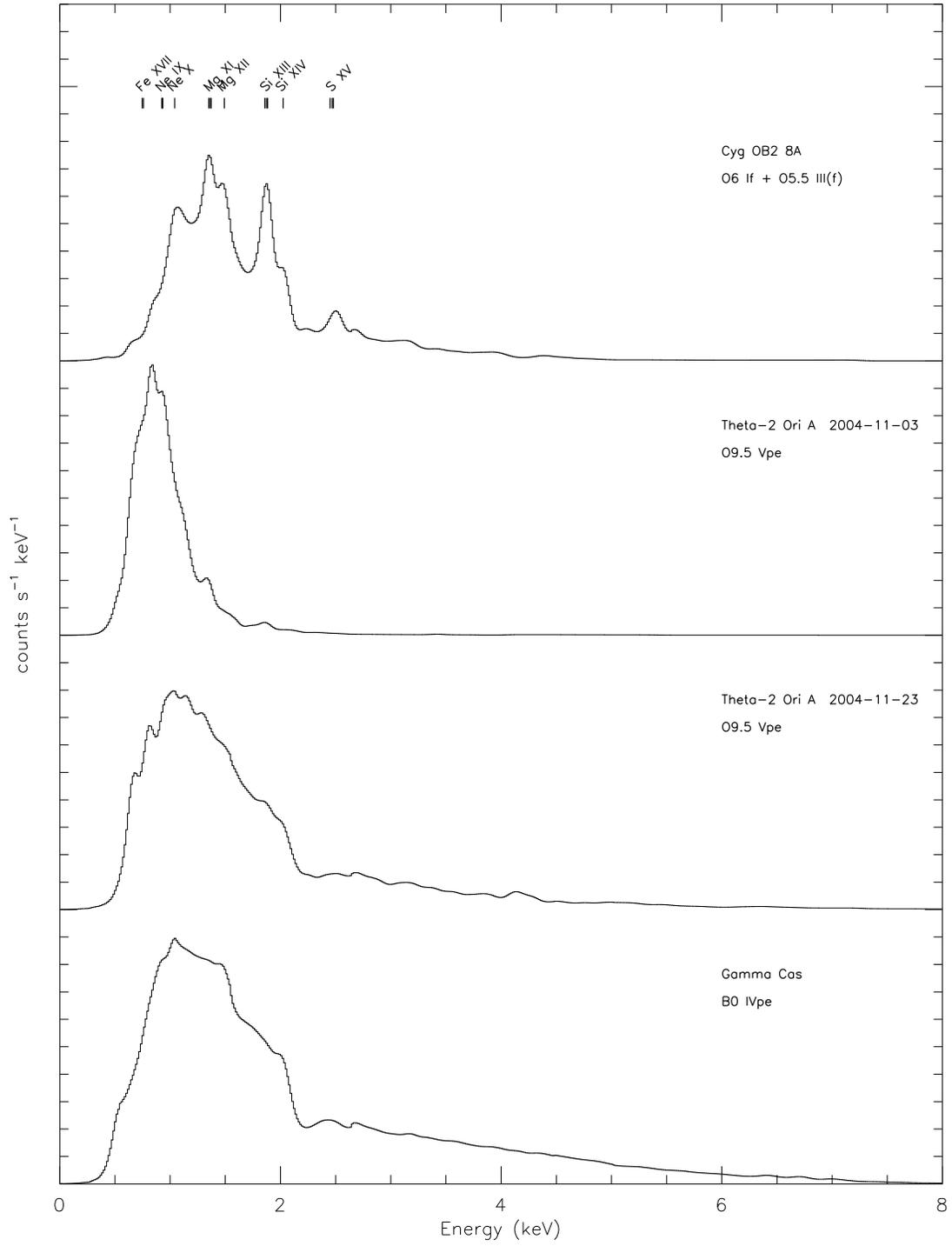}
\caption{\scriptsize As in Figure~\ref{figobsgl}, for
the three peculiar OB stars: Cyg OB2 8A, $\theta^{2}$ Ori A, and $\gamma$
Cas.  \label{figobpecl}}
\end{figure}

\begin{figure}
\includegraphics[width=6.2in]{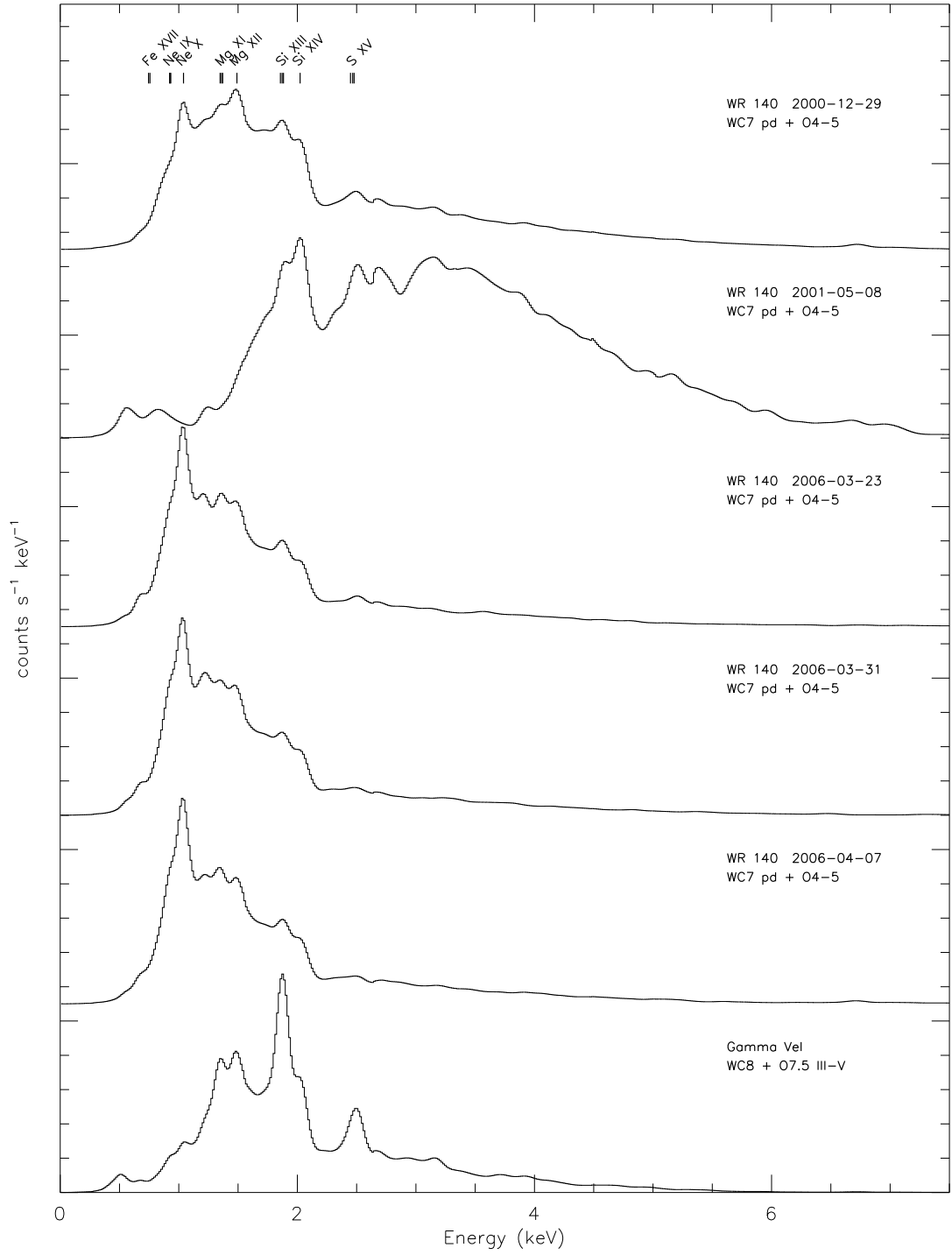}
\caption{\scriptsize As in Figure~\ref{figobsgl}, for
the two Wolf-Rayet stars observed with HETG. \label{figwrl}}
\end{figure}

\begin{figure}
\includegraphics[width=6.2in]{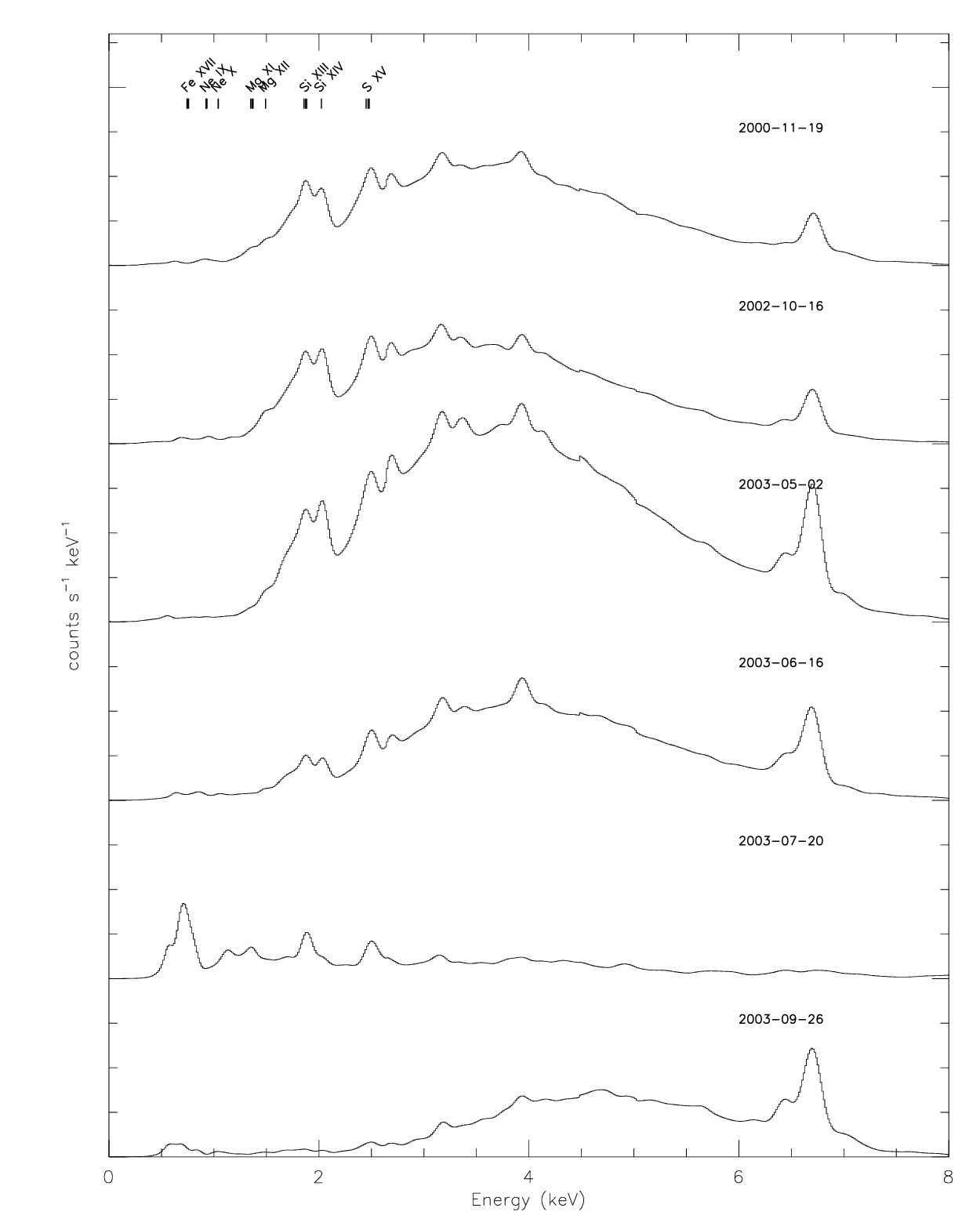}
\caption{\scriptsize As in Figure~\ref{figobsgl}, for
the six HETG observations of Eta Carinae. $\eta$ Carinae underwent an
X-ray eclipse in mid-2003: the 2003-05-02 observation was made close to
X-ray maximum, and the 2003-07-02 observation was taken near X-ray
minimum \citep{cor05}. \label{figetal}}
\end{figure}

\begin{figure}
\includegraphics[width=6.2in]{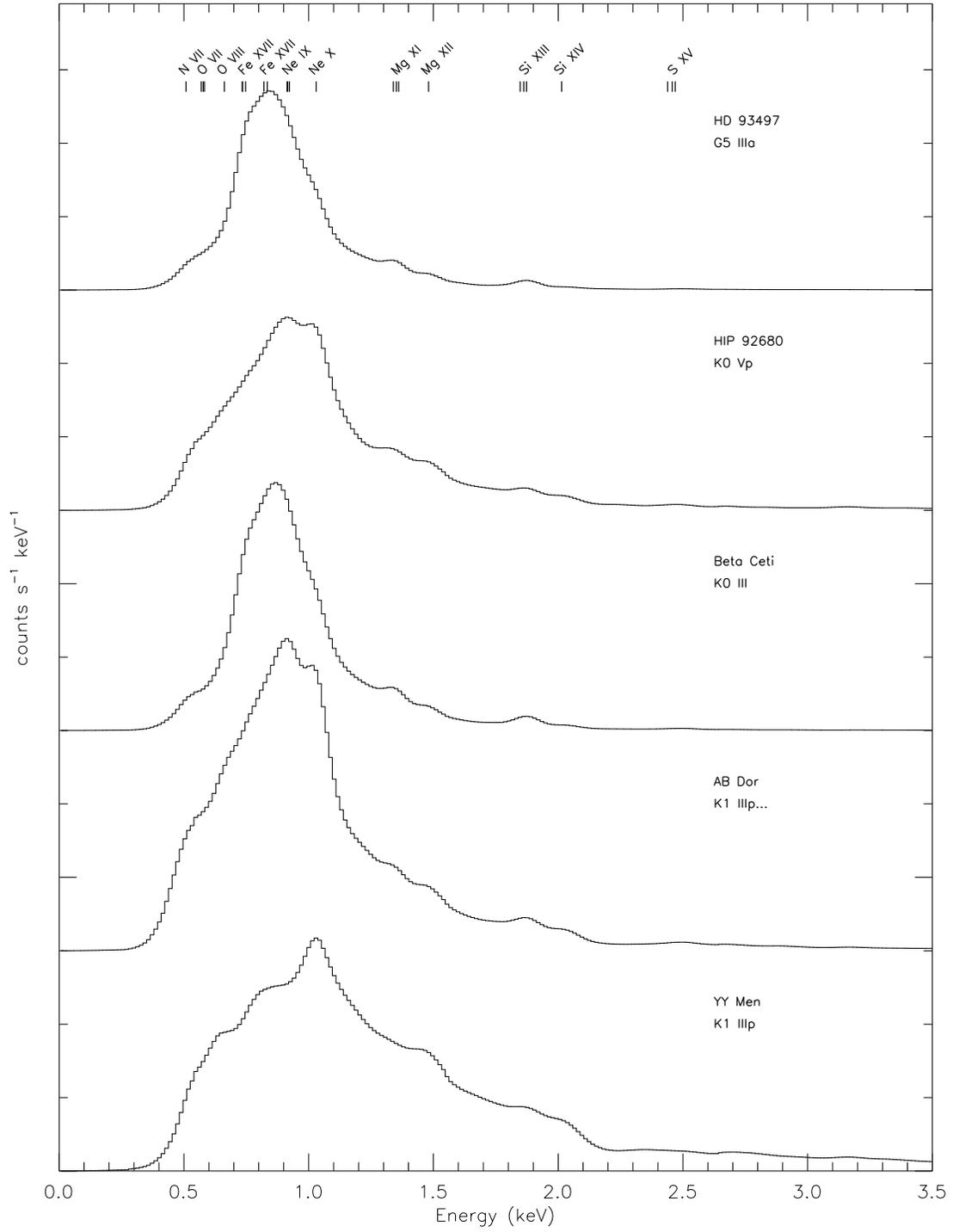}
\caption{\scriptsize As in Figure~\ref{figobsgl}, for
normal low-mass stars.  Capella is
not included here but is featured in a separate montage available on
the X-Atlas website. \label{figcoolnorml}}
\end{figure}

\begin{figure}
\includegraphics[width=6.2in]{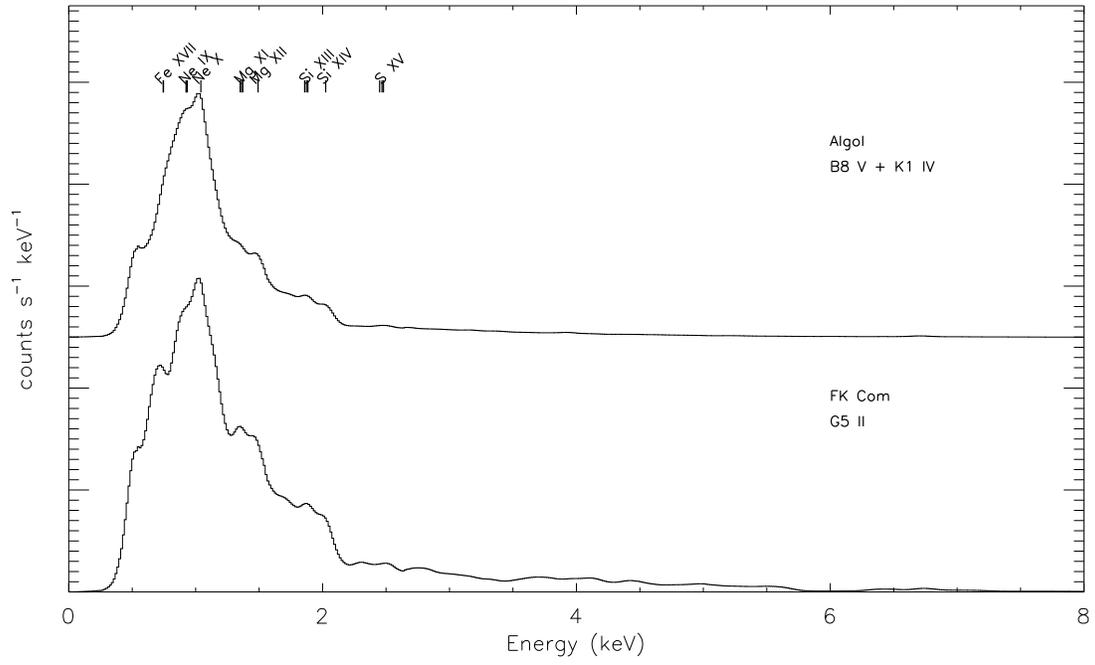}
\caption{\scriptsize As in Figure~\ref{figobsgl}, for
two unique systems, Algol and FK Com. \label{figcoolpecl}}
\end{figure}

\begin{figure}
\includegraphics[width=6.2in]{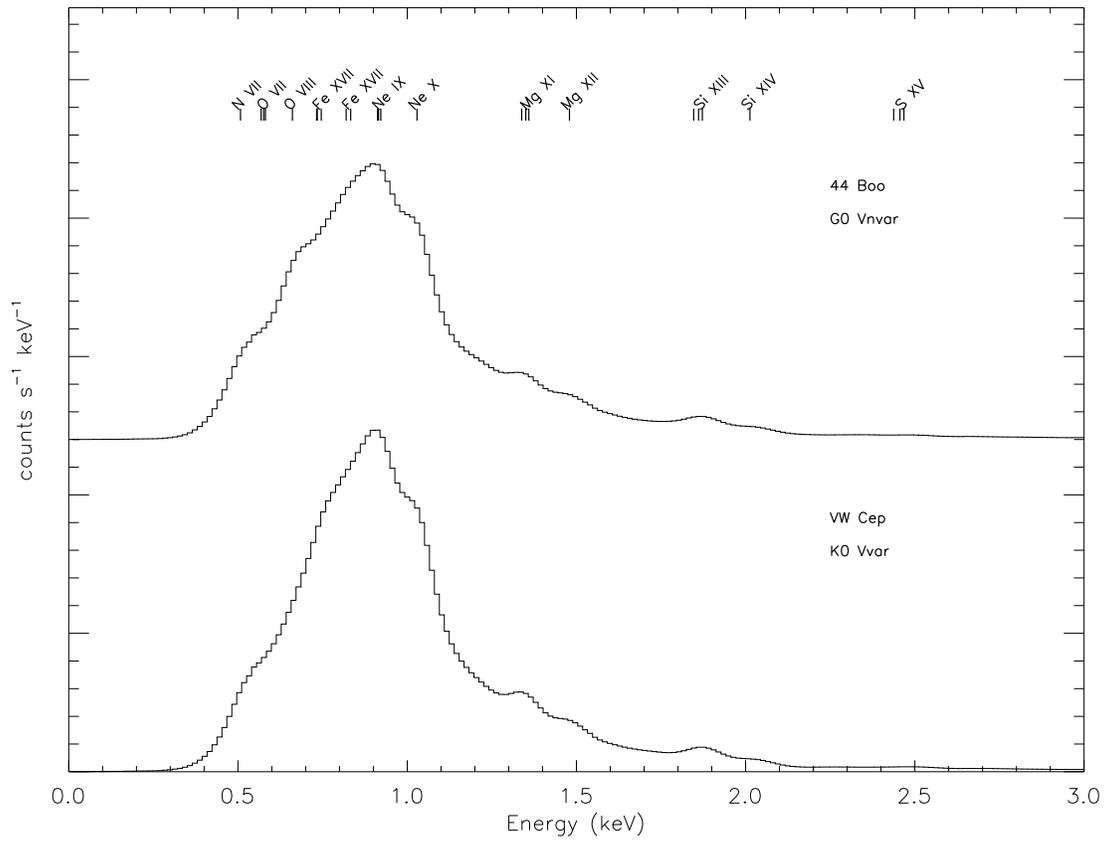}
\caption{\scriptsize As in Figure~\ref{figobsgl}, for
W UMa--type variables in the atlas. \label{figwumal}}
\end{figure}

\begin{figure}
\includegraphics[width=6.2in]{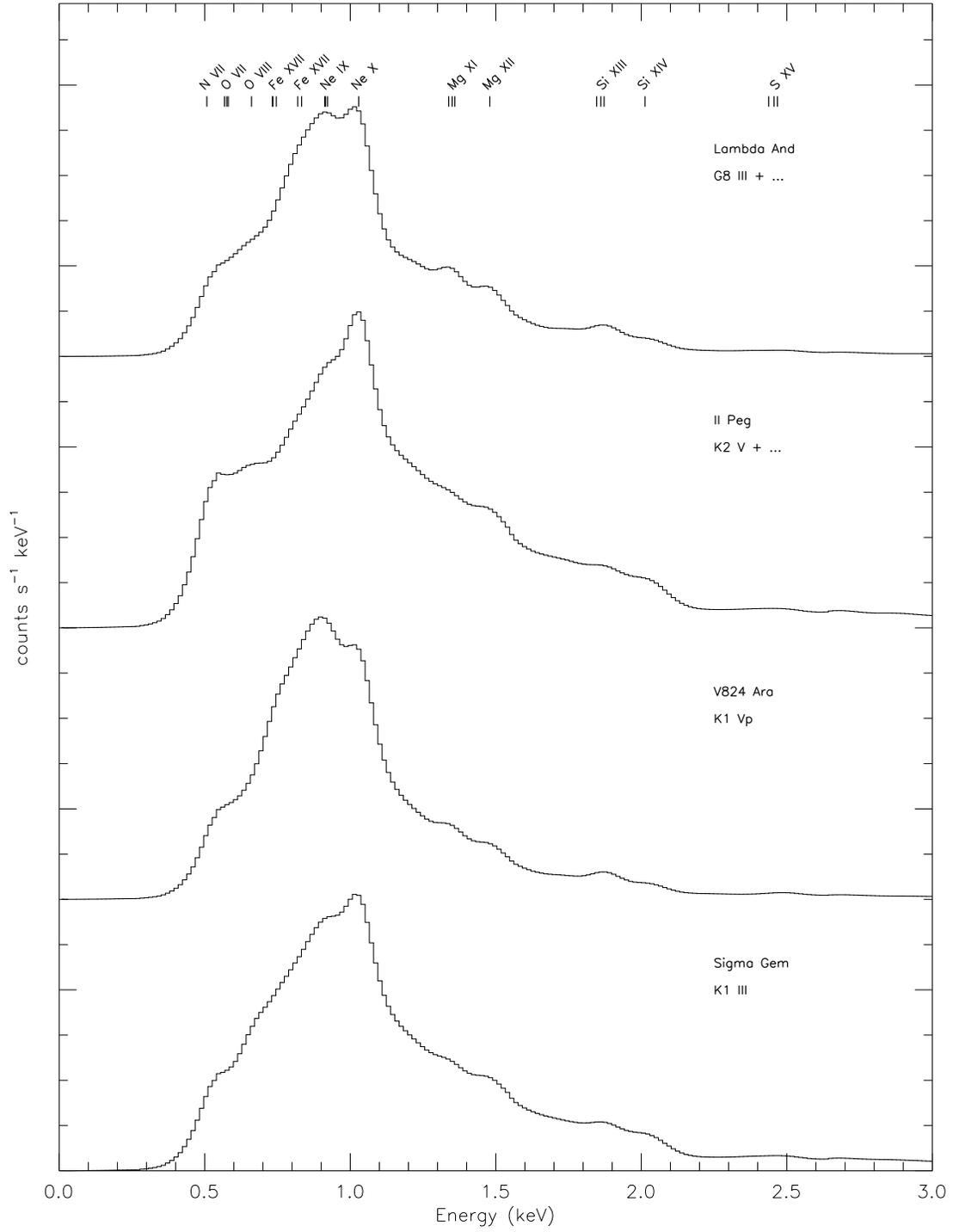}
\caption{\scriptsize As in Figure~\ref{figobsgl}, for
eight RS CVn stars.  AR Lac and IM
Peg are not included here but are featured in separate montages on the
X-Atlas website. \label{figrscvnl}}
\end{figure}

\begin{figure}
\includegraphics[width=6.2in]{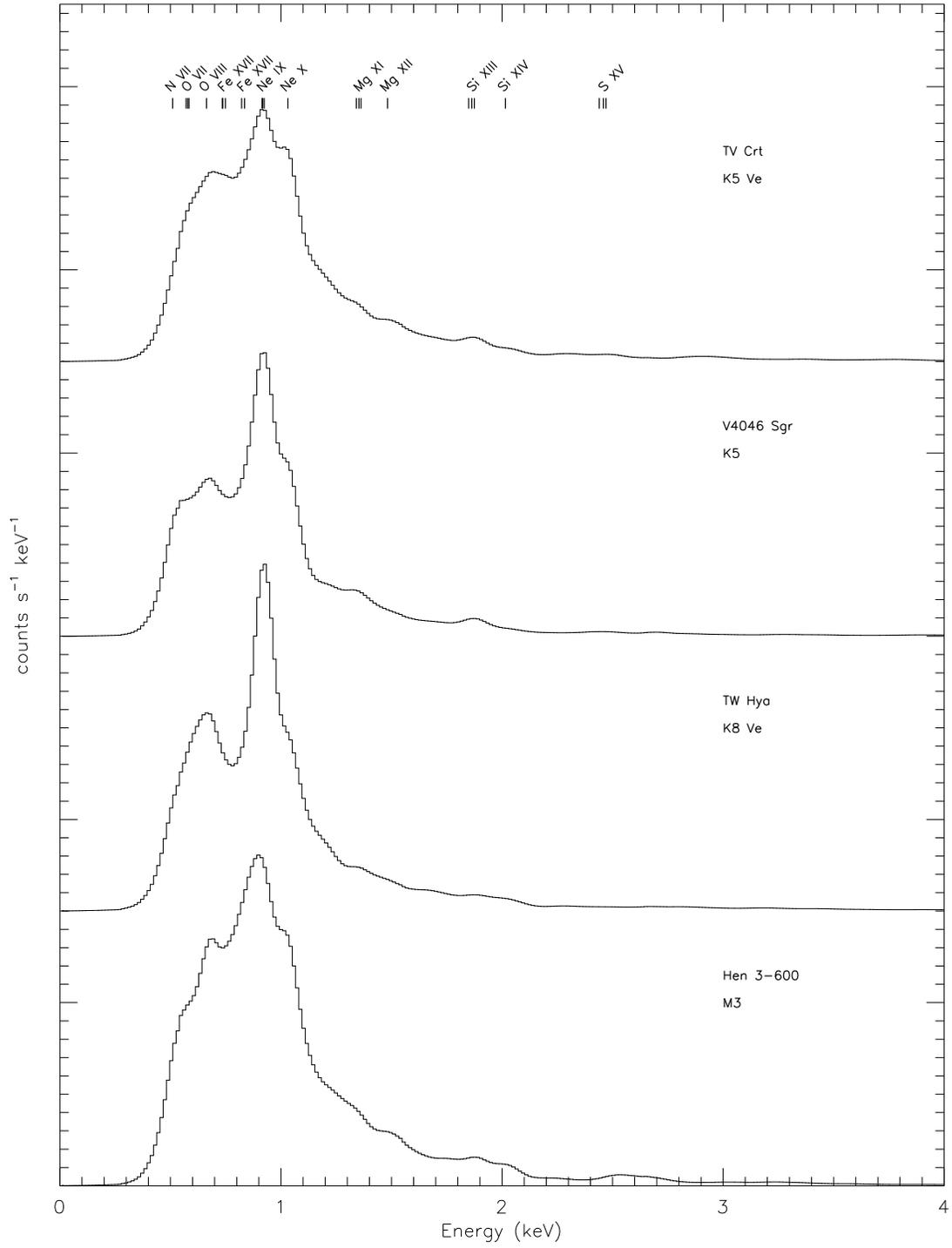}
\caption{\scriptsize As in Figure~\ref{figobsgl}, for
nine T Tauri stars. \label{figttsl}}
\end{figure}

\begin{figure}
\includegraphics[width=6.2in]{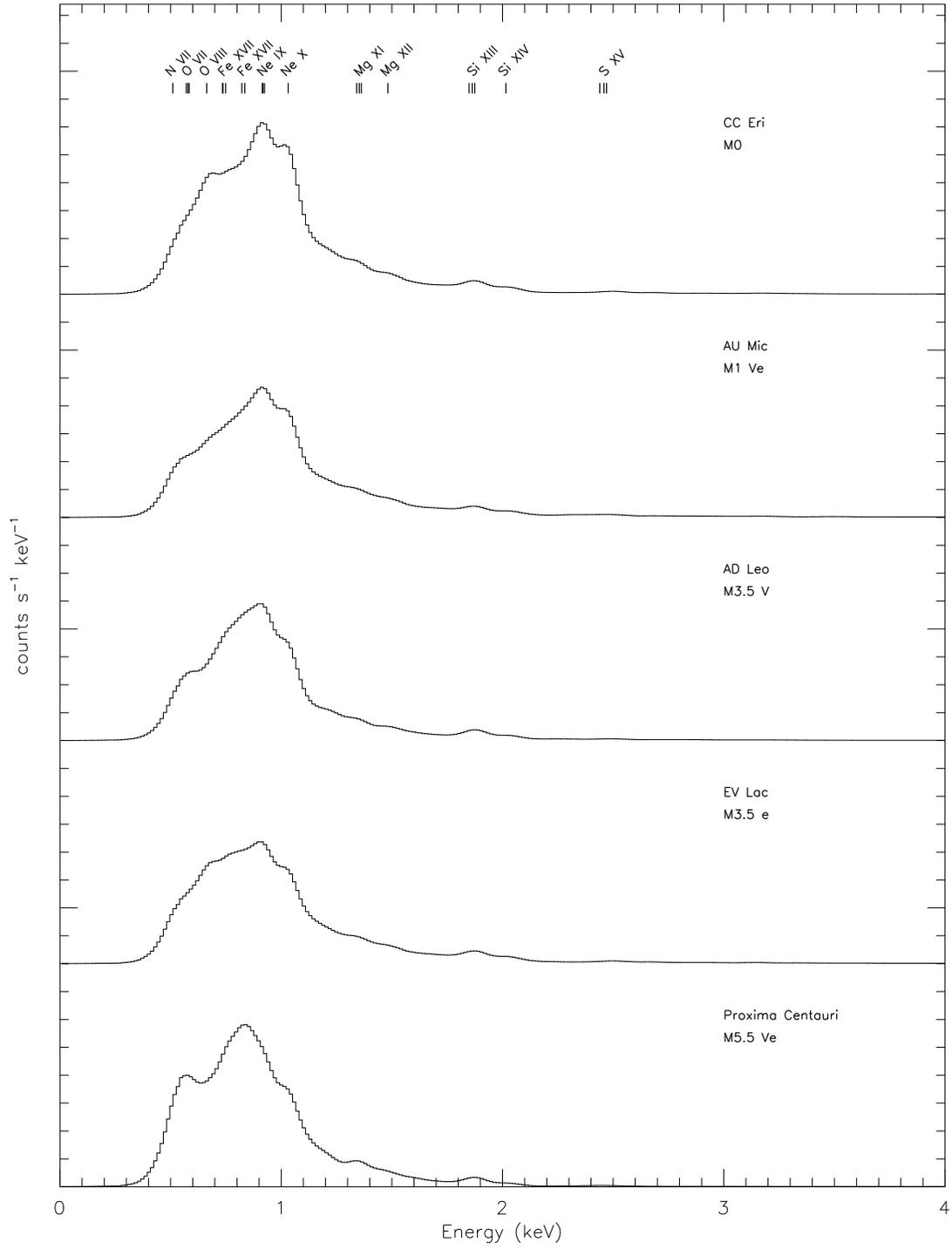}
\caption{\scriptsize As in Figure~\ref{figobsgl}, for
four low-mass flare stars. \label{figflarel}}
\end{figure}

For every high-resolution spectral montage we have produced (see section
2.4), we have also created a montage of the predicted low-resolution
spectra of the same targets.  Only predicted ACIS-S spectra are
displayed here, but montages of the predicted ACIS-I spectra will also
be made available online.  As with the HETG montages, all spectra
shown here are normalized for viewing purposes (Figures \ref{figobsgl}
$-$ \ref{figflarel}).

\section{DISCUSSION\label{s:discuss}}

We proceed with a discussion of the properties of the ensemble of
stellar spectra, continuing from the summary of
high-resolution spectra in Section 2.7.  We discuss the hardness ratios and quantiles
derived from the predicted low-resolution spectra in section 4.1 and
the X-ray variability of individual observations in Section 4.2.

\subsection{The X-ray Color of Stars: Quantiles and Hardness ratios}

While high-resolution spectral studies by $Chandra$ have proven
uniquely powerful in their ability to uncover the detailed
astrophysics in the source, the time required by these observations
limits them to no more than about 20 stellar observations per year.  A database such as
X-Atlas enhances the high-resolution dataset by providing
capabilities for comparison of the low-resolution spectra of the
vast numbers of stars in ACIS images in the $Chandra$ archive with the
relatively few well-studied examples.  Two techniques for reducing
spectral data to X-ray "color" are hardness ratios and quantiles.

Hardness ratios are comparisons of the total number of counts observed
in different X-ray passbands.  Using a Bayesian estimation approach
developed by \citet{par06}, we calculated projected hardness ratios
for each star in the atlas, using the predicted ACIS-S and ACIS-I
low-resolution spectra derived from the original HETG observations.
For each target, the predicted ACIS counts were divided into soft,
medium, and hard passbands, with energies of 0.3-0.9 keV, 0.9-1.5 keV,
and 1.5-8.0 keV respectively.

Then, for each star, two hardness ratios were calculated (see Table \ref{tblqh}):
\begin{equation}
HR_{1}=\frac{H-M}{H+M}
\end{equation}
and
\begin{equation}
HR_{2}=\frac{M-S}{M+S}
\end{equation}
S, M, and H are the total number of predicted counts in the
soft, medium, and hard energy passbands, respectively.  Error bars
were also calculated using the methods described in \citet{par06}.
For most observations, the number of predicted counts was high enough
that the error bars were insignificant compared to the 
scatter of the hardness plots.  Capella, for instance, exhibits relatively
consistent hardness over twelve observations spanning 1999 to 2006.
However, the dispersion in the hardness ratios among observations
exceeds the 1-sigma error bars, implying that even a relatively
unvarying star will undergo statistically significant shifts in hardness over
time.  Capella's variability is discussed further in Section 4.3.

\begin{figure}
\includegraphics[width=6.2in]{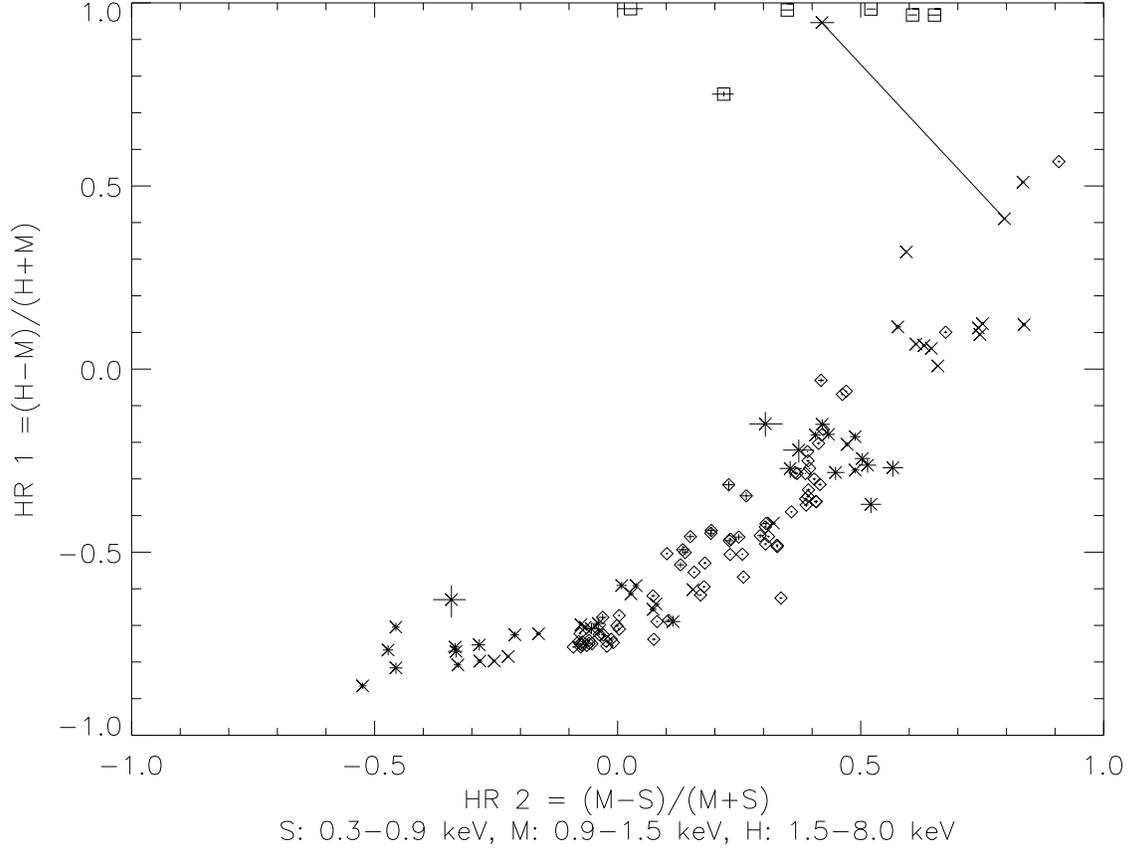}
\caption{\scriptsize A comparison of predicted ACIS-I hardness ratios for each stellar
observation in X-Atlas.  $HR_{1}$ =($\frac{H-M}{H+M}$) and $HR_{2}$=
($\frac{M-S}{M+S}$) were calculated from the predicted ACIS-I aimpoint
spectra (ACIS-S version, not shown here, is similar) of all stars in
X-Atlas.  H, M, and S are the predicted number of counts from the
predicted ACIS-I spectra in the
passbands designated on the plots.  The softest X-ray emitters fall in
the lower left, with hardness increasing towards the upper right.
Errors were calculated using a Bayesian estimation technique (Park et al.\
2006) and are
typically smaller than the symbols. $\times$ designates an observation
of a hot star; $\Diamond$ designates a cool star; $\sq$ designates an
observation of $\eta$ Car.  The outliers include six HETG observations
of $\eta$ Car and one, at approximately (0.42, 0.95), of WR 140, observed at high extinction.  The lines on the plot connects that
observation to another observation of WR 140 taken five months earlier
at low extinction, indicating the direction of the reddening vector in
the hardness ratio plot.  Three observations of WR 140 made about five
years later are also included, clustered around (0.75, 0.1).  
\label{fighr}}
\end{figure}

When the values of each star are plotted with $HR_{1}$ vs. $HR_{2}$, a
clear trend emerges (Figure \ref{fighr}). Both high-mass
and low-mass stars appear to follow a common, confined locus arcing
from softer emission near $\sim$ (-0.6,-0.8) through (0.0,-0.7) to
harder emission at (0.9, 0.6). The high-mass stars actually occupy the
extrema of the locus and the cool stars occupy a smaller range
extending from $\sim$ (0.0,-0.2) to (0.5, 0.0).  The outliers, which
include a $\sim$ 25 ks observation of the Wolf-Rayet star WR 140
and several observations of $\eta$ Carinae, will be discussed further below.

The observed hardness of a source is a function of temperature and
interstellar absorption.  To model this dependence, we used a Mekal
emissivity model \citep{mew85,mew86} to predict the expected relationship
between $HR_{1}$ and $HR_{2}$.  Model photon fluxes in each hardness
band were calculated from the integrated Mekal emissivities for a
range of temperatures and $N_{H}$ values, assuming interstellar medium
(ISM) absorption cross-sections.  Normalization of the emissivities
was not necessary since hardness ratios depend solely on the ratio of the
defined energy bands.  We then used the Cycle 06 ACIS-S and ACIS-I
response functions to predict the X-ray counts in each passband,
yielding hardness ratios derived from the Mekal fluxes for targets
observed at the ACIS-S and ACIS-I aimpoints.

\begin{figure}
\includegraphics[width=6.2in]{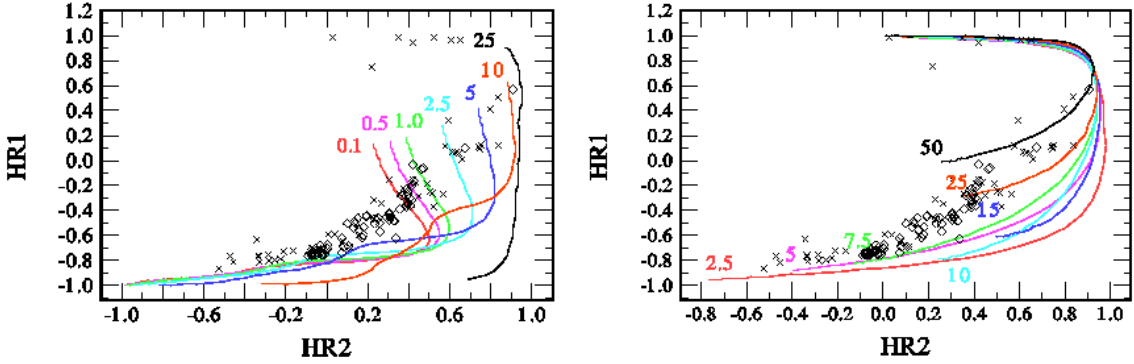}
\caption{\scriptsize The predicted relation of $HR_{1}$ to $HR_{2}$
for ACIS-I count spectra (Cycle 06) generated with a Mekal emissivity
model using solar abundances.  The left plot shows lines of constant
$N_{H}$ in units of 10$^{21}$/cm$^{2}$ for temperatures ranging from
Log $T$ = 6 to Log $T$ = 8 starting from the bottom left along each contour.  The
right plot displays lines of constant $T_{x}$ from 2.5 to 50 MK, with
absorption increasing from 0.01 to 2.5$\times$ 10$^{22}$/cm$^{2}$ starting from
the bottom left along each contour.  While this model is adequate to
explain the high-mass star distribution (indicated by $\times$), it
fails to account for the low-mass stars (indicated by $\Diamond$). \label{fighrmodel1}}

\end{figure}

\begin{figure}
\includegraphics[width=6.2in]{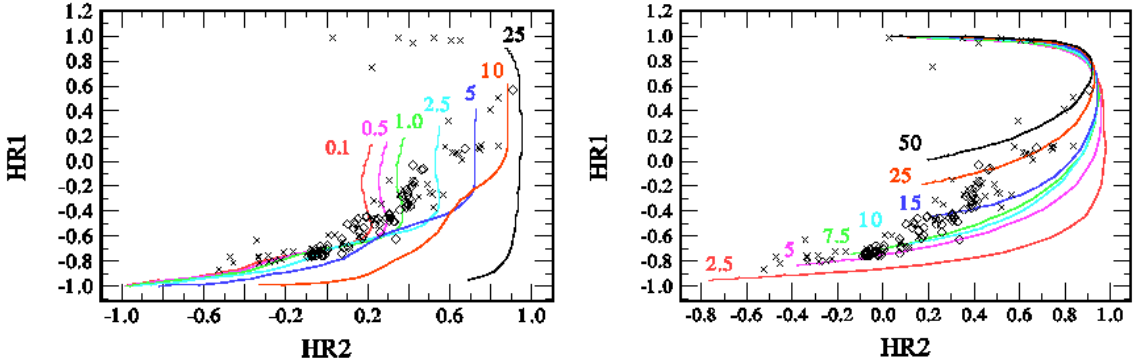}
\caption{\scriptsize The predicted relation of $HR_{1}$ to $HR_{2}$
for ACIS-I count spectra (Cycle 06) generated with a Mekal emissivity
model using an abundance of 0.1 of solar for elements except H.
Contour lines are shown as in Figure \ref{fighrmodel1}, with lines of
constant absorption in the left plot and constant temperature in the
right plot.  Both the high-mass and the low-mass stars fall on the
model contours when the abundance is lowered to 0.1 of solar. \label{fighrmodel01}}
\end{figure}

In Figures \ref{fighrmodel1} and \ref{fighrmodel01}, we show the
results of modeling the predicted dependence of $HR_{1}$ on
$HR_{2}$ for ACIS-I count spectra.  Using Mekal emissivities for
typical coronal temperatures and absorptions, we overlay lines of constant
temperature and absorption on separate plots for two abundance
values: solar abundance in Figure \ref{fighrmodel1} and 0.1 times
solar abundance (for all elements except H) in Figure
\ref{fighrmodel01}.   The lines of constant temperature range from 2.5
to 50 MK (0.16 to 3 keV) for $N_{H}$ values between 0.01 and 2.5$\times$ 10$^{22}$/cm$^{2}$, with absorption increasing from left to right
along each contour.  Similarly, the lines of constant $N_{H}$ include
values between 0.01 and 2.5 $\times$ 10$^{22}$/cm$^{2}$, with
temperature increasing from 2.5 to 50 MK left to right on each
contour.  According to this model, sources with low absorption and
temperature will appear in the lower left of the hardness plot, and
hotter, more highly absorbed stars will fall at the harder end of the
plot towards the upper right.  In addition, at the soft end of the
hardness plot, both the temperature and $N_{H}$ contours exhibit
degeneracy, indicating that we cannot uniquely determine
a given star's temperature and $N_{H}$ from its hardness ratios with
this model.  However, since the majority of the stars observed
with HETG have low absorption, we expect most of the stars in X-Atlas to
fall close to the leftmost endpoints of the appropriate temperature
contour.  Under this assumption of low absorption, we can use the
$HR_{2}$ value of a star as a rough measure of temperature up to
around (0.5, -0.6), or $\sim$15 MK ($\sim$1 keV), where the constant
temperature contours fall perpendicular rather than parallel to lines
of constant $N_{H}$.

As noted previously, the high-mass stars show in Figure \ref{fighr} occupy two
distinct ranges at the endpoints of the locus sweeping from $\sim$
(-0.6,-0.8) to (0.9, 0.6), with the majority at the soft end of the
plot between approximately (-0.5,-0.8) and (0.0,-0.7) and the
remainder falling at the hard end from around (0.4, -0.3) to
(0.9, 0.5).  The low-mass stars lie in the phase space between these
two groups on the hardness plot, with some overlap between the groups.
Assuming low absorption, we can identify the group of
high-mass stars in the lower left of the plot as corresponding to
single-temperature fits of below $\sim$7.5 MK ($\sim$0.5 keV), while
the low-mass stars reside in a warmer region between roughly 7.5 and 15 MK
(0.5 to 1 keV).  The second grouping of high-mass stars is indicative
of harder emission and therefore some combination of higher temperatures and $N_{H}$ values than are seen in
the first group of high-mass stars and the low-mass stars.  While the first, cooler
high-mass star grouping consists largely of normal OB stars, the
second, harder grouping of high-mass stars includes several
unusual objects, including the magnetic O star $\theta^{1}$ Ori C, the
Wolf-Rayet stars WR 140 and HD 68273, and peculiar stars Cyg OB2 8A
and Gamma Cas.  $\tau$ Sco, another known magnetic O star, appears in
the cooler group of high-mass stars, though at the harder end of that
subset.  Intriguingly, one star, $\theta^{2}$ Ori A, has moved
from the first high-mass star group to the second in two observations
separated by 20 days.  This shift between two X-ray emission states is
further corroborated by differences in the HETG spectra, quantile
measurements and variability tests and is explored further below.  For
a more thorough discussion of $\theta^{2}$ Ori A's variability see \citet{sch06}.

Both high-mass star groups fall roughly within the contour lines for
both solar and 0.1 times solar abundance.  This correspondence
suggests that hardness ratios do not provide a useful means of
distinguishing high-mass coronal abundances.  The low-mass stars,
however, do not show good agreement with the Mekal model calculated
with solar abundance (Figure \ref{fighrmodel1}) but the fit improves
dramatically when the non-H abundances are lowered by a factor of
0.1.  To some extent, this relatively low abundance value may
accurately reflect low coronal abundances in low-mass stars.
\citet{sce07}, for example, found that the coronal abundances of 20
X-ray bright pre-main sequence stars were significantly depleted in
Fe, with an average of less than 0.2 of the solar photosphere.  In
addition, spectral fits of the predicted ACIS spectra of low-mass
stars generally found abundances in the 0.1 to 0.4 times solar range.  However, lowering the abundance may
treat the symptoms of deviations between the model and the
observations without telling the whole story.  For OB stars,
\citet{wal94} has shown that if a point on the hardness ratio plot
does not fall within the predicted range using ISM opacities, then
stellar wind absorption, not a variation in abundance, is most likely
responsible for the discrepancy.  The presence of multiple X-ray
emission components could similarly skew the calculated hardness
ratios and imitate a decrease in abundance.  In addition, our hardness
model depends on the Mekal model's treatment of abundances in calculating
emissivities.  Therefore, although the coronae of low-mass stars might
conceivably have depleted abundances relative to solar, our model
lacks the sophistication to differentiate between actual variations in
abundance and other effects. 

The models shown in Figures \ref{fighrmodel1} and \ref{fighrmodel01}
do account for the outlying observations of WR 140 at approximately
(0.42, 0.95) and of $\eta$ Car between $\sim$(0.0, 0.95) and
$\sim$(0.6,0.95).  WR 140 was observed pre-- and post--periastron with
observations separated by five months.  The hardness model suggests that some
combination of a large shift in column density and a rise in
temperature could be responsible for the observed change in hardness.
The model does not enable us to characterize the change in
temperature, since thermal information is lost in the hardness plots
at high absorption.  We can confirm, however, that the HETG
spectrum of the post-periastron observation appears strongly absorbed
between 1-2 keV in a way not present in the first observation.  This
discrepancy is most likely due to absorption by the cool WC material
along the line of sight between $Chandra$ and the region of wind
interaction.  This observation, then, is the only one that appears
strongly affected by absorption besides the six observations of $\eta$
Car.  $\eta$ Car is a massive, highly-luminous star thought to produce
X-rays through wind collisions with a massive companion
\citep{ham07}.  The spectra from series of {\it XMM-Newton} and {\it Chandra}
observations have been fit by \citet{ham07}, who report kT values
ranging from 2.9 to 5.4 keV (33 to 63 MK) and $N_{H}$ between 4.8 and
42 $\times$ 10$^{22}$/cm$^{2}$.  The combination of high temperature
and high absorption values places $\eta$ Car at the extreme hard end
of the temperature contours in Figures \ref{fighrmodel1} and
\ref{fighrmodel01}.  The appropriate $N_{H}$ contours for $\eta$ Car
are not shown on the plots.

Both ACIS-I and ACIS-S versions of the hardness plots are available on
the X-Atlas web page for each observation, 
although only the ACIS-I plot is shown here.
To see the plots, use the main search page to find an observation, 
and then click on ``Object Spectral Type".
The web page offers the additional capability of quickly identifying
and highlighting individual stars of interest for both ACIS-I and
ACIS-S plots.  

Hardness ratio plots for ACIS-I and ACIS-S are very similar.
Any systematic shifts between the ACIS-I and ACIS-S
versions of the hardness plots are due to the increased sensitivity at
low energy of the back--illuminated S3 chip.  This characteristic
leads to more counts for soft sources in the "S" passband and hence
lower values of $HR_{2}$ for the predicted ACIS-S data. While S3
provides a somewhat broader range in $HR_{2}$, the majority of stars
observed without benefit of the gratings have been observed on
front-illuminated CCDs.

Hardness ratio calculations are most useful if the spectra in question have
multiple counts in each passband.  While that criterion is certainly
met for any of the predicted ACIS spectra in X-Atlas, actual ACIS
observations can contain thousands of detectable sources with of order
one hundred counts.  When a source has few counts or is absorbed,
there is a strong likelihood of low counts in one of the three
pre-determined passbands.  These circumstances can lead to very large
and non-uniform errors in hardness calculations due to the location or
type of the source.

For such sources, quantile analysis provides an alternative technique
to hardness ratios for classifying X-ray sources through X-ray color
measurements \citep{hon04}.  Since the method uses the quantile (in
this case quartile) of the source's own photons, the statistics are
uniform across all bands.  We have applied the quantile analysis
methods to X-Atlas to enable comparisons between faint X-ray sources
found in ACIS images and the bright stars observed with HETG.

Like the hardness ratio calculations, quantile analysis is performed on the
predicted ACIS-S or ACIS-I aimpoint spectra.  Using the methods
described in \citet{hon04}, we take the total predicted counts within a
0.3-8.0 keV range and determine the energies that divide those
predicted counts into fractions of 25\%, 50\%, and
75\%.  Upon finding $E_{x\%}$, the energy below which x\% of the total
counts fall, we calculate normalized quantiles $Q_{x}$:
\begin{equation}
Q_{x}=\frac{E_{x\%}-E_{lo}}{E_{hi}-E_{lo}},
\end{equation}
where $E_{lo}$ = 0.3 keV and $E_{hi}$ = 8.0 keV.

\begin{figure}
\includegraphics[width=6.2in]{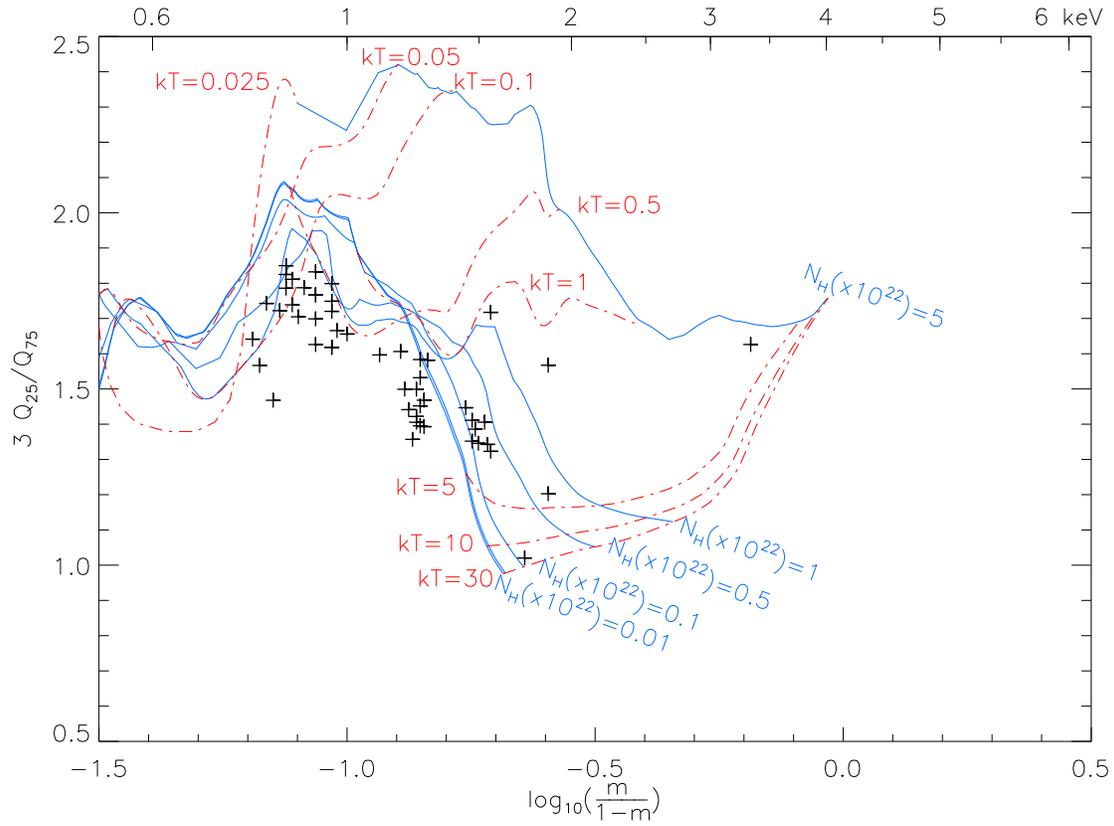}
\caption{\scriptsize Quantile-based color-color diagram (see \citet{hon04}) for the
predicted ACIS-I aimpoint spectra of the observations of high-mass
stars in X-Atlas.  The quantile grids were generated assuming solar
metallicity.  
$m$ = $Q_{50}$ = ($E_{50}$ - $E_{lo}$)/($E_{hi}$ - $E_{lo}$); $E_{lo}$ = 0.3
keV; $E_{hi}$ = 8.0 keV). The upper scale on the plots is the median
energy  in keV.  The six HETG observations of the unique
object $\eta$ Car are not included in the plot. \label{fighquant}}
\end{figure}

\begin{figure}
\includegraphics[width=6.2in]{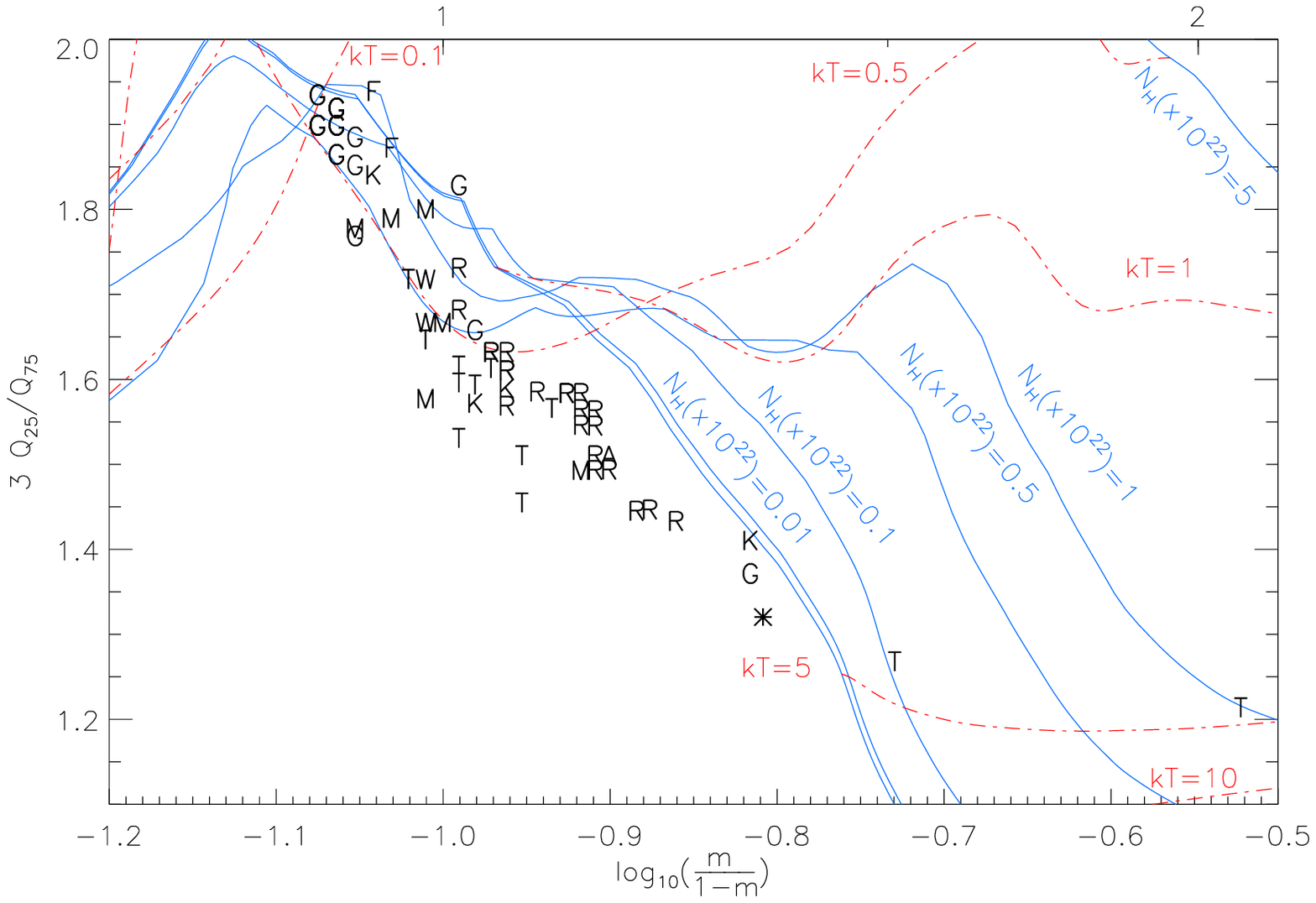}
\includegraphics[width=6.2in]{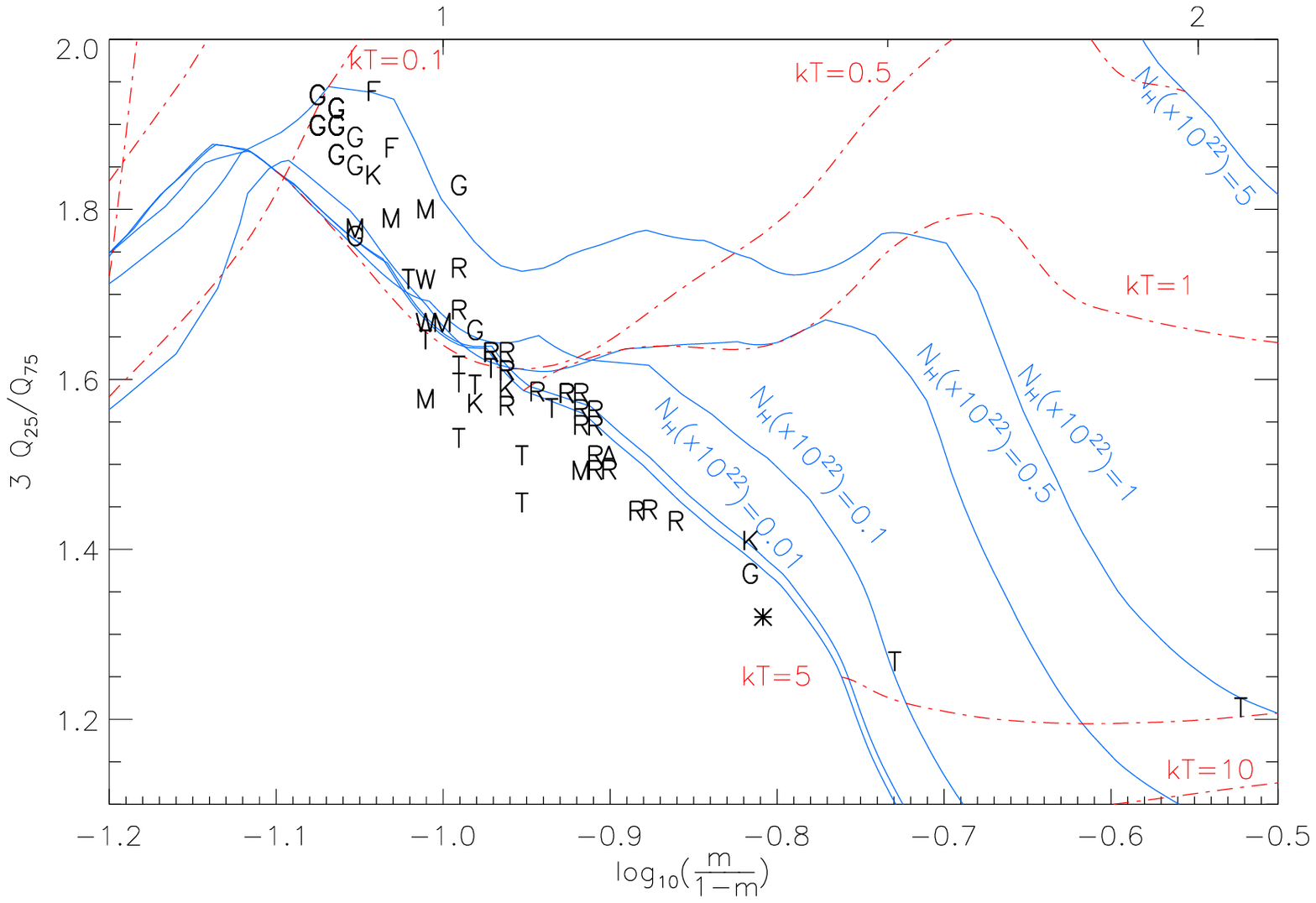}
\caption{\scriptsize Quantile-based color-color diagrams for the
predicted ACIS-I aimpoint spectra of the low-mass stars in X-Atlas.
The quantile grids were calculated using metallicities of 0.3 times
solar ({\it top}) and 0.1 times solar ({\it bottom}).  The symbols on
the plot correspond to stellar types as follows:
F = normal F star; G = normal G star; K = normal K star; M = normal M
star; R = RS CVn star; T = T Tauri star; W = W UMa star; A = Algol; *
= FK Com. $m$ = $Q_{50}$ = ($E_{50}$ - $E_{lo}$)/($E_{hi}$ - $E_{lo}$); $E_{lo}$ = 0.3
keV; $E_{hi}$ = 8.0 keV.  The upper scale on the plots is the median  
energy  in keV. \label{figcquant}}
\end{figure}

The typical quantile based color-color diagram (QCCD) plots
3*$Q_{25}$/$Q_{75}$, a measure of the emission from the middle half of
the spectrum, against the median ($m$=$Q_{50}$), a measure of the
overall hardness of the source, in the form $log_{10}$[$m$/(1-$m$)].
The particular form of the x-axis median expression is chosen to avoid
compressing the regions of phase space corresponding to harder X-ray
emission \citep{hon04}.  The ratio of the $75^{th}$ percentile photon
to the $25^{th}$ percentile photon is a function of both the temperature
and the absorption of the source.  This dependence means that as the absorption
varies, the quantile ratio ($Q_{75}/Q_{25}$) for a given temperature
may be different.  In order to interpret the resulting QCCD, we
overlay a model grid of hydrogen column depth ($N_{H}$) and
temperature (kT).  This grid is determined from the ACIS-S or ACIS-I
aimpoint ARF and combined interstellar absorption and Raymond-Smith
(1977) thermal models (Figures \ref{fighquant}, \ref{figcquant}). 
We used  Raymond--Smith plasma (Raymond \& Smith 1977) for general                   
purpose fitting of                                                                   
low resolution coronal spectrum. While more recent models including                  
MeKa (Mewe  et al.\ 1985)                                                            
and APEC (Smith et al.\  2001) have more lines included,                             
Raymond-Smith models were found by                                                   
Wolk et al.\ (2006) to have the lowest residual $\chi^2$                                  
(degree~of~freedom) and/or less outliers                                             
in a direct comparison of fitting about 80 moderately bright stellar                 
sources using Sherpa.                                                                
The same study found that all three models gave similar results for                  
kT and nH for stars with between 100 and 1000 counts.                                
 On the quantile plots Figures \ref{fighquant} and \ref{figcquant} 
one can distinguish changes in temperature from extinction and
even thermal and non-thermal changes.  Further, the grids shown in
Figures \ref{fighquant} and \ref{figcquant} are sensitive to
metallicity.

Two things become immediately clear from these plots. First, as noted
previously, the vast majority of stars observed with HETG had
relatively low absorption.  No low-mass star exceeds $\sim N_{H}\ 5
\times 10^{21}$ cm$^{-2}$ and only about five high-mass stars exceed
that value.  Second, while low-mass stars are smoothly distributed
along the temperaturea axis in this plot, with temperatures ranging from
about 1 to 3 keV, the high-mass stars are clearly congregated at the
edges.  The appropriate one-temperature representations appear to be
either below about 1 keV or above about 3 keV with few high-mass stars
occupying the middle ground dominated by the low-mass stars.  This
distribution has two clear implications. First, the X-ray generation
mechanism is clearly different between the high and low-mass stars.
This result is not unexpected, as many other lines of argument
have led to the development of various dynamo-driven models for the
generation of X-rays in low-mass stars.  More surprising is the bimodal distribution of
high-mass stars, which suggests that two different physical states are
at work.  A similar bimodality was noted in the hardness ratios of the
high-mass stars (Figure \ref{fighr}), and as expected, the same
stars that appear at the warmer, harder end of the hardness ratio
plot generally lie in the high-temperature group on the quantiles
plots.  Once again, $\theta^{2}$ Ori A has moved from one group to the
other between the two observations.

$Chandra$ grating spectra have provided material for deliberation about
complexities and puzzles in high-mass star interpretation.  There are a
number of possible explanations for high-mass stellar X-ray emission relevant
to X-Atlas.  Colliding winds in binaries provide high temperature
plasma in some cases (e.g. \citet{ham07}), while magnetic activity
plays a role in others.  Quantile measurements of stars in the atlas
reveal that about 50\% of the high-mass sample is cool
with one-temperature component fits below $\sim$ 1 keV.  This result
is consistent with the standard model of X-rays originating in shocked
material in line-driven spherically symmetric winds \citep{mac91}.  Typical temperatures in these models are about 10MK ($\sim$ 900 eV).  Hot
X-ray emission from OB stars has been the subject of recent investigations.
\citet{sch00,sch01,sch03} reported temperatures of about 5 keV from
$\theta^{1}$ Ori C. Schulz et al. conclude that the component of the
emission at this temperature must be formed ``near the terminal
velocity of the wind,'' at about seven stellar radii from the
photosphere.  \citet{gag05} show that such emission can be fitted by a
two-dimensional MHD magnetospheric wind model due to \citet{udd02}.
Several other examples of X-ray hot O and B stars have since been
found \citep{lop06,rak06}.  \citet{mul06} have proposed a
two-component scenario by which the cooler X-ray emission is generated
by line-driven, spherically-symmetric winds, and a second component of
the wind emerges from magnetically active regions in polar caps that
may extend as low as 45$^{\rm o}$.  In such a scenario, whether an O
star is found to be a relatively hot or soft source may simply be a
function of inclination.

We note that in the star-forming region RCW 108, about 75\% of the OB
star candidates in the field are found to have X-ray temperatures
above 3 keV (Wolk et al. 2007).  The fraction of high-temperature stars was
greatest in the RCW 108--IR embedded cluster, which is still actively
forming stars.  The inner group indicates an evolutionary aspect to
the observed temperature, in contradiction of models that propose that
the observed temperature may be a function of the observation's line--of--sight
 unless the extent of the polar cap decreases in time.  X-Atlas
contains a sample of relatively old high-mass stars in which the
cooler variety are in the majority.  This characteristic may indicate
that X-ray temperatures in high-mass stars decline with time, just as
they do in cool stars, although the X-ray generation mechanisms are different.

The various types of low-mass/cool stars studied with $Chandra$ seem to be
very tightly correlated in Figure \ref{figcquant}.  For example, the bulk of the F and G
stars lie between y--axis values of 1.75 and 1.95, while the general
RS CVn  and T Tauri star populations have y--axis  values between 1.4 and 1.7.
We ran model grids at several metallicities ranging from 0.001 to 5 times solar.
While no particular metallicity seemed to improve the relation between
the positions of the high-mass/hot stars and the model grids, we found
that for low-mass/cool stars with cool coronae at $<$ 1 keV, a best fit
to the data was a metallicity of about 0.3.  For the warmer coronae,
the best fit was a metallicity of about 0.1 (See Figure
\ref{figcquant}).  The fact that the stars are still tightly
correlated despite the high metal sensitivity of the model grids is
indicative of very uniform metallicity among the observed coronae.
This leads us to believe that the handling of the metallicity by the
models as a function of temperature, not a change in the metallicity, is the root of the
apparent metallicity discontinuity. Indeed, given the sensitivity of
quantiles to metallicity shown here, the metallicity must be very
similar in all the observed low-mass/cool star coronae.

\subsection{Variability Analysis}

Variability studies allow us to assess the plausibility of different X-ray
generation mechanisms.  These can be constrained by the timescales and
flux changes observed in the variability.  Specifically, 
for low mass stars, the cooling
time depends on the loop length, plasma density, conductivity and
emissivity (See \citet{rea02} for a review).  \citet{tes04} undertook
one such                                           
study based on 26 observations of 22 low-mass stars observed with                    
HETG during the first                                                                
three years of the Chandra mission. They concur with the findings of                 
 \citet{hue01} who found that while the continuum emission changes                           
significantly during flares, the                                                     
line emission does not.  However, this is not uniformly the case.                    
Audard et al.\  (2001) found that in the case of the                                 
RS~CVn HR 1099 Ne stays constant at the quiescent level during a                     
flare but the detected a flux increase in the                                        
Fe XXIV  lines during a flare.  G{\"u}del et al. (2001) also found                   
enhancment in the Fe XXIV lines during a flare of the rapidly  rotating              
ZAMS star AB~Dor.                                                                    
                                                                                     
A large proportion of the stars in X-Atlas appear to be variable.  To test for
variability in the flux level over the course of an observation, we constructed
light curves and performed variability analysis on the observations
using two methods. The first method was simple binning.  {\it ACIS
Grating Light Curve
(aglc)}\footnote{http://space.mit.edu/cxc/analysis/aglc/index.html}
was used to create light curves from the HEG and MEG HETG data.  Light
curves with bin sizes of 4000 and 1200 seconds were extracted.

In a more quantitative approach, all observations were analyzed using the
algorithms developed by \citet{gl92} as adapted specifically for
$Chandra$ data \citep{rots06}.  This algorithm uses maximum-likelihood
statistics and evaluates a large number of possible break points
against the prediction of constancy.

An overview of the Gregory--Loredo analysis is presented in Table
\ref{tblvar}. At the time of this analysis there were 131 observations
of 64 sources (24 hot stars and 40 cool stars) in the database.   The
criteria for variability was the probability returned by the
Gregory--Loredo test.   If the probability exceeded 99\% the star was
considered "variable."  If the probability was between 90\% and 99\%,
the star was deemed "probably variable."  Stars with variability
probability below 90\% were considered "not variable."  A source was
considered variable if any observation of that source was noted to be
variable. Epoch to epoch comparisons were not undertaken in a
systematic way for this study. 

Table \ref{tblvar} is designed only to give an example of the content
of the database.  The high-mass star classification combines single OB
stars, multiple OB stars and Wolf-Rayet stars.  We find a variability
rate of $\sim$ 40\% among the high-mass stars. This is a remarkably
high rate since the X-ray production mechanisms of high-mass stars are
expected to be wind driven and hence more stable than dynamo-driven
low-mass stars, which are subject to loop reconnection at all scales.
A closer examination of the data reveals that of the variables, several
are unique objects, including Algol, $\gamma$ Cas, $\theta^{1}$ Ori C,
and $\eta$ Carina.

Although it contains a B8 star, Algol also contains a K1 subgiant.
The latter tends to be active.  The flare seen from Algol {\em is of
particular interest}, though, as it appears to have induced a change
in the apparent characteristic flux level.  HD 150136, 9 Sgr, $\tau$ Sco,
$\tau$ CMa and HD 206267 are generally well-behaved OB stars that all
exhibited small deviations on short time scales (10-20 ks).  $\gamma$
Cas is a unique hot star that shows fluctuations of about 25\% on
short time scales.  The variability of $\theta^{1}$ Ori C has been
previously reported \citep{sch01,gag05} and is thought to
be related to its relatively high temperature and strong variable
magnetic field.  Support for this hypothesis comes from the
observations of $\theta^{2}$ Ori A (discussed in further detail in
\citet{sch06}).  This star shows variability in only one of its two
HETG observations but is also seen to shift from a cool to a hot state
as evidenced by the two-temperature fit in which the kT value jumps
from a single temperature of around 0.7 keV in epoch one to two
temperatures of approximately 17 and 2 keV in epoch two
(Table \ref{tblfit}).  The flux is seen to drop steadily and by 25\%
over the course of the hotter, but generally brighter, epoch. 

About 85\% of the low-mass stars are found to be variable.  This
proportion is well in excess of the typical fraction reported in star
forming regions, which is about 30\% in a 100 ks observation
(Cf. \citet{get02,wol06}). This apparent discrepancy is undoubtedly 
due to counting statistics and
selection effects.  Wolk et al.\ (2007) report that in a 90~ks
observation of RCW 108, 66\% of stars with over 200 counts are
detected as variable but only 21\% of stars with between 100 and 200
net counts.  In the X-Atlas database almost all the stars are bright
and have over 1000 $0^{th}$-order counts. The results for low-mass stars
here are consistent with the results of the COUP survey \citep{get05}
which found the majority of the $\sim$1600 stars monitored for $\sim$10
days to be variable at some level. 

With the exception of the composite system Algol, no high-mass star
was seen to flare. Among the low-mass stars, TZ CrB, CC Eri, TW Hya,
and DoAr 21 showed the strongest flares, more than doubling their flux
in less than one hour with cooling times in excess of 10 ks.  The
cooling time of stars with disks, such as TW Hya, is of particular
interest since the cooling time is related to the loop length. Such
systems can provide direct evidence of star--disk coupling via the
magnetic field.  While these are the longest cooling times in the
group, 10-30 ks is not unusually long (cf. \citet{fav05}).  The
X-Atlas database is a very biased sample for flare studies.  Many
stars were targeted precisely because they were known flare stars
(e.g. AD Leo, EV Lac, and CC Eri) or rotationally variable (e.g. HD
223460, FK Com and AB Dor).

\subsection{Variability of Capella}

Over twelve observations from 1999 to 2006, Capella's X-ray hardness
ratios are clustered at $(HR_1,HR_2)=(-0.75\pm0.01,-0.058\pm0.025)$
(see Figure~\ref{fighr}).  These are significantly overdispersed compared to the typical
errors in the individual hardness ratios, by factors ranging
from 5x to 12x.  Over this same period, the overall X-ray count
rate undergoes variations of $\approx17$\%.  However, the hardness
ratios are uncorrelated with the count rates: we find correlations
of Spearman's $\rho=-0.033\ (p=0.9)$ for $HR_1$ versus rate, and
$\rho=0.11\ (p=0.7)$ for $HR_2$ versus rate.

Within each observation, there is no evidence for any variability,
as is confirmed by an application of the Gregory-Loredo test
\citep{gl92}: the odds that the star varies during any of the
individual observation is $<90$\%.  In contrast, combining 15 of the
16 available observations (excepting ObsID 1199, a 2 ks observation)
to construct a concatenated event list spanning over 350~ks, and
applying the same test to this concatenated events list, we find
that the odds of Capella being variable are 10:1, indicating a probability
of over 0.99 that Capella is variable on timescales of years.

Capella, as the strongest coronal source, has been a common X-ray
calibration target for many instruments.  It has remained stable
in its total X-ray output even as the detailed emission structure
has shown considerable change (see e.g., \citet{you01}).  For
instance, \citet{dup96} have shown based on EUVE data that the high-temperature component is
variable, a finding confirmed by \citet{raa07} based on
LETG data.  There is also considerable evidence for the
dominant emission to change between the G1\,III primary and the
G8\,III secondary (e.g., \citet{lin98,joh02,ish06}).

We confirm the primary conclusions of \citet{raa07}, who
found no evidence of changes in intensity over timescales of $<100$~ks,
but a large timescale variation that resulted in an increased flux
early 2006.  We find no evidence for variations at timescales $<50$~ks,
and conclusively find that the source is variable over larger
timescales.

\section{SUMMARY\label{s:summary}}

By developing a uniformly processed stellar database,
we intend to provide a powerful yet accessible tool for the study of
stellar X-ray emission.  As the early science results from X-Atlas
demonstrate, the ability to compare the high-mass and low-mass stars
observed with HETG opens up intriguing avenues of investigation.  The
spectral montages and line-profile plots created from the
high-resolution HETG data offer a detailed glimpse at the physical
mechanisms for X-ray generation among stellar types.  In addition, by
convolving the high-resolution spectra to low-resolution, we can
compare the limited number of stars studied in high-resolution to the
thousands of sources in $Chandra$ images.  Spectral fits and X-ray
color analyses performed on the predicted low-resolution spectra
reveal differences in the X-ray temperature distributions of high-mass
and low-mass stars and provide evidence for at least two distinct
classes of X-ray emission in high-mass stars.  Variability studies
help to further constrain the causes of stellar X-ray emission and
suggest that the proportion of high-mass stars exhibiting X-ray variability may
be greater than previously suspected.  We also definitively establish
the X-ray variability of Capella over timescales of years.  We have made
available the means to conduct systematic studies of the stellar HETG
dataset, and therefore hope that these early results will represent only the
beginnings of the scientific return of the database.  Finally, we have
applied the reduction and analysis methods described above to
all non-stellar point sources observed with HETG, including neutron
stars, X-ray binaries, and active galatic nuclei.  Future releases of
X-Atlas will adapt these methods to LETG point source observations as
well, extending the atlas to all $Chandra$ gratings observation of point
sources.

\acknowledgments

We are happy to thank the referee John Pye for a careful reading 
and a number of improvements.
Funding for this work was provided by $Chandra$ Grants GO5-6006E and
GO5-6007A, Chandra X-ray Center NASA Contract NAS8-39073, and by
ANCHORS (AN archive of CHandra Observations of Regions of Star
formation), $Chandra$ archival project AR5-6002A.  Wayne L. Waldron
was supported in part by $Chandra$ grant GO5-6006A.

We would like to thank N. S. Schulz, D. P. Huenemoerder,
and P. Testa for feedback during the testing stages of the project.

\clearpage

\clearpage


\clearpage

\begin{figure}
\includegraphics[width=6.2in]{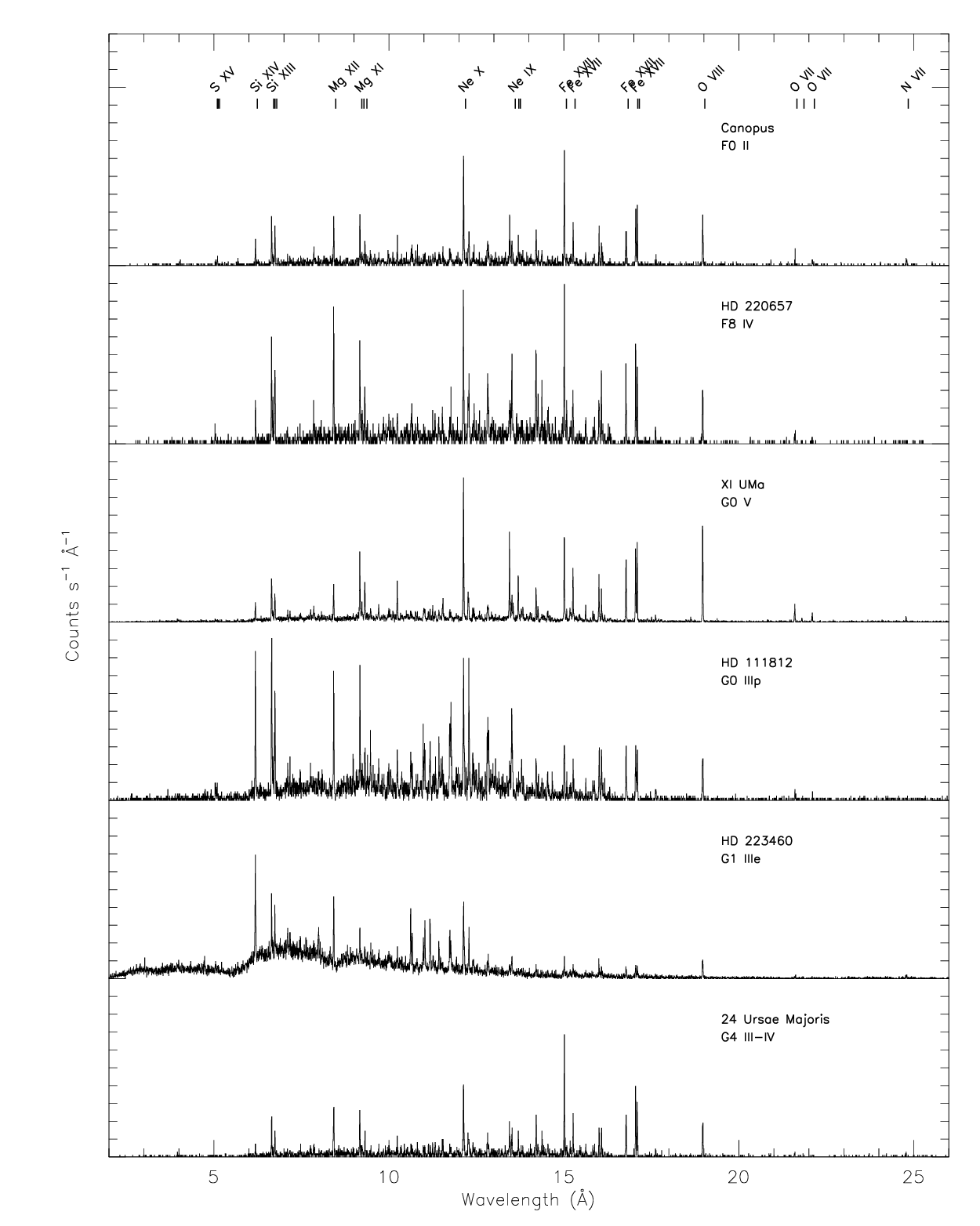}
\end{figure}

\begin{figure}
\includegraphics[width=6.2in]{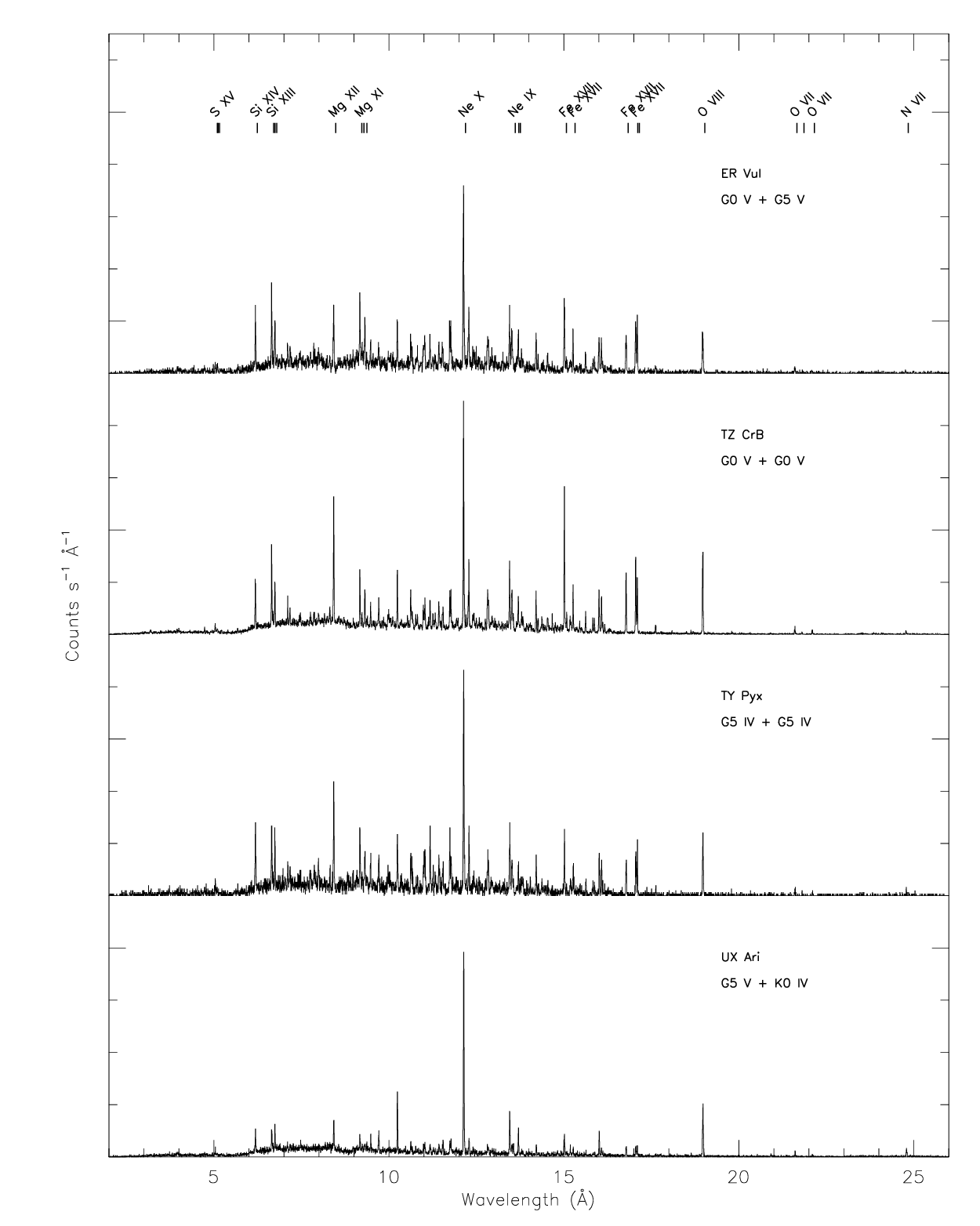}
\end{figure}

\begin{figure}
\includegraphics[width=6.2in]{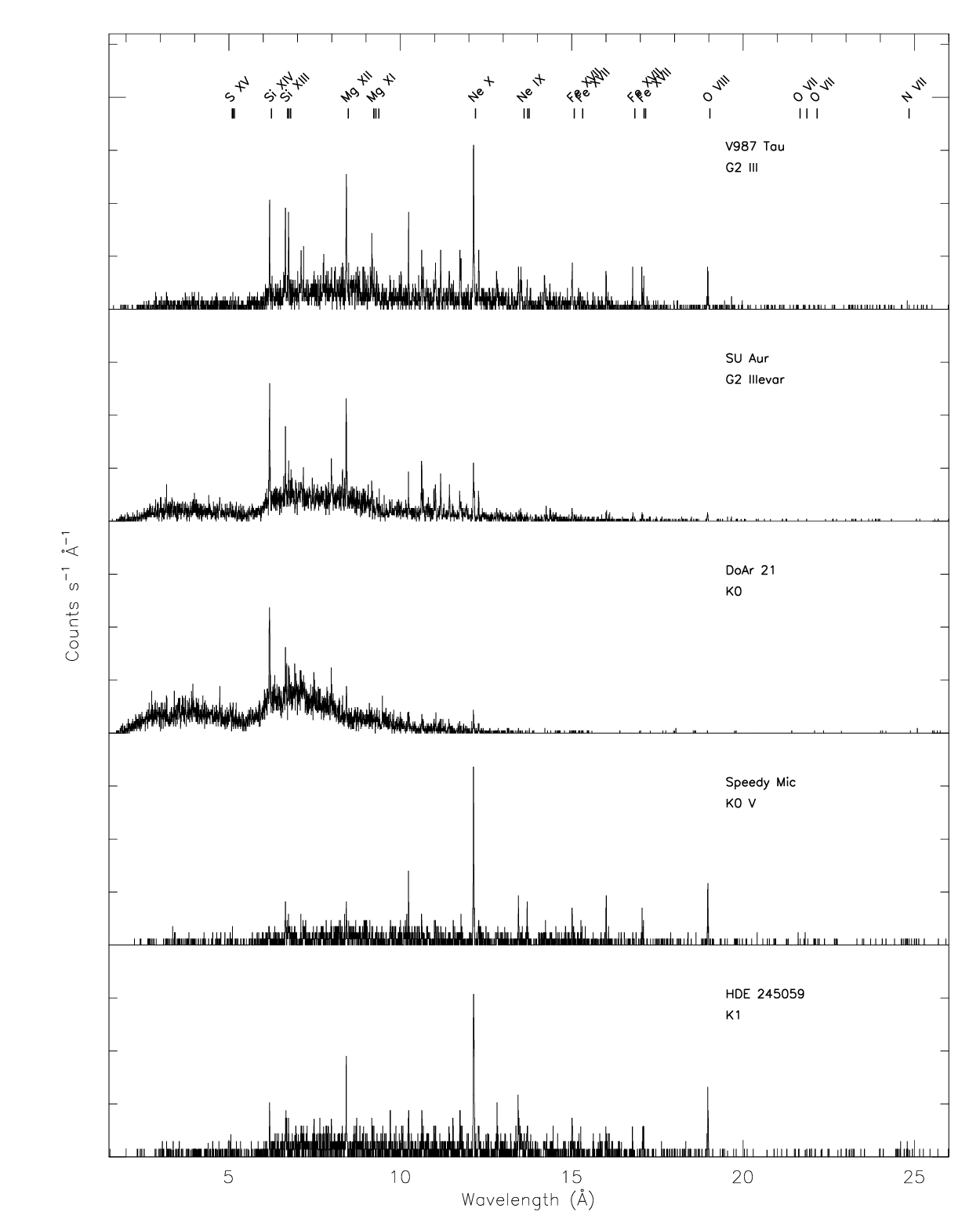}
\end{figure}

\clearpage


\clearpage

\clearpage

\begin{figure}
\includegraphics[width=6.2in]{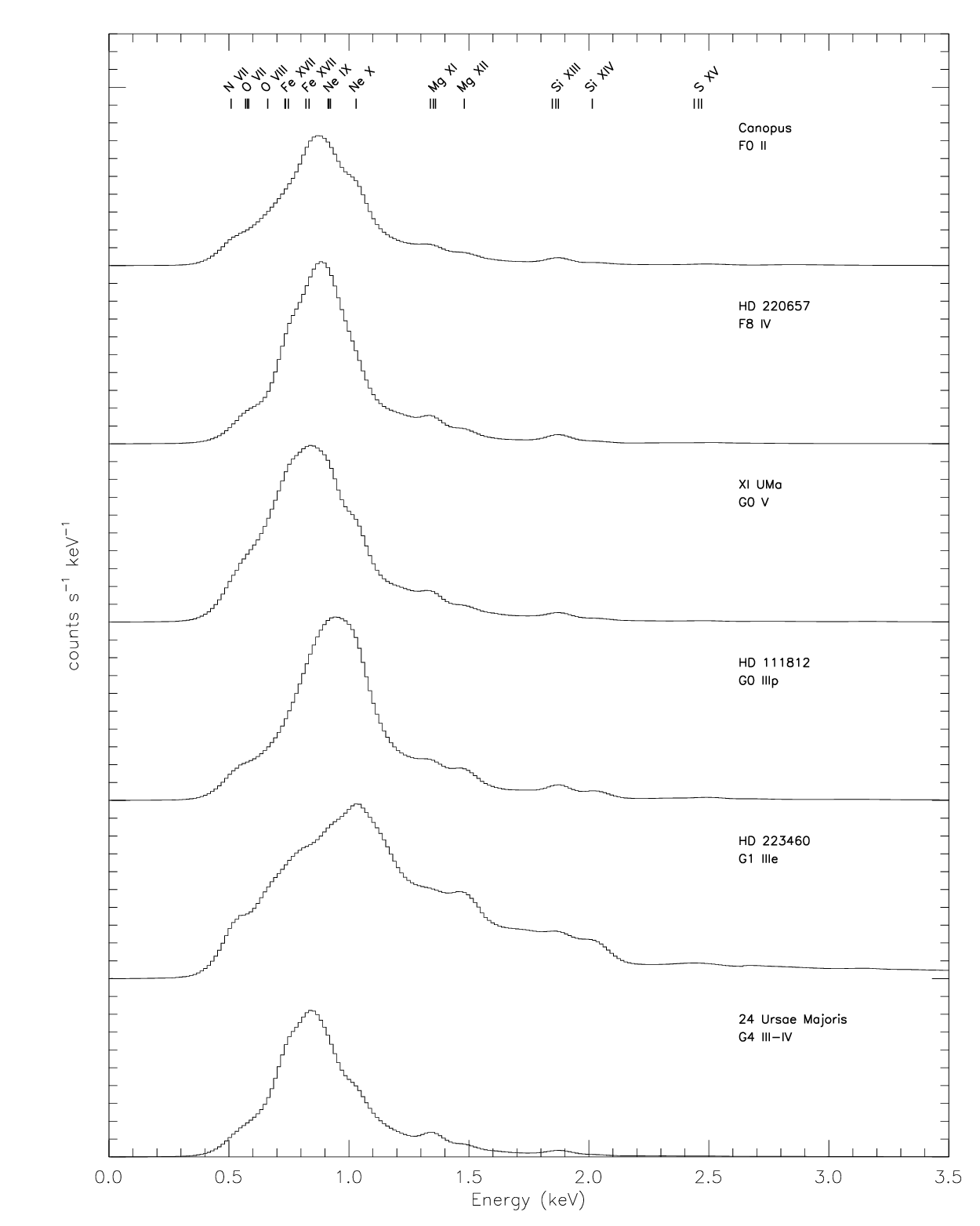}
\end{figure}

\begin{figure}
\includegraphics[width=6.2in]{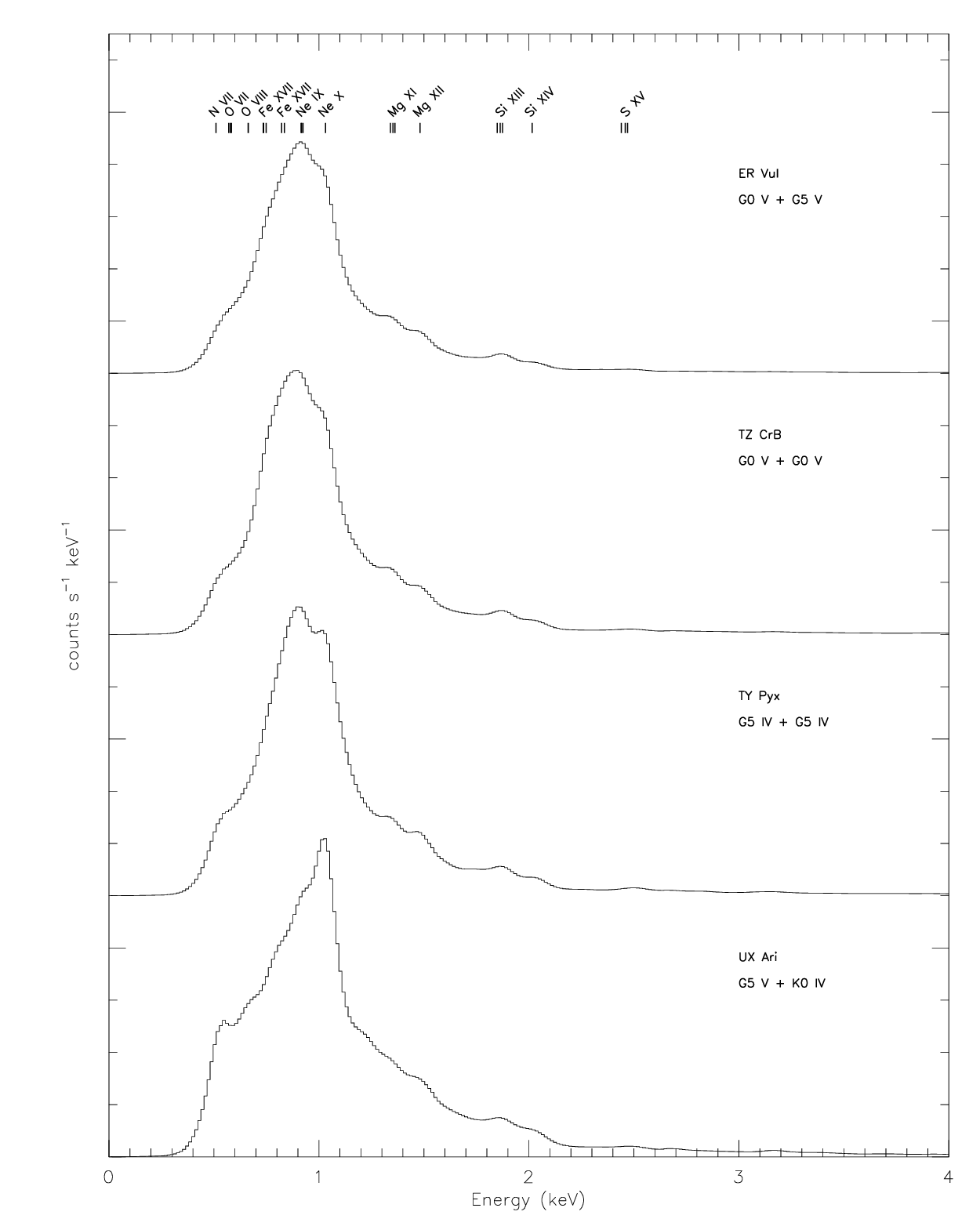}
\end{figure}

\begin{figure}
\includegraphics[width=6.2in]{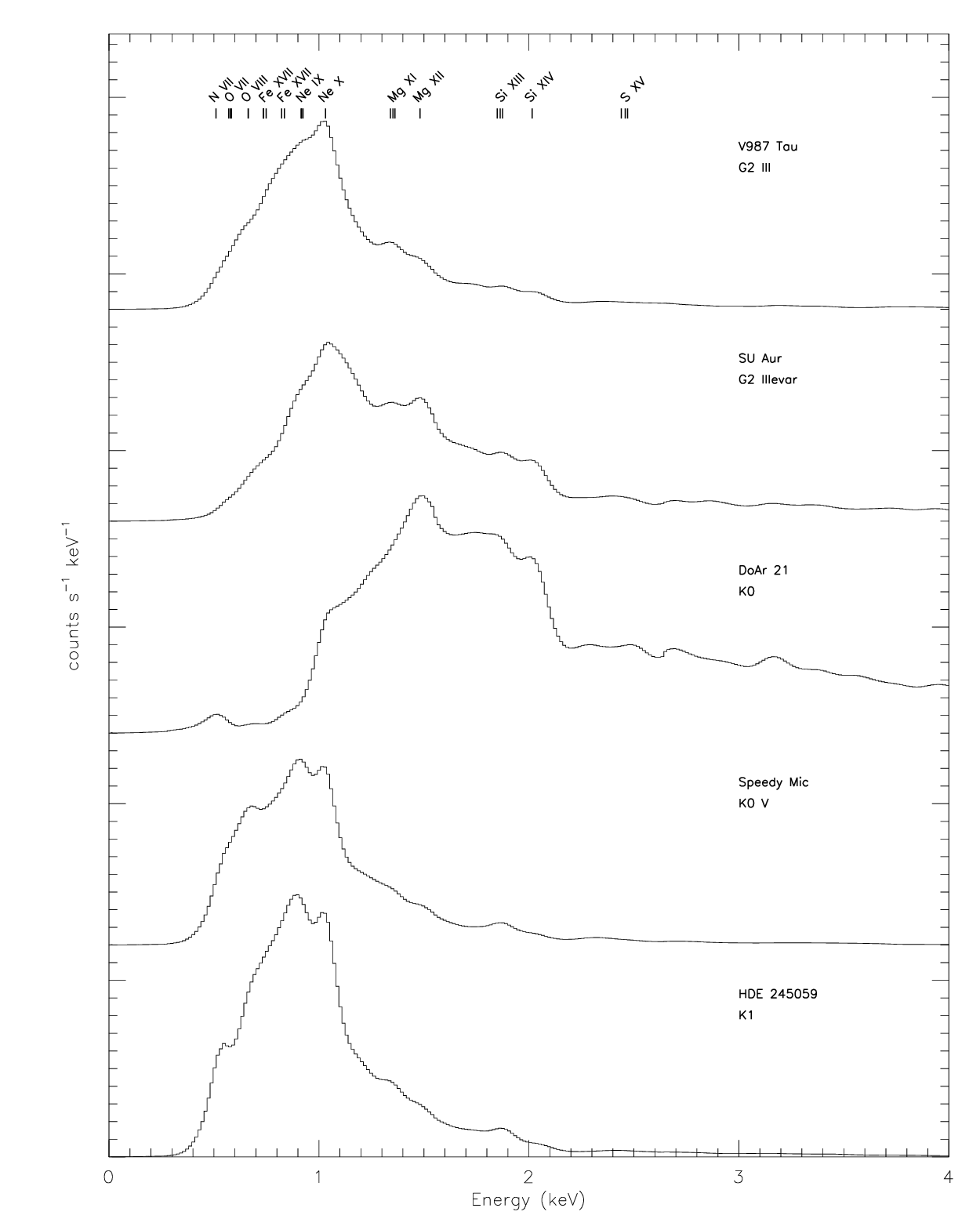}
\end{figure}

\clearpage


\clearpage

\clearpage

\begin{figure}
\includegraphics[width=3.1in]{f35a.ps}
\includegraphics[width=3.1in]{f35b.ps}
\newline
\newline
\includegraphics[width=3.1in]{f35c.ps}
\includegraphics[width=3.1in]{f35d.ps}
\newline
\newline
\includegraphics[width=3.1in]{f35e.ps}
\includegraphics[width=3.1in]{f35f.ps}
\newline
\newline
\includegraphics[width=6.2in]{f35g.ps}
\end{figure}
\begin{figure}
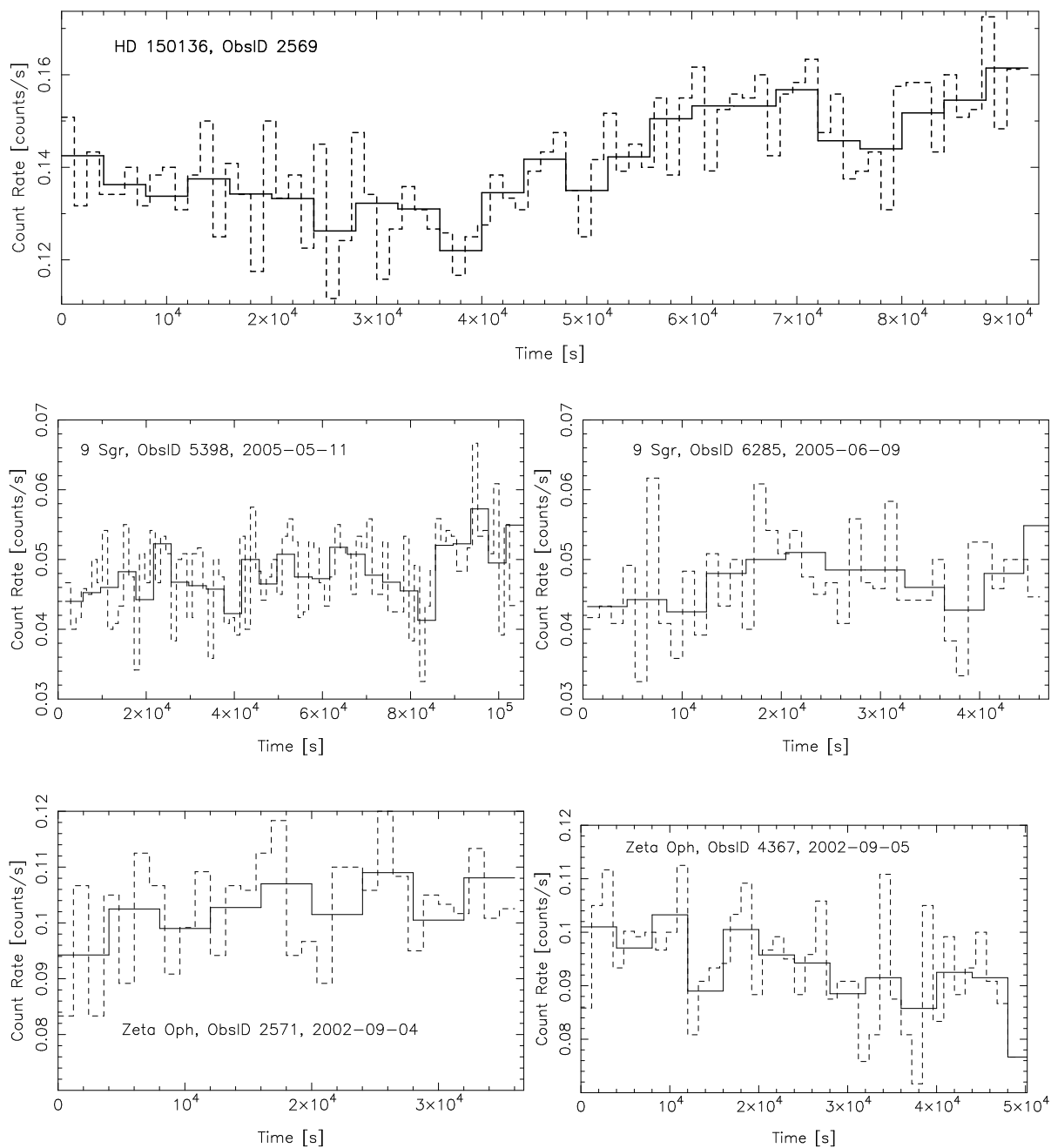

\includegraphics[width=6.2in]{f35h.ps}
\newline
\newline
\includegraphics[width=3.1in]{f35i.ps}
\includegraphics[width=3.1in]{f35j.ps}
\newline
\newline
\includegraphics[width=3.1in]{f35k.ps}
\includegraphics[width=3.1in]{f35l.ps}
\caption{\scriptsize Light curves obtained from the combined
dispersed counts in the HEG and MEG arms
of selected HETG observations of
high-mass stars.  The light curves were binned at time intervals of 1200 s {\it (dashed line)} and 4000 s {\it
(solid line)}:
Row 1-2: Four observations of $\theta^{1}$ Ori C.  Row 3: Two
observations of $\theta^{2}$ Ori A. Row 4: $\gamma$ Cas.  Row 5: HD 150136.  Row 6: Two observations of 9 Sgr.  Row 7:
Two observations of $\zeta$ Oph.\label{figlchm}}
\end{figure}

\begin{figure}
\includegraphics[width=6.2in]{f36a.ps}
\newline
\newline
\includegraphics[width=6.2in]{f36b.ps}
\newline
\newline
\includegraphics[width=6.2in]{f36c.ps}
\newline
\newline
\includegraphics[width=6.2in]{f36d.ps}
\newline
\newline
\end{figure}
\begin{figure}
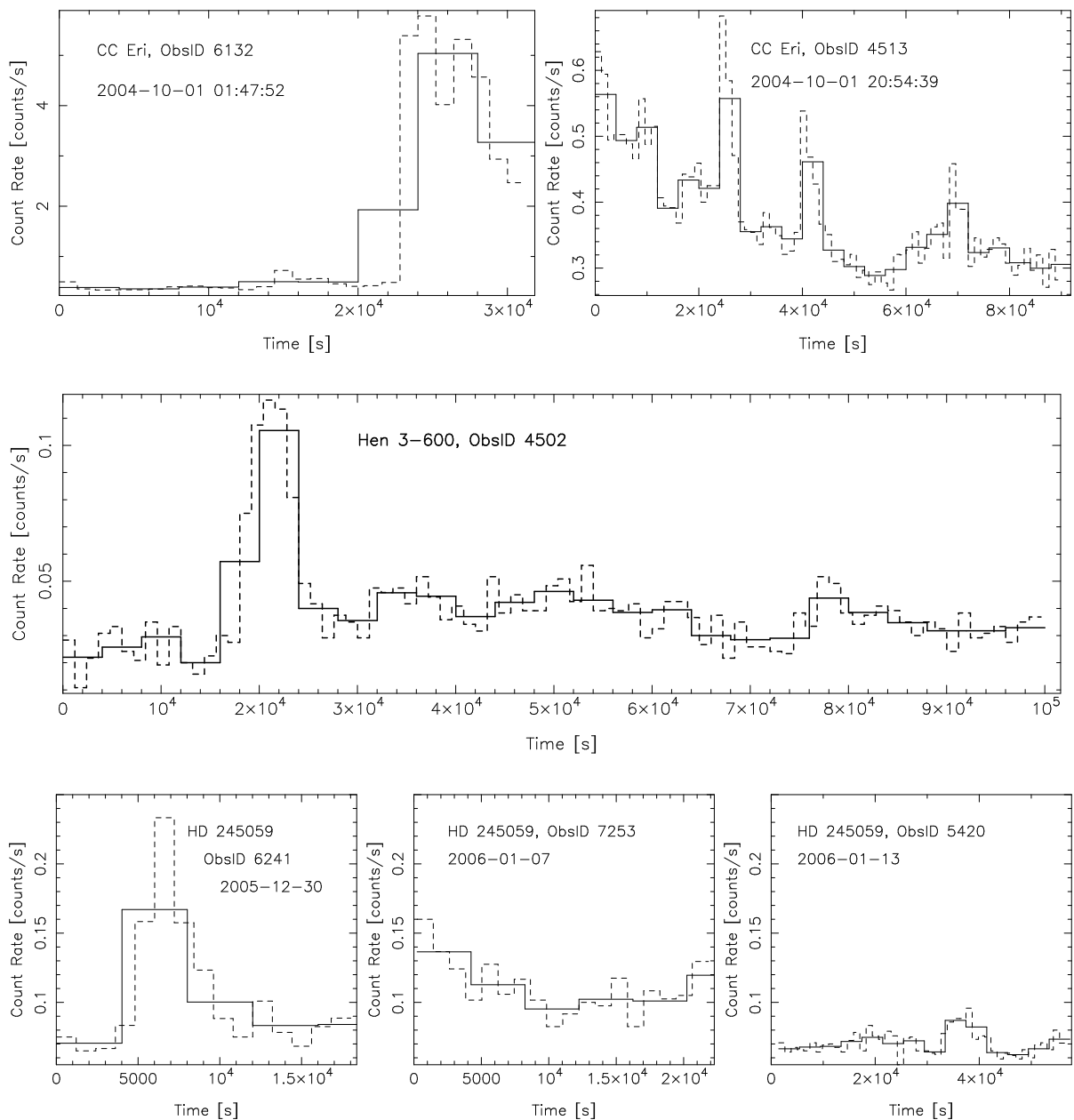

\includegraphics[width=3.1in]{f36e.ps}
\includegraphics[width=3.1in]{f36f.ps}
\newline
\newline
\includegraphics[width=6.2in]{f36g.ps}
\newline
\newline
\includegraphics[width=2.05in]{f36h.ps}
\includegraphics[width=2.05in]{f36i.ps}
\includegraphics[width=2.05in]{f36j.ps}
\caption{\scriptsize As in Figure~\ref{figlchm}, for
selected low-mass stars.
Row 1: Algol.  Row 2: TW Hya.  Row 3: DoAr 21.  Row 4: Proxima
Centauri.  Row 5: Two observations of CC Eri.  Row 6: Hen 3-600.  Row
7: Three observations of HD 245059. \label{figlclm}}
\end{figure}
\begin{figure}
\includegraphics[width=2.05in]{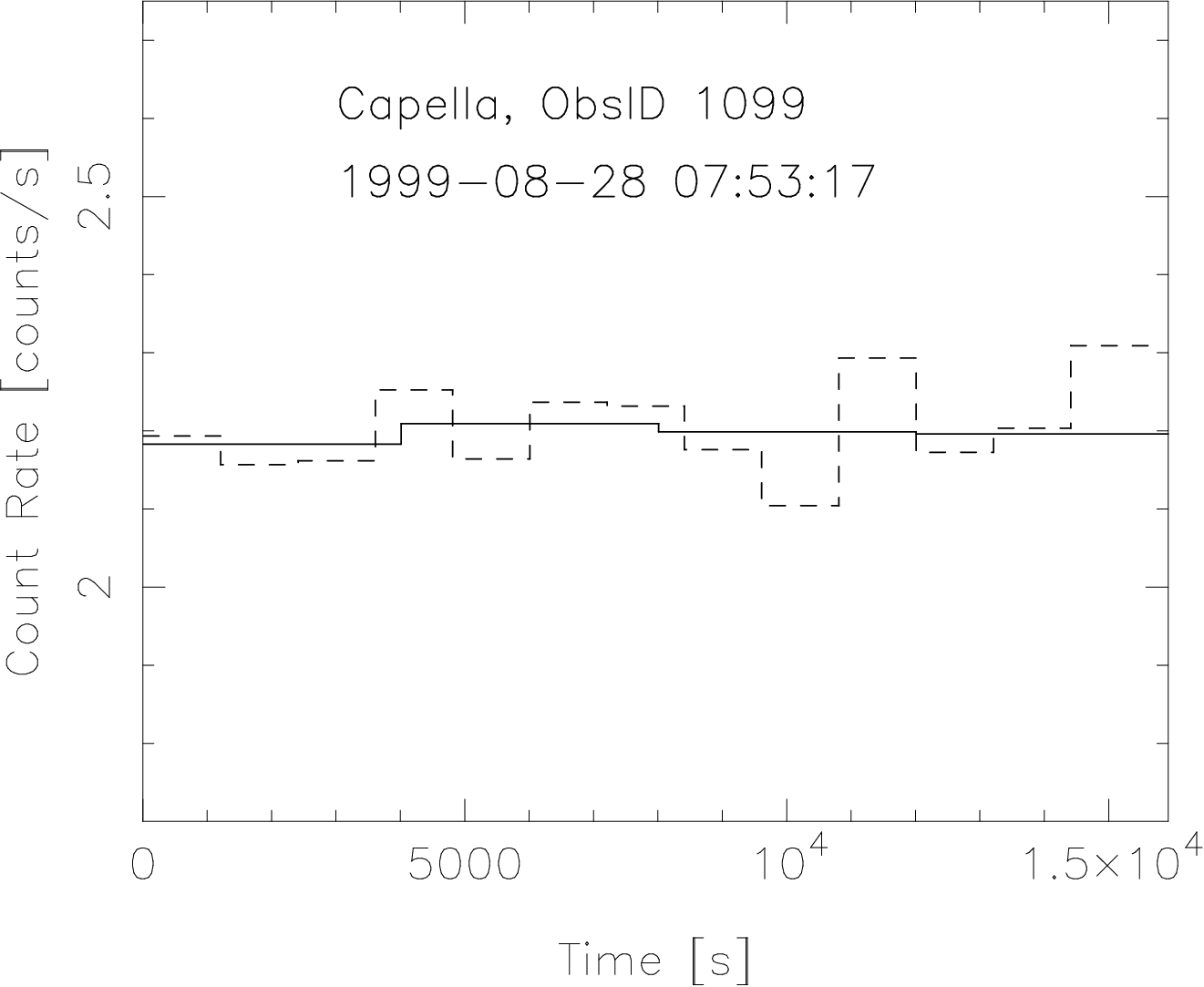}
\includegraphics[width=2.05in]{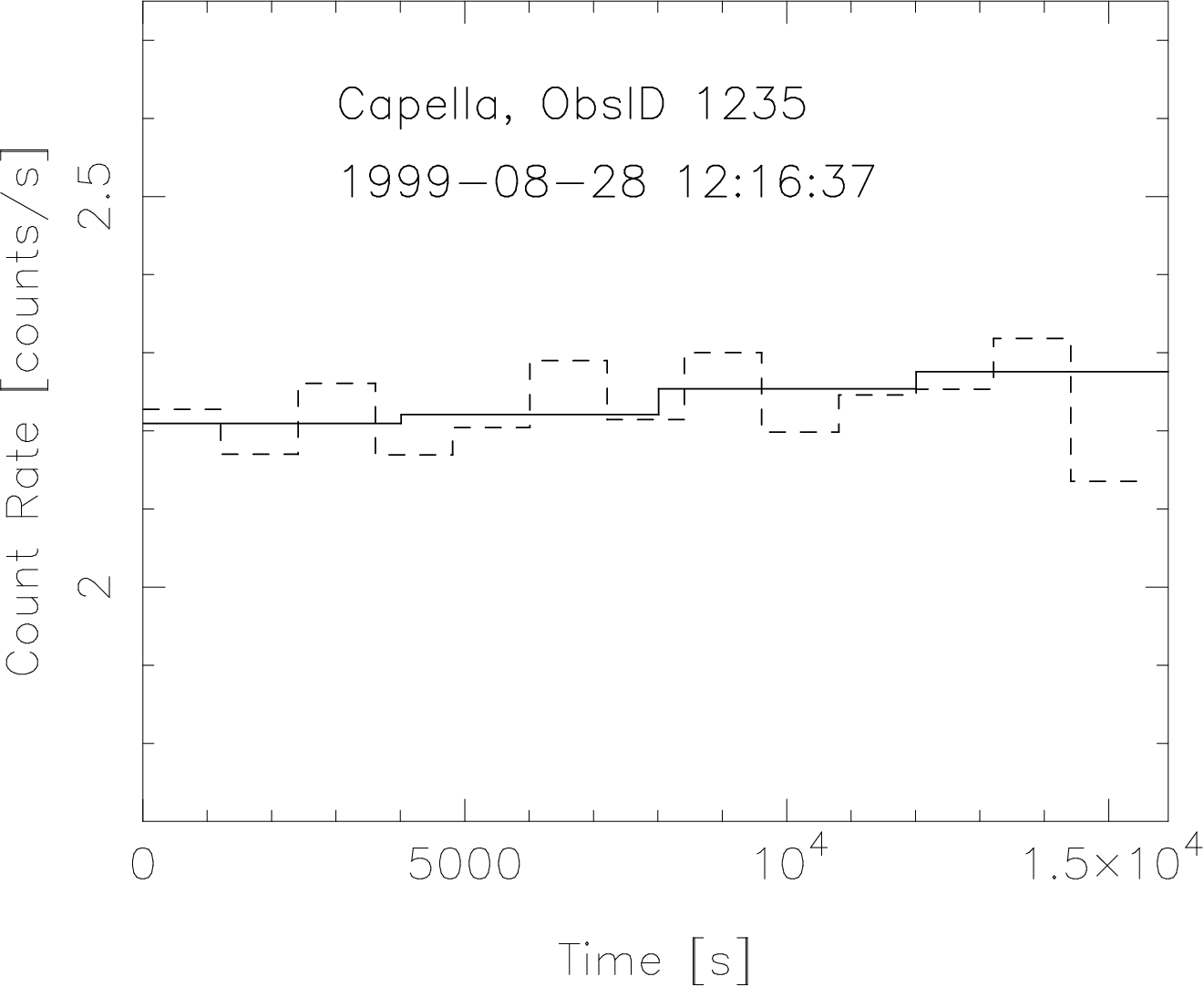}
\includegraphics[width=2.05in]{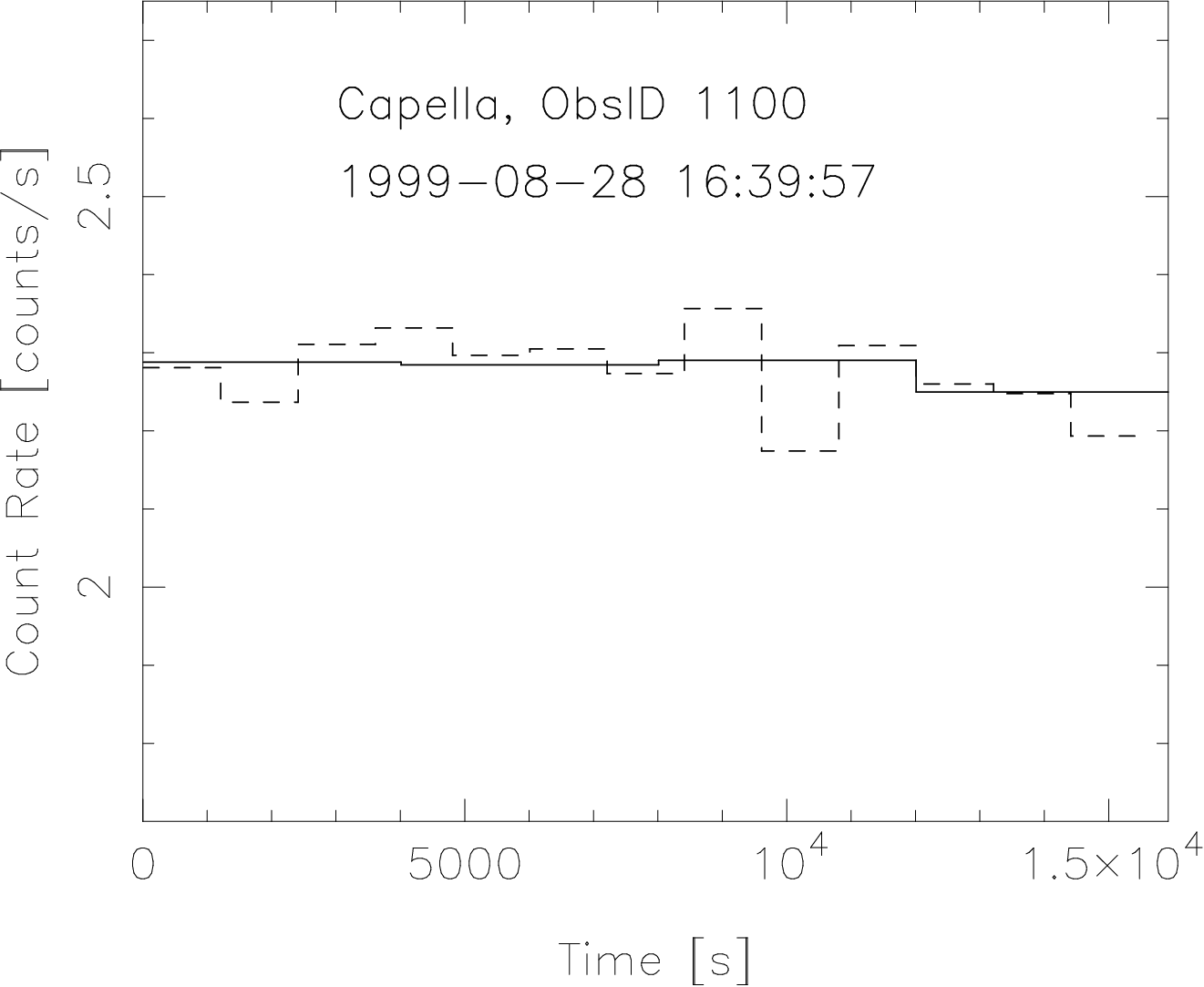}
\newline
\newline
\includegraphics[width=2.05in]{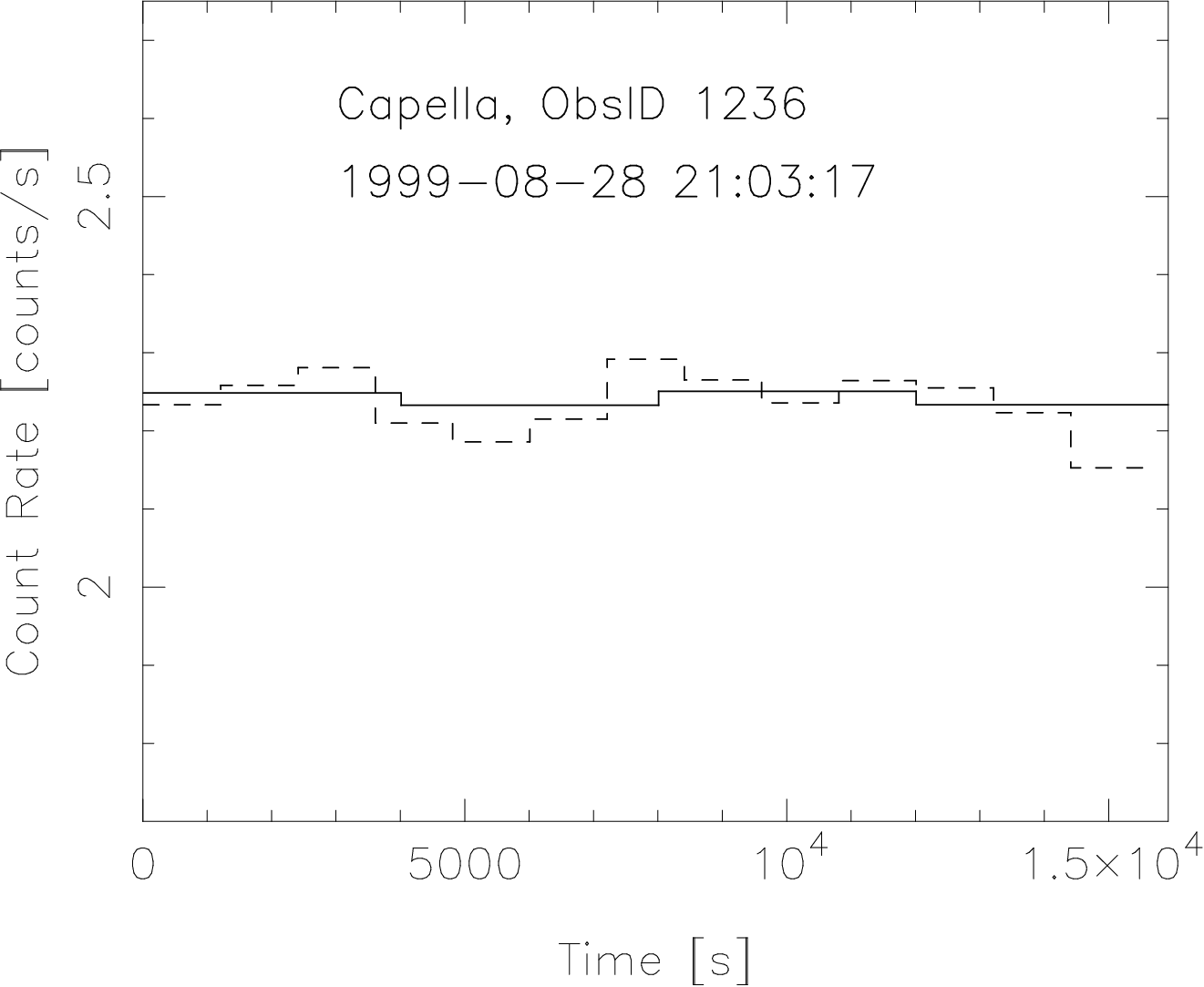}
\includegraphics[width=2.05in]{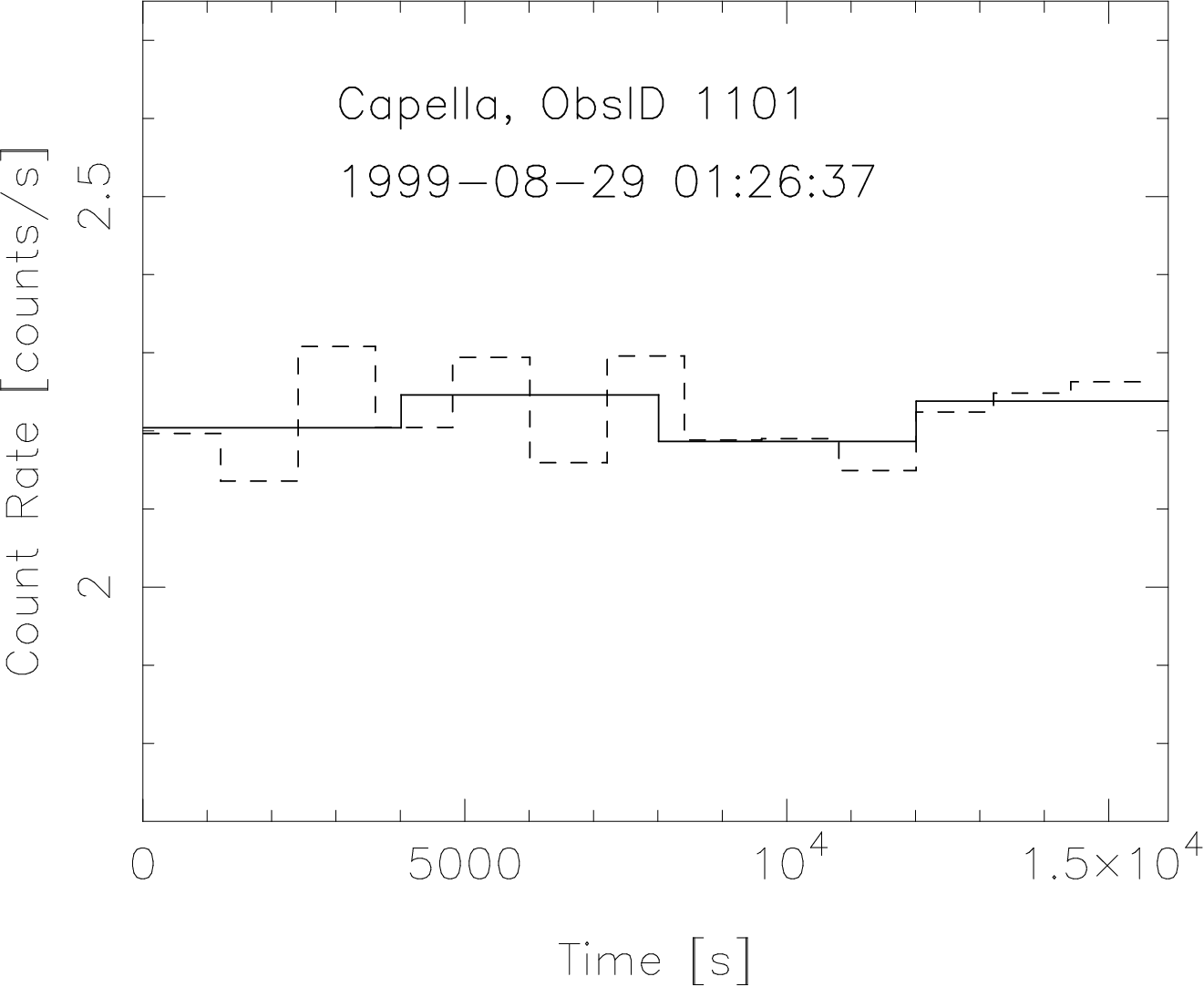}
\includegraphics[width=2.05in]{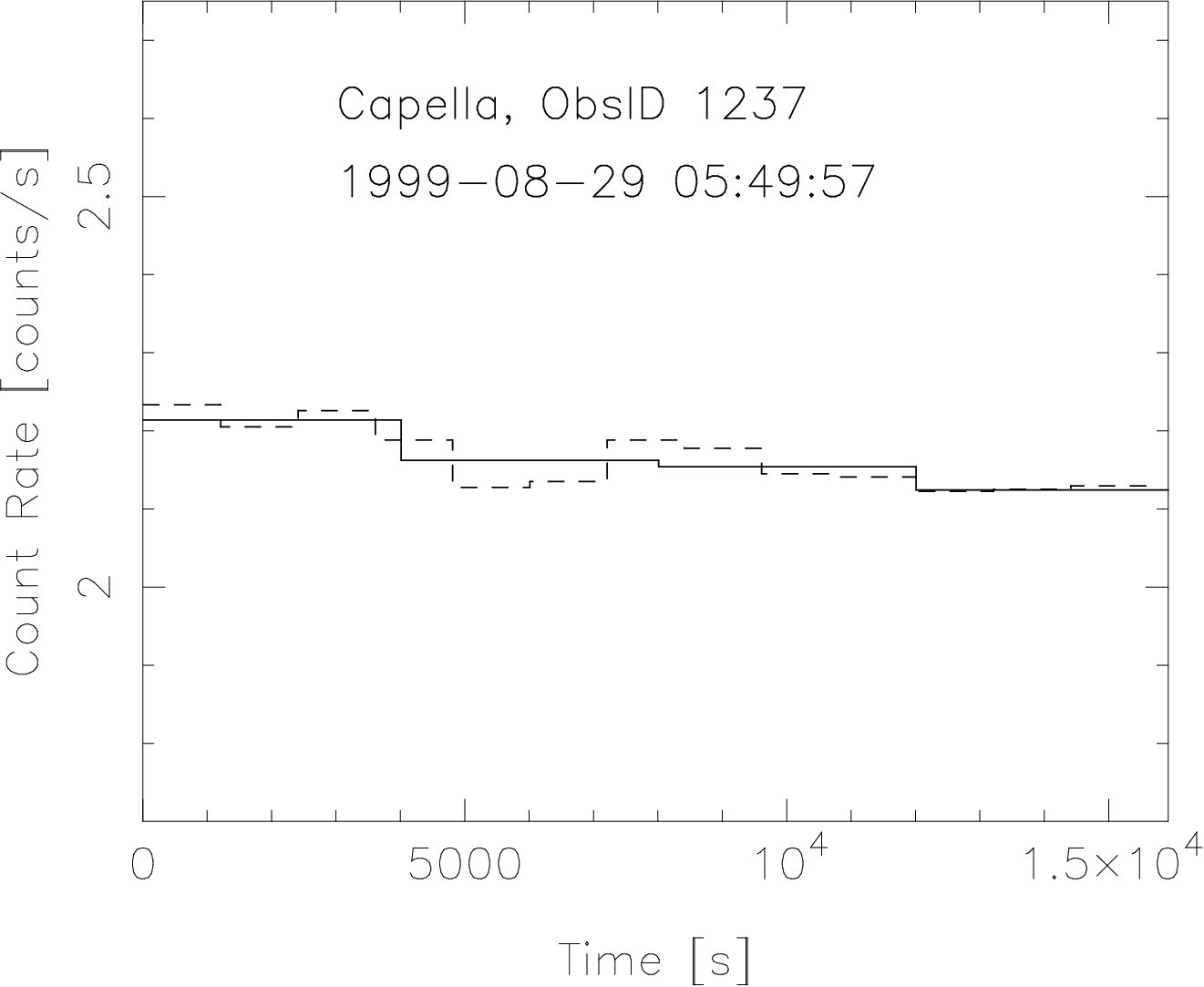}
\newline
\newline
\includegraphics[width=2.05in]{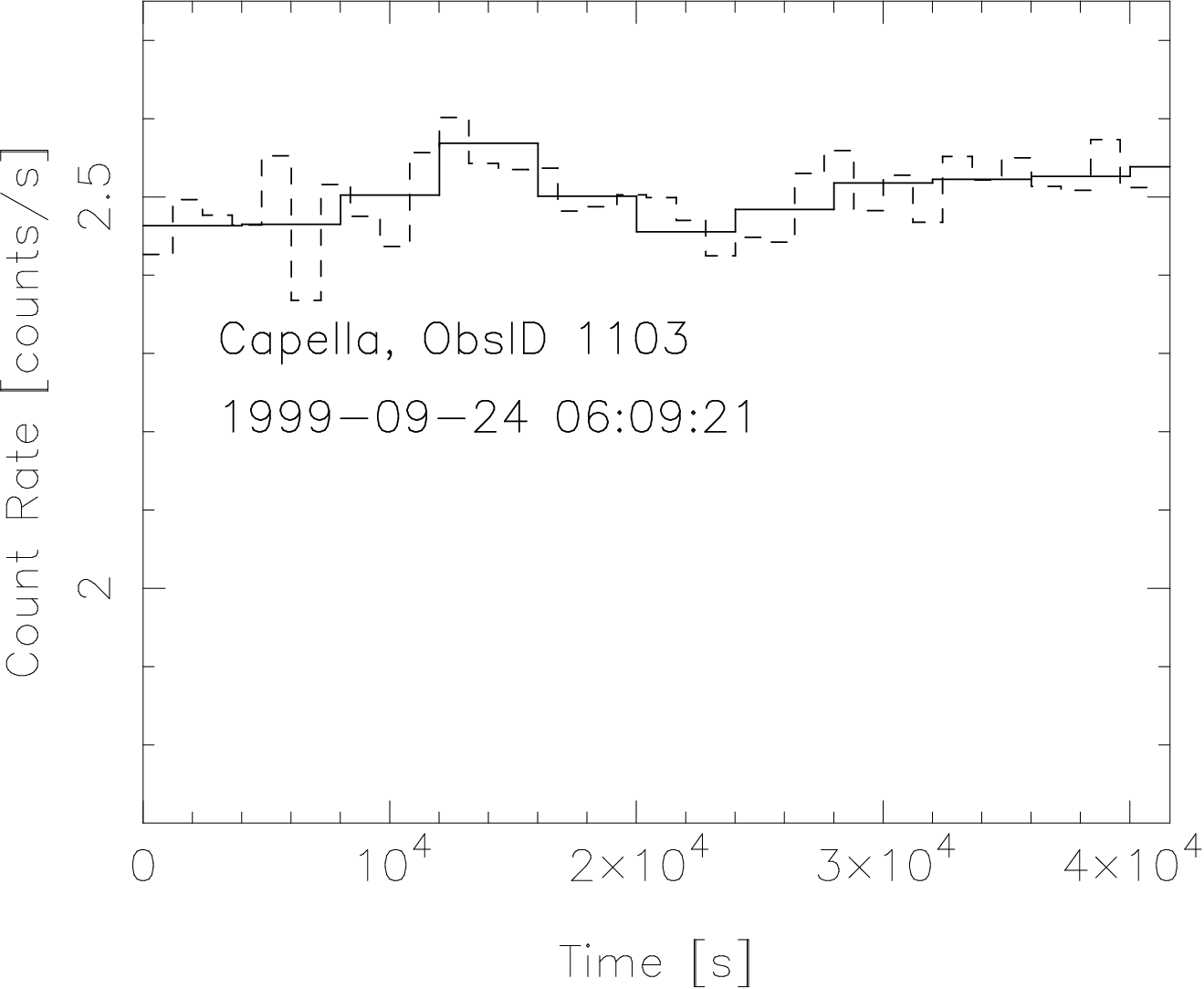}
\includegraphics[width=2.05in]{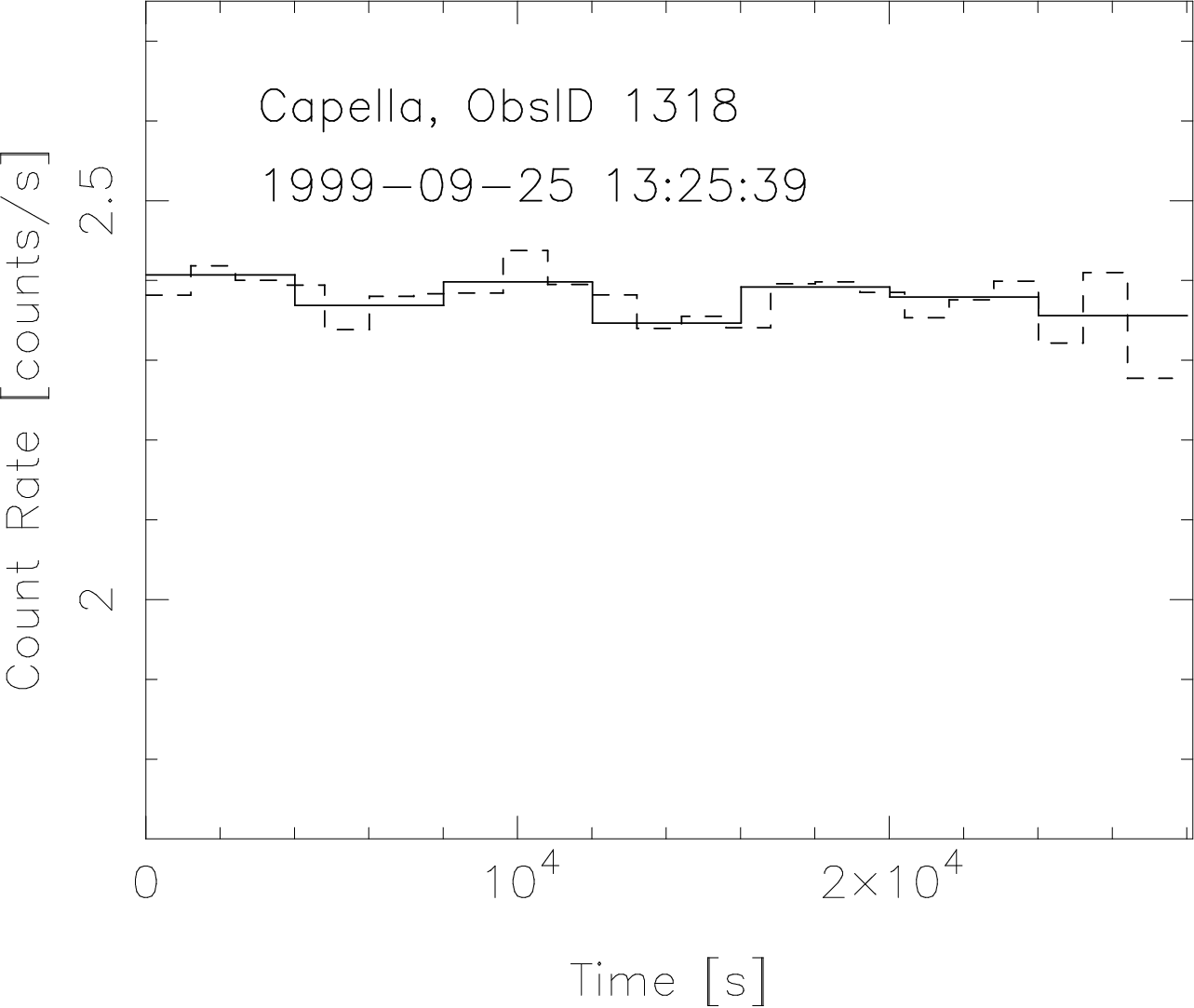}
\includegraphics[width=2.05in]{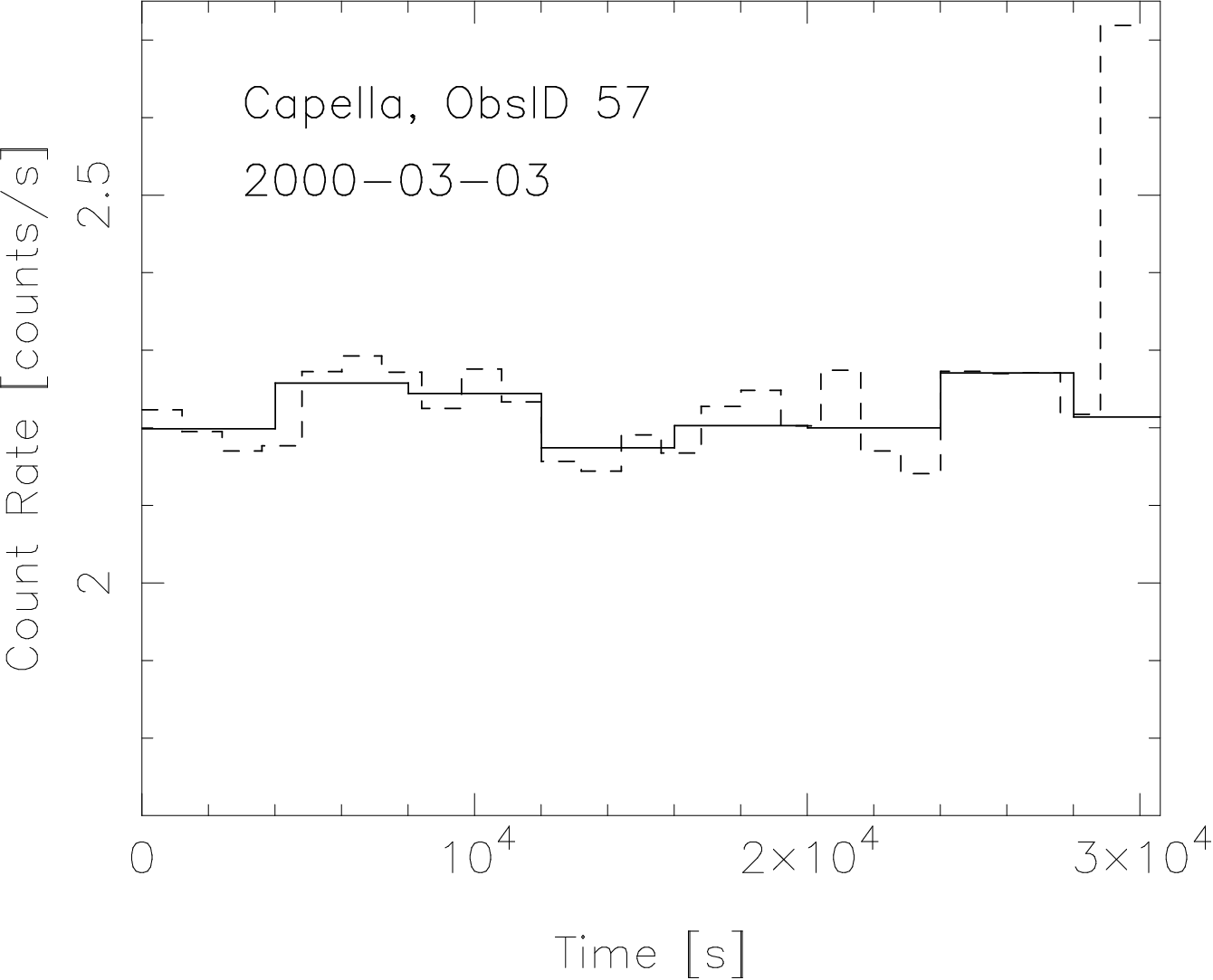}
\newline
\newline
\includegraphics[width=2.05in]{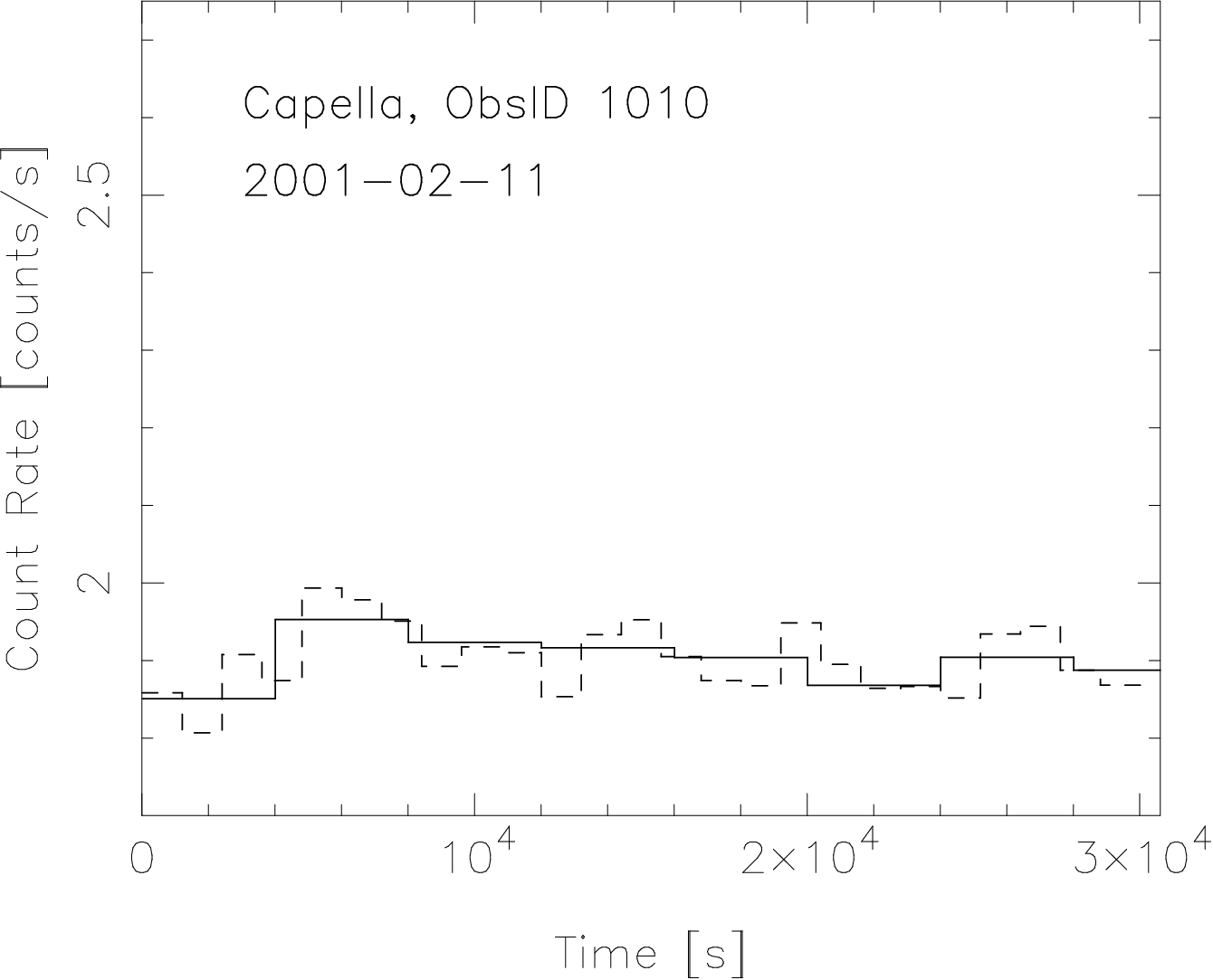}
\includegraphics[width=2.05in]{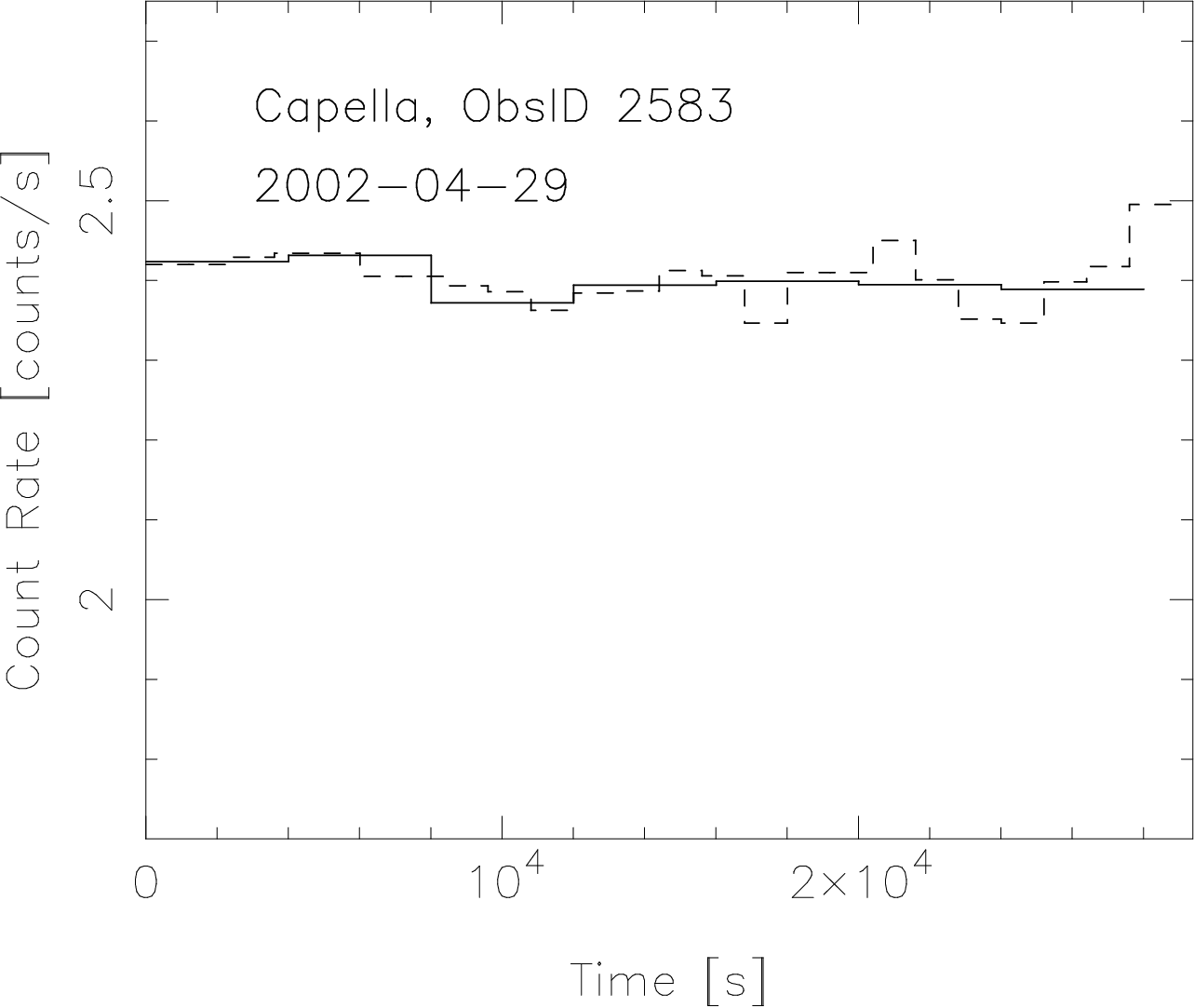}
\includegraphics[width=2.05in]{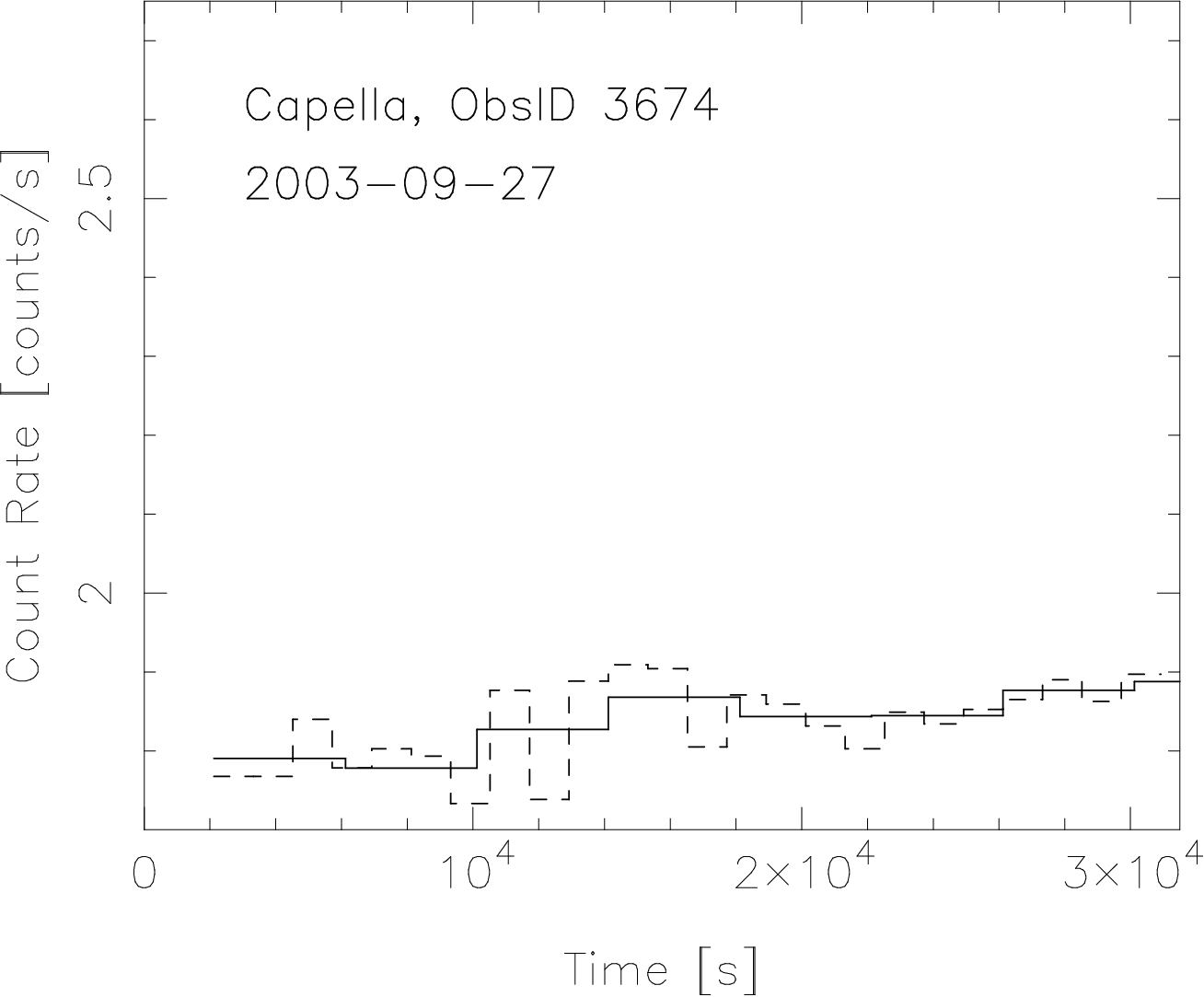}
\newline
\newline
\includegraphics[width=2.05in]{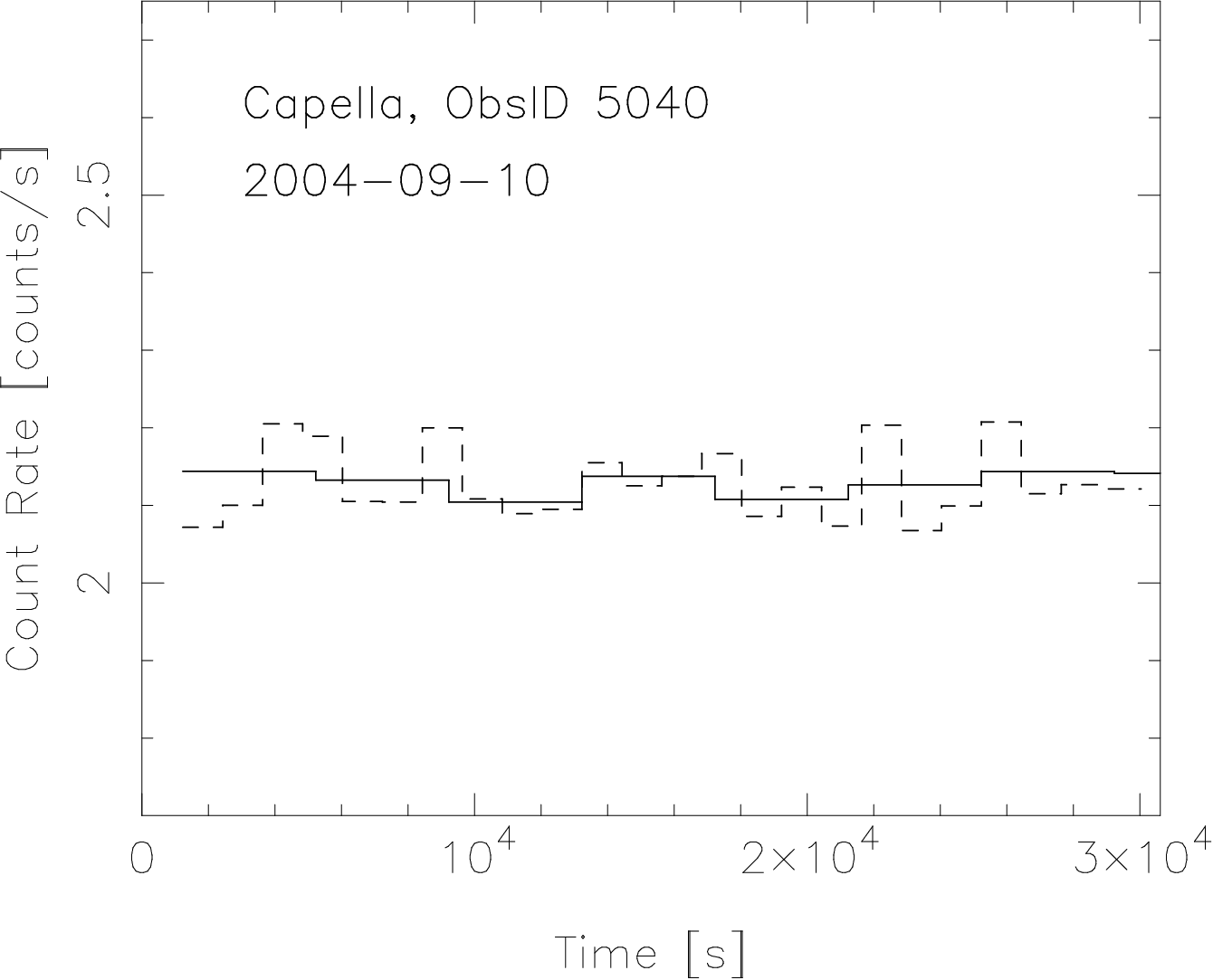}
\includegraphics[width=2.05in]{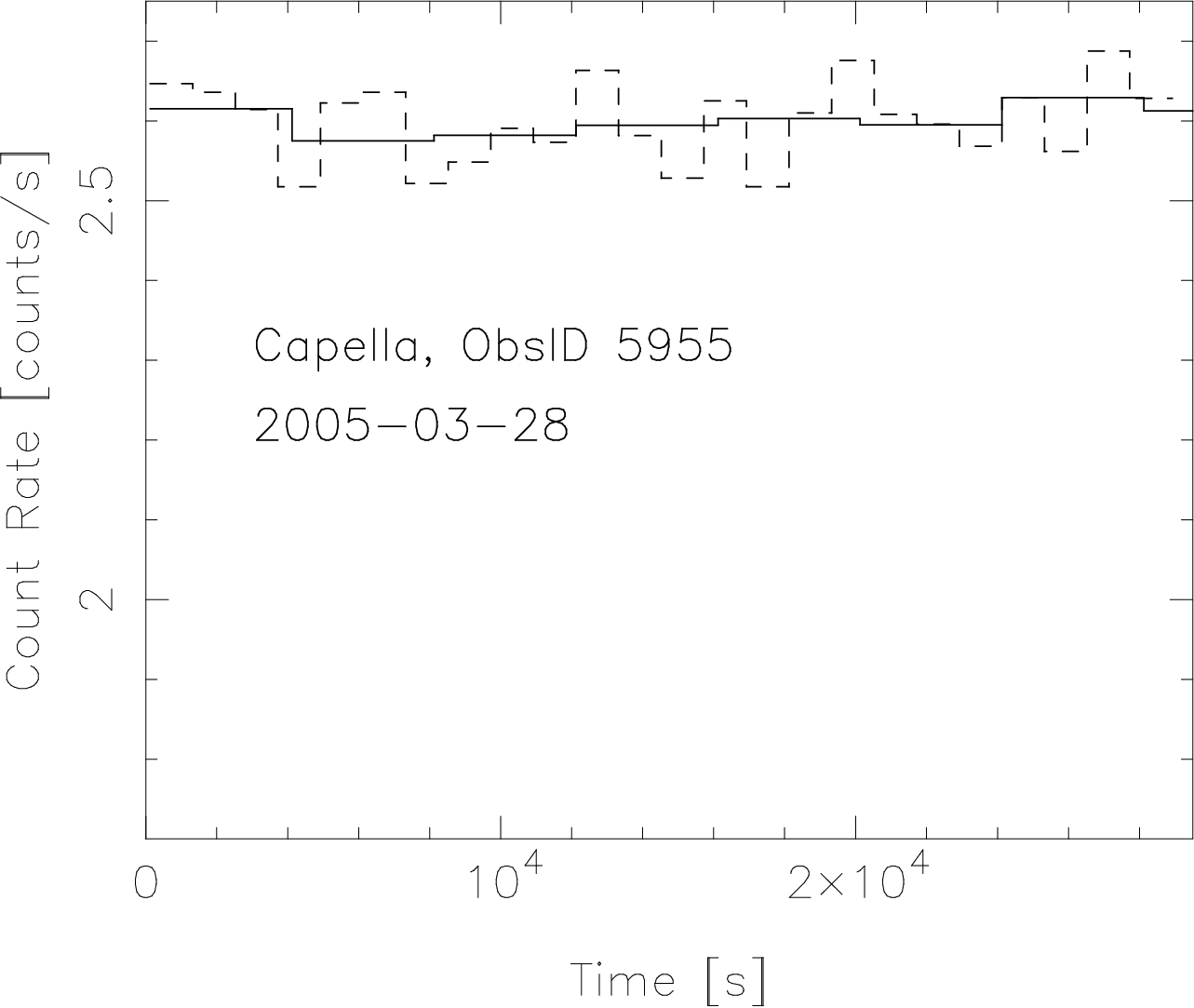}
\includegraphics[width=2.05in]{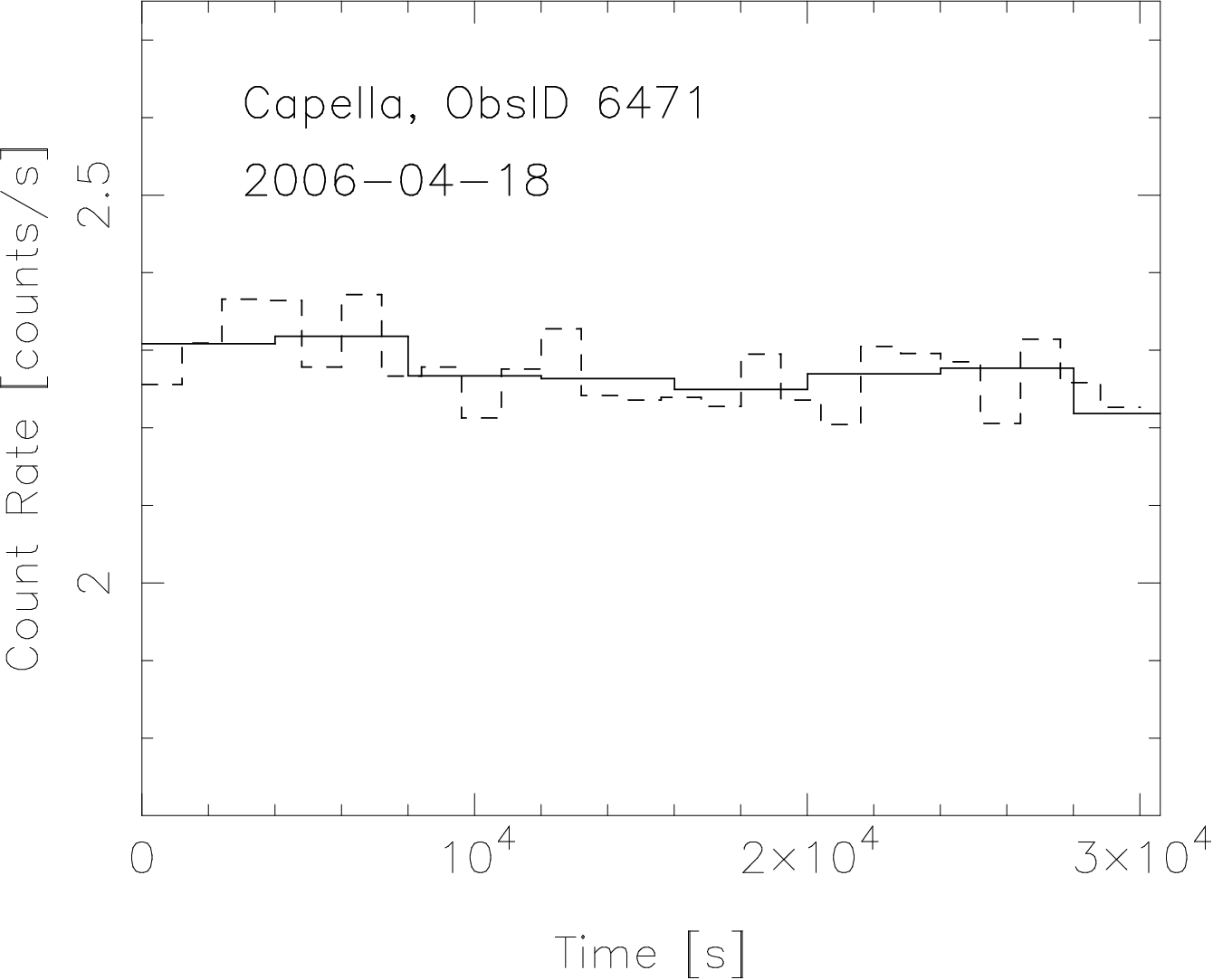}
\caption{\scriptsize As in Figure~\ref{figlchm}, for
a sequence of twelve light curves of Capella.
\label{figlccap}}
\end{figure}

\clearpage

\begin{deluxetable}{lll}
\tabletypesize{\scriptsize}
\tablewidth{0pt}
\tablecaption{Spectral Line Metrics\label{tblmetrics}}
\tablehead{\colhead{Metric} & \colhead{Flux ratio} & \colhead{Comment} \tabletypesize{5pt} \tablecolumns{3} }
\startdata
\tableline
NeO & Ne\,X~$\lambda$12.13 / O\,VIII~$\lambda$18.97 & tracks Ne abundance \\
Oif & O\,VII(i)~$\lambda$21.8 / O\,VII(f)~$\lambda$22.1 & O\,VII is density sensitive \\
O78 & O\,VII(r)~$\lambda$21.6 / O\,VIII~$\lambda$18.97 & tracks temperature \\
OFe & O\,VIII~$\lambda$18.97 / Fe\,XVII~$\lambda\lambda$15.01,17.1 & tracks high FIP element abundances \\
MgFe & Mg\,XII$\lambda$8.419 / Fe\,XVII~$\lambda\lambda$15.01,17.1 & high-FIP elements, possibly tracks fractionation \\
SiFe & Si\,XIV$\lambda$6.18 / Fe\,XVII~$\lambda\lambda$15.01,17.1 & high-FIP elements, possibly tracks fractionation \\
SFe & S\,XV~$\lambda$5.1 / Fe\,XVII~$\lambda\lambda$15.01,17.1 & S is a high FIP element \\
short line/cont & $f_{\rm line}(\lambda<12.13)$ / $f_{\rm continuum}$ & sensitive to high temperatures and metallicities \\
long line/cont & $f_{\rm line}(\lambda>12.13)$ / $f_{\rm continuum}$ & sensitive to low temperatures \\
line/cont & $f_{\rm line}$ / $f_{\rm continuum}$ & tracks metallicity \\
cont long/short & $f_{\rm continuum}(13<\lambda<25)$ / $f_{\rm continuum}(3<\lambda<12)$ & categorizes continuum shape \\
cont long/medium & $f_{\rm continuum}(13<\lambda<25)$ / $f_{\rm continuum}(6<\lambda<12)$ & categorizes continuum shape \\
cont medium/short & $f_{\rm continuum}(6<\lambda<12)$ / $f_{\rm continuum}(3<\lambda<6)$ & categorizes continuum shape \\
\enddata
\end{deluxetable}

\clearpage

\begin{deluxetable}{lllll}
\tablewidth{0pt}
\tablecaption{Coadded observations in X-Atlas.  'Num. Obs.' indicates
the number of observations that were combined.  The $\chi^{2}$ values
listed are derived from the comparison of the observed combined $0^{th}$-order
spectrum and the predicted combined $0^{th}$-order, with an extra 10\%
systematic error added to the observed spectrum to account for
uncertainties in the effective area.  'Num. Bins' is the
number of non-zero bins in the observed $0^{th}$-order spectrum used
to calculate the $chi^{2}$ value.  \label{tblcoadd}}
\tablehead{
\colhead{Star} & \colhead{Num. Obs.} &
\colhead{Observation IDs} & \colhead{$\chi^{2}$} & \colhead{Num. Bins}
\tabletypesize{\scriptsize}
\tablecolumns{3}
}
\startdata
\tableline
HD 93129AB & 7 & 5397, 7201, 7202, 7203, 7204, 7228, 7229 & 471.4 & 298\\
HD 93250 & 5 & 5399, 5400, 7189, 7341, 7342 & 865.5 & 630 \\
9 Sgr & 2 & 5398, 6285 & 401.8 & 461 \\
15 Mon & 3 & 5401, 6247, 6248 & 445.8 & 236 \\
\tableline
\enddata
\end{deluxetable}

\clearpage

\begin{deluxetable}{lllllllll}
\tabletypesize{\scriptsize}
\tablewidth{0pt}
\tablecaption{Stars in the X-Atlas database, ordered approximately by
spectral type.  The 'Star' column lists
the common name (if applicable) of the primary target of the observation.  'Alt. Name'
gives a catalog listing for that star, usually the HD number.  'ObsID'
is the $Chandra$ Observation ID.  'Obs. Date' is the date the
observation was made.  '$t_{exp}$' is the exposure time of the
observation, in ks.  'Spec. Type' is the spectral type of the
star.  'B$-$V' is the B$-$V color of the star.  '$d$' is the
distance in parsecs from the star to Earth.  'Figure (h/l)' indicates the
spectral montage figures in which the spectra from the
observation can be found: the first number refers to the figure with
the high-resolution MEG spectrum, and the second number refers to the
figure with the projected low-resolution ACIS-S spectrum.  Spectral types, B$-$V
values, and distances were retrieved from the {\it SIMBAD} database,
operated in Strasbourg, France, unless otherwise noted. \label{tbl1}}
\tablehead{
\colhead{Star}           & \colhead{Alt. Name}      &
\colhead{ObsID} & \colhead{Obs. Date} & \colhead{$t_{exp}$ (ks)} & \colhead{Spec. Type}  &
\colhead{$\bv$}          & \colhead{$d$ (pc)}  & \colhead{Figure (h/l)}
\tablecolumns{9}
}
\startdata
\tableline
HD 93129 AB\tablenotemark{a} & ... & 5397 & 2005-11-29 & 10.2 & (2x)O2If*+O3.5V((f+))\tablenotemark{e} & 0.206\tablenotemark{e} & 2500\tablenotemark{n} & \ref{figobsgh}/\ref{figobsgl} \\
 &  & 7201 & 2005-11-08 & 18.2 &  &  &  & \ref{figobsgh}/\ref{figobsgl} \\
 &  & 7202 & 2005-11-10 & 9.5 &  &  &  & \ref{figobsgh}/\ref{figobsgl} \\
 &  & 7203 & 2005-11-12 & 27.3 &  &  &  & \ref{figobsgh}/\ref{figobsgl} \\
 &  & 7204 & 2005-11-09 & 33.2 &  &  &  & \ref{figobsgh}/\ref{figobsgl} \\
 &  & 7228 & 2005-12-01 & 19.7 &  &  &  & \ref{figobsgh}/\ref{figobsgl} \\
 &  & 7229 & 2005-12-02 & 19.6 &  &  &  & \ref{figobsgh}/\ref{figobsgl} \\
HD 93250\tablenotemark{b} & ... & 5399 & 2005-10-22 & 39.7 & O3.5V((f+))\tablenotemark{f} & 0.165\tablenotemark{e} & 2500\tablenotemark{n} & \ref{figobmsh}/\ref{figobmsl} \\
 &  & 5400 & 2005-09-05 & 33.7 &  &  &  & \ref{figobmsh}/\ref{figobmsl} \\
 &  & 7189 & 2006-06-23 & 17.7 &  &  &  & \ref{figobmsh}/\ref{figobmsl} \\
 &  & 7341 & 2006-06-19 & 53.3 &  &  &  & \ref{figobmsh}/\ref{figobmsl} \\
 &  & 7342 & 2006-06-22 & 49.3 &  &  &  & \ref{figobmsh}/\ref{figobmsl} \\
9 Sgr & HD 164794 & 5398 & 2005-05-11 & 101.2 & O4V((f))\tablenotemark{f} & 0.017\tablenotemark{e} & 1500 & \ref{figobmsh}/\ref{figobmsl} \\
 & & 6285 & 2005-06-09 & 44.6 & & &  & \ref{figobmsh}/\ref{figobmsl} \\
$\zeta$ Pup & HD 66811 & 640 & 2000-03-28 & 67.6 & O4I(n)f\tablenotemark{f} & -0.276 & 710\tablenotemark{e} & \ref{figobsgh}/\ref{figobsgl} \\
HD 150136 & ... & 2569 & 2002-06-27 & 90.3 & O5III:n(f)\tablenotemark{e} & 0.161\tablenotemark{e} & \nodata & \ref{figobsgh}/\ref{figobsgl} \\
$\theta^{1}$\ Ori\ C & HD 37022 & 3 & 1999-10-31 & 49.5 &
O5.5V\tablenotemark{g} & 0.021\tablenotemark{e} & \nodata & \ref{figobmagh}/\ref{figobmagl} \\
 & & 4 & 1999-11-24 & 30.9 & & & & \ref{figobmagh}/\ref{figobmagl} \\
 & & 2567 & 2001-12-28 & 46.4 & & &  & \ref{figobmagh}/\ref{figobmagl} \\
 & & 2568 & 2002-02-19 & 46.3 & & &  & \ref{figobmagh}/\ref{figobmagl} \\
Cyg OB2 8A & BD +40 4227 & 2572 & 2002-07-31 & 65.1 & O6I(f)+O5.5III(f)\tablenotemark{h} & 1.09 & \nodata & \ref{figobpech}/\ref{figobpecl} \\
HD 206267 & ... & 1888 & 2001-02-18 & 34.1 & O6.5V((f))\tablenotemark{e} & 0.205\tablenotemark{e} & 591\tablenotemark{e} & \\
 & & 1889 & 2001-02-09 & 39.5 & & & & \ref{figobmsh}/\ref{figobmsl} \\
15 Mon & HD 47839 & 5401 & 2005-11-20 & 54.8 & O7V((f))\tablenotemark{e}+O9.5Vn\tablenotemark{i} & -0.248 & 447\tablenotemark{e} & \ref{figobmsh}/\ref{figobmsl} \\
& & 6247 & 2006-01-31 & 37.8 & & &  & \ref{figobmsh}/\ref{figobmsl} \\
& & 6248 & 2005-12-13 & 7.6 & & &  & \ref{figobmsh}/\ref{figobmsl} \\
$\xi$ Per & HD 24912 & 4512 & 2004-03-22 & 158.8 & O7.5III(n)((f))\tablenotemark{f} & 0.014\tablenotemark{e} & 2130\tablenotemark{e} & \ref{figobsgh}/\ref{figobsgl} \\
$\iota$ Ori & HD 37043 & 599 & 2001-02-07 & 37.2 & O9III+?\tablenotemark{f} & -0.244\tablenotemark{e} & 1110\tablenotemark{e} & \ref{figobmsh}/\ref{figobmsl} \\
 & & 2420 & 2001-02-08 & 12.8 & & & & \\
$\tau$ CMa & HD 57061 & 2525 & 2002-03-28 & 43.2 & O9II\tablenotemark{e} & -0.155\tablenotemark{e} & 980. & \\
 & & 2526 & 2002-04-24 & 43.9 & & & & \\
$\zeta$ Oph & HD 149757 & 2571 & 2002-09-04 & 35.4 & O9.5Vnn\tablenotemark{e} & 0.019 & 146\tablenotemark{e} & \\
 & & 4367 & 2002-09-05 & 48.3 & & & & \ref{figobmsh}/\ref{figobmsl} \\
$\theta^{2}$\ Ori\ A & \nodata & 4473 & 2004-11-03 & 49.1 & O9.5Vpe\tablenotemark{e} & -0.084\tablenotemark{e} & 581.  & \ref{figobpech}/\ref{figobpecl} \\
 & & 4474 & 2004-11-23 & 50.6 & & &  & \ref{figobpech}/\ref{figobpecl} \\
$\sigma$ Ori & HD 37468 & 3738 & 2003-08-12 & 91.0 & O9.5V+B0.5V\tablenotemark{f} & -0.240\tablenotemark{e} & 381\tablenotemark{e} & \ref{figobmsh}/\ref{figobmsl} \\
$\delta$ Ori & HD 36486 & 639 & 2000-01-13 & 49.0 & O9.5II+B0.5III\tablenotemark{f} & -0.219\tablenotemark{e} & 310\tablenotemark{e} & \ref{figobsgh}/\ref{figobsgl} \\
$\zeta$ Ori & HD 37742J & 610 & 2000-04-08 & 59.7 &
O9.7Ib\tablenotemark{f} & -0.208\tablenotemark{e} &
278\tablenotemark{e} & \ref{figobsgh}/\ref{figobsgl} \\
 & & 1524 & 2000-04-09 & 13.8 & & & & \\
$\gamma$ Cas & HD 5394 & 1895 & 2001-08-10 & 51.2 & B0IVpe & -0.10 & 188 & \ref{figobpech}/\ref{figobpecl} \\
$\epsilon$ Ori & HD 37128 & 3753 & 2003-12-12 & 91.6 & B0Ia & -0.19 & 412 & \ref{figobsgh}/\ref{figobsgl} \\
$\tau$ Sco & HD 149438 & 638 & 2000-09-17 & 59.2 & B0.2V & -0.201 & 132 & \ref{figobmagh}/\ref{figobmagl} \\
 & & 2305 & 2000-09-18 & 13.0 & & &  & \ref{figobmagh}/\ref{figobmagl} \\
$\beta$ Cru & HD 111123 & 2575 & 2002-02-01 & 74.4 & B0.5III\tablenotemark{f} & -0.1520 & 108 & \ref{figobmsh}/\ref{figobmsl} \\
WR 140 & HD 193793 & 2337 & 2000-12-29 & 45.5 & WC7pd+O4-5\tablenotemark{j} & 0.27\tablenotemark{j} & 1100\tablenotemark{j} & \ref{figwrh}/\ref{figwrl} \\
 & & 2338 & 2001-05-08 & 24.5 & & &  & \ref{figwrh}/\ref{figwrl} \\
 & & 5419 & 2006-03-31 & 45.1 & & &  & \ref{figwrh}/\ref{figwrl} \\
 & & 6286 & 2006-03-23 & 46.9 & & &  & \ref{figwrh}/\ref{figwrl} \\
 & & 6287 & 2006-04-07 & 49.3 & & &  & \ref{figwrh}/\ref{figwrl} \\
$\gamma$ Vel & HD 68273 & 629 & 2000-03-15 & 64.8 & WC8+O7.5III-V\tablenotemark{j} & -0.32\tablenotemark{j} & 368\tablenotemark{k} & \ref{figwrh}/\ref{figwrl} \\
$\eta$ Car & HD 93308 & 632 & 2000-11-19 & 89.5 & ?p & 0.61 & 2500\tablenotemark{n} & \ref{figetah}/\ref{figetal} \\
 &  & 3745 & 2003-05-02 & 94.5 & & & & \ref{figetah}/\ref{figetal} \\
 &  & 3746 & 2003-07-20 & 90.3 & & & & \ref{figetah}/\ref{figetal} \\
 &  & 3747 & 2003-09-26 & 70.1 & & & & \ref{figetah}/\ref{figetal} \\
 &  & 3748 & 2003-06-16 & 97.2 & & & & \ref{figetah}/\ref{figetal} \\
 &  & 3749 & 2002-10-16 & 91.3 & & & & \ref{figetah}/\ref{figetal} \\
Algol & HD 19356 & 604 & 2000-04-01 & 51.7 & B8V+K1IV\tablenotemark{l} & -0.05 & 28.5 & \ref{figcoolpech}/\ref{figcoolpecl} \\
Canopus & HD 45348 & 636 & 2000-07-21 & 94.6 & F0II & 0.15 & 95.9 & \ref{figcoolnormh}/\ref{figcoolnorml} \\
$\upsilon$ Peg & HD 220657 & 3731 & 2003-08-26 & 106.8 & F8IV & 0.61 & 53.1 & \ref{figcoolnormh}/\ref{figcoolnorml} \\
TZ CrB & HD 146361 & 15 & 2000-06-18 & 83.7 & G0V+G0V\tablenotemark{l} & 0.51 & 21.7 & \ref{figrscvnh}/\ref{figrscvnl} \\
44 Boo & HD 133640 & 14 & 2000-04-25 & 59.1 & G1V+G2V\tablenotemark{l} & 0.65 & 12.8 & \ref{figwumah}/\ref{figwumal} \\
ER Vul & HD 200391 & 1887 & 2001-03-29 & 112.0 & G0V+G5V\tablenotemark{l} & 0.57 & 49.9 & \ref{figrscvnh}/\ref{figrscvnl} \\
$\xi$ UMa & HD 98230J & 1894 & 2000-12-28 & 70.9 & G0V+G0V\tablenotemark{l} & 0.59 & 7.7\tablenotemark{l} & \ref{figcoolnormh}/\ref{figcoolnorml} \\
LS Com & HD 111812 & 1891 & 2001-03-12 & 130.2 & G0IIIp & 0.627 & 94.2 & \ref{figcoolnormh}/\ref{figcoolnorml} \\
OU And & HD 223460 & 1892 & 2001-08-11 & 95.7 & G1IIIe & 0.79 & 135 & \ref{figcoolnormh}/\ref{figcoolnorml} \\
AR Lac & HD 210334 & 6 & 2000-09-11 & 32.1 &
G2IV+K0IV\tablenotemark{l} & 0.72 & 42.0 & \\
 & & 7 & 2000-09-16 & 7.4 & & & & \\
 & & 8 & 2000-09-15 & 7.5 & & & & \\
 & & 9 & 2000-09-17 & 32.2 & & & & \\
 & & 10 & 2000-09-20 & 7.2 & & & & \\
 & & 11 & 2000-09-19 & 7.2 & & & & \\
V987 Tau & HD 283572 & 3756 & 2003-10-20 & 100.5 & G2III & 0.77 & 128 & \ref{figttsh}/\ref{figttsl} \\
SU Aur & HD 282624 & 3755 & 2003-11-27 &  98.5 & G2IIIevar & 0.80 & 152 & \ref{figttsh}/\ref{figttsl} \\
24 Ursae Majoris & HD 82210 & 2564 & 2002-03-26 & 44.4 & G4III-IV & 0.753 & 32.4 & \ref{figcoolnormh}/\ref{figcoolnorml} \\
 & & 3471 & 2002-03-29 & 44.5 & & & & \\
TY Pyx & HD 77137 & 601 & 2001-01-03 & 49.1 & G5IV+G5IV & 0.664 & 56 & \ref{figrscvnh}/\ref{figrscvnl} \\
UX Ari & HD 21242 & 605 & 2000-01-12 & 48.5 & G5IV & 0.9 & 50.2 & \ref{figrscvnh}/\ref{figrscvnl} \\
Capella & HD 34029 & 57 & 2000-03-03 &  28.8 & G5IIIe+... & 0.8 & 13.2 & \\
 & & 1010 & 2001-02-11 &  29.5 & & & & \\
 & & 1099 & 1999-08-28 &  14.6 & & & & \\
 & & 1100 & 1999-08-28 &  14.6 & & & & \\
 & & 1101 & 1999-08-29 &  14.6 & & & & \\
 & & 1103 & 1999-09-24 &  40.5 & & & & \\
 & & 1199 & 1999-08-30 &   2.0 & & & & \\
 & & 1235 & 1999-08-28 &  14.6 & & & & \\
 & & 1236 & 1999-08-28 &  14.6 & & & & \\
 & & 1237 & 1999-08-29 &  14.6 & & & & \\
 & & 1318 & 1999-09-25 &  26.7 & & & & \\
 & & 2583 & 2002-04-29 &  27.6 & & & & \\
 & & 3674 & 2003-09-27 &  28.7 & & & & \\
 & & 5040 & 2004-09-10 &  28.7 & & & & \\
 & & 5955 & 2005-03-28 &  28.7 & & & & \\
 & & 6471 & 2006-04-18 &  29.6 & & & & \\
$\mu$ Vel & HD 93497 & 1890 & 2001-09-24 &  19.7 & G5IIIa & 0.914 & 35.5 & \\
 & & 3403\tablenotemark{c} & 2001-10-29 &  57.4 & & & & \\
 & & 3410 & 2001-12-18 &  57.0 & & &  & \ref{figcoolnormh}/\ref{figcoolnorml} \\
FK Com & HD 117555 & 614 & 2000-03-27 &  41.4 & G5II & 0.87 & 234 & \ref{figcoolpech}/\ref{figcoolpecl} \\
$\lambda$ And & HD 222107 & 609 & 1999-12-17 &  81.9 & G8III & 1.08 & 25.8 & \ref{figrscvnh}/\ref{figrscvnl} \\
V2246 Oph & DoAr 21 & 3761 & 2003-05-05 & 91.1 & K0\tablenotemark{m} & 0.33 & \nodata & \ref{figttsh}/\ref{figttsl} \\
II Peg & HD 224085 & 12\tablenotemark{d} & 1999-10-17 & 0.073 & K0V... & 1.01 & 42.3 & \\
 & & 1451 & 1999-10-17 &  42.7 & & &  & \ref{figrscvnh}/\ref{figrscvnl} \\
Speedy Mic & HD 197890 & 2536 & 2002-04-26 & 34.4 & K0V & 0.94 & 44.4 & \ref{figttsh}/\ref{figttsl} \\
 & & 3491 & 2002-04-27 &  34.6 & & & & \\
HIP 92680 & HD 174429 & 3729 & 2003-06-07 &  73.9 & K0Vp & 0.77 & 49.7 & \ref{figcoolnormh}/\ref{figcoolnorml} \\
VW Cep & HD 197433 & 3766 & 2003-08-29 & 116.6 & K0Vvar & 0.75 & 27.7 & \ref{figwumah}/\ref{figwumal} \\
$\beta$ Ceti & HD 4128 & 974 & 2001-06-29 &  86.1 & K0III & 1.02 & 29.4 & \ref{figcoolnormh}/\ref{figcoolnorml} \\
$\lambda$ Ori X-1 & HD 245059 & 5420 & 2006-01-13 & 54.8 & K1 & 0.7 & \nodata & \ref{figttsh}/\ref{figttsl} \\
 & & 6241 & 2005-12-30 & 17.7 & & & & \\
 & & 7253 & 2006-01-07 & 20.9 & & & & \\
V824 Ara & HD 155555 & 2538 & 2002-07-10 & 94.2 & K1Vp & 0.78 & 31.4 & \ref{figrscvnh}/\ref{figrscvnl} \\
AB Dor & HD 36705 & 16 & 1999-10-09 &  52.3 & K1IIIp... & 0.8 & 14.9 & \ref{figcoolnormh}/\ref{figcoolnorml} \\
YY Men & HD 32918 & 2557 & 2002-02-01 &  74.2 & K1IIIp & 1.01 & 292 & \ref{figcoolnormh}/\ref{figcoolnorml} \\
$\sigma$ Gem & HD 62044 & 5422 & 2005-05-16 &  62.8 & K1III & 1.12 & 37.5 & \ref{figrscvnh}/\ref{figrscvnl} \\
 & & 6282 & 2005-05-17 &  57.9 & & & & \\
IM Peg & HD 216489 & 2527 & 2002-07-01 & 24.6 & K1.5II-IIIe & 1.134 & 96.8 & \\
& & 2528 & 2002-07-08 &  24.8 & & & & \\
& & 2529 & 2002-07-13 &  24.8 & & & & \\
& & 2530 & 2002-07-18 &  23.9 & & & & \\
& & 2531 & 2002-07-25 &  23.9 & & & & \\
& & 2532 & 2002-07-31 &  22.5 & & & & \\
& & 2533 & 2002-08-08 &  23.7 & & & & \\
& & 2534 & 2002-08-15 &  23.9 & & & & \\
TV Crt & HD 98800 & 3728 & 2003-03-07 &  98.9 & K5Ve & 1.16 & 46.7 & \ref{figttsh}/\ref{figttsl} \\
V4046 Sgr & HD 319139 & 5423 & 2005-08-06 & 97.7 & K5 & 0.9 & \nodata & \ref{figttsh}/\ref{figttsl} \\
 & & 6265 & 2005-08-09 &  45.9 & & & & \\
TW Hya & HIP 53911 & 5 & 2000-07-18 &  47.7 & K8Ve & 0.7 & 56.4 & \ref{figttsh}/\ref{figttsl} \\
CC Eri & HD 16157 & 4513 & 2004-10-01 & 89.5 & M0 & 1.36 & 11.5 & \ref{figflareh}/\ref{figflarel} \\
 & & 6132 & 2004-10-01 &  30.6 & & & & \\
AU Mic & HD 197481 & 17 & 2000-11-12 &  58.8 & M1Ve & 1.44 & 9.94 & \ref{figflareh}/\ref{figflarel} \\
Hen 3-600 & GSC 07726 & 4502 & 2004-03-16 &  99.4 & M3 & 1.52 & \nodata & \ref{figttsh}/\ref{figttsl} \\
AD Leo & SAO 81292 & 2570 & 2002-06-01 &  45.2 & M3.5V & 1.54 & 4.70 & \\
EV Lac & HIP 112460 & 1885 & 2001-09-19 & 100.0 & M3.5e & 1.36 & 5.05 & \ref{figflareh}/\ref{figflarel} \\
Proxima Centauri & V645 Cen & 2388 & 2001-09-13 &  42.4 & M5.5Ve & 1.97 & 1.30 & \ref{figflareh}/\ref{figflarel} \\
\tableline
\enddata
\tablenotetext{a}{HD 93129AB was observed with HD 93129A at the
aimpoint.  $Chandra$ has sufficient spatial resolution to
differentiate HD 93129A from HD 93129B.  However, components A and B
are about 3 arcsec apart along the dispersion direction, and the ACIS
CCD energy resolution is not high enough to separate the events from
these two sources.  A third X-ray source, Trumpler 14 MJ 176, also lies in the dispersion
direction $\sim$5 arcsec away from HD 93129A.  The extracted spectrum
includes events from all three sources.}
\tablenotetext{b}{$\eta$ Car lies directly on the spectral arm in all
spectra of HD 93250.  Some events from $\eta$ Car may be included in
the extracted spectra, and the background counts level will be
anomalously high in this region.}
\tablenotetext{c}{ObsID 3403, a $\sim$60 ks observation of HD 93497,
is not included in X-Atlas.  During the observation, the ACIS CCD was
suffering from a problem known as threshold-plane (T-plane) lockup in
chips S4 and S2, causing the science run to be stopped prematurely and
replanned as ObsID 3410, which was carried out successfully.}
\tablenotetext{d}{ObsID 12, a $\sim$73 s observation of II Peg, is not
included in X-Atlas.  Like ObsID 3403, ObsID 12 also suffered ACIS
T-plane lockup, rendering the data useless.  The observation was
repeated successfully as ObsID 1451.}
\tablenotetext{e}{\citet{mai04}}
\tablenotetext{f}{\citet{walb06}}
\tablenotetext{g}{\citet{sch06}}
\tablenotetext{h}{\citet{deb04}}
\tablenotetext{i}{\citet{gie93}}
\tablenotetext{j}{\citet{huc01}}
\tablenotetext{k}{\citet{mil07}}
\tablenotetext{l}{\citet{tes04}}
\tablenotetext{m}{\citet{pre99}}
\tablenotetext{n}{\citet{eva03}}
\end{deluxetable}

\clearpage

\begin{deluxetable}{lclcc}
\tablecaption{$\chi^{2}$ values for the comparison of the observed and
predicted $0^{th}$-order spectra for selected X-Atlas targets with low
count rates.  An extra 10\% systematic error was added to the observed
$0^{th}$-order spectra to account for uncertainties in the effective area.
The rightmost column lists the number of non-zero bins in the observed
$0^{th}$-order spectrum. \label{tblpa_zero}}
\tablewidth{0pt}
\tablehead{
\colhead{Target}           & \colhead{Observation ID}      &
\colhead{Count Rate}          & \colhead{$\chi^{2}$} & \colhead{Num. Bins}
\tabletypesize{5pt}
\tablecolumns{5}
}
\startdata
\tableline
HD 93250 & 7189 & 0.0337 & 174.8 & 257 \\
$\xi$ Per & 4512 & 0.0418 & 381.7 & 340 \\
HD 37468 & 3738 & 0.348 & 323.1 & 153 \\
15 Mon & 5401 & 0.0260 & 280.7 & 173 \\
HD 206267 & 1888 & 0.0252 & 203.8 & 152 \\
\tableline
\enddata
\end{deluxetable}

\clearpage

\begin{deluxetable}{llllllll}
\tabletypesize{\small}
\tablewidth{0pt}
\tablecaption{Selected spectral fit results from the predicted ACIS-S aimpoint
spectra.  Fits for $\theta^{1}$ Ori C, $\theta^{2}$ Ori A, and
$\gamma$ Cas were calculated using a known, fixed $N_{H}$, while for
IM Peg and Capella, $N_{H}$ fit values were allowed to float.  All
stars were fit with a two-temperature Mewe-Kaastra-Liedahl thermal
plasma model.  Fits that have only one temperature listed had a flux
emission ratio (FER) of 0, meaning that all of the flux was calculated to
have originated from the $kT_{1}$ component of the plasma.  All
abundances are with respect to solar. $\chi^{2}$ in the final column is 
reduced $\chi^{2}$.\label{tblfit}}
\tablehead{
\colhead{Star}           & \colhead{ObsID}      &
\colhead{$N_{H}\times 10^{22}/cm^{2}$}    &
\colhead{$kT_{1}$ (keV)}  &  \colhead{$kT_{2}$ (keV)}    &
\colhead{Abun.}  &
\colhead{F.E.R. ($kT_{2}$/$kT_{1}$)}  &
\colhead{$\chi^{2}$}
\tablecolumns{8}
}
\startdata
\tableline
$\theta^{1}$ Ori C & 3    & 0.207 & $>$15 & 1.140 & 0.12 & 0.89 & 1.67\tablenotemark{a} \\
                   & 4    & 0.207 & 10.329 & 1.286 & 0.14 & 0.97 & 1.06\tablenotemark{a} \\
                   & 2567 & 0.207 &  9.928 & 1.426 & 0.23 & 1.61 & 1.26\tablenotemark{a} \\
                   & 2568 & 0.207 & 11.852 & 1.140 & 0.07 & 0.78 & 1.34\tablenotemark{a} \\
$\theta^{2}$ Ori A & 4473 & 0.148 &  0.794 & 0.335 & 0.52 & 1.21 & 1.29\tablenotemark{b} \\
                   & 4474 & 0.148 &  3.533 & 1.142 & 0.05 & 0.05 & 1.28\tablenotemark{b} \\
$\gamma$ Cas & 1895 & 0.136 &  $>$15 & ... & 1.25 & 0 & 1.19\tablenotemark{a} \\
Capella &   57 & 0.014 & 0.602 & ... & 0.31 & 0 & 1.80\tablenotemark{b} \\
        & 1010 & 0.057 & 0.568 & ... & 0.31 & 0 & 1.71\tablenotemark{b} \\
	& 1099 & 0.022 & 0.568 & ... & 0.30 & 0 & 1.08\tablenotemark{b} \\
	& 1100 & 0.034 & 0.561 & ... & 0.30 & 0 & 1.14\tablenotemark{b} \\
	& 1101 &  0.01 & 0.597 & ... & 0.27 & 0 & 1.75\tablenotemark{b} \\
	& 1103 & 0.031 & 4.098 & 0.561 & 0.32 & 9.79 & 1.42\tablenotemark{b} \\
	& 1235 & 0.047 & $>$10 & 0.565 & 0.35 & 20.85 & 1.33\tablenotemark{b} \\
	& 1236 & 0.033 & 0.574 & ... & 0.33 & 0 & 1.89\tablenotemark{b} \\
	& 1237 &  0.01 & 0.606 & ... & 0.53 & 0 & 1.14\tablenotemark{b} \\
	& 1318 & 0.038 & 0.561 & ... & 0.51 & 0 & 1.49\tablenotemark{b} \\
	& 2583 & 0.003 & $>$10 & 0.576 & 0.52 & 5.77 & 0.99\tablenotemark{b} \\
	& 3674 & 0.076 & 0.561 & ... & 0.51 & 0 & 1.25\tablenotemark{b} \\
	& 5040 & 0.066 & 0.577 & ... & 0.44 & 0 & 1.73\tablenotemark{b} \\
	& 5955 & 0.081 & 0.592 & ... & 0.40 & 0 & 1.30\tablenotemark{b} \\
	& 6471 & 0.081 & 0.590 & ... & 0.31 & 0 & 1.26\tablenotemark{b} \\
IM Peg & 2527 & 0.029 & 3.028 & 0.989 & 0.37 & 0.54 & 0.99\tablenotemark{a} \\
       & 2528 & 0.026 & 3.350 & 1.035 & 0.27 & 1.43 & 1.29\tablenotemark{a} \\
       & 2529 & 0.048 & 1.753 & 0.728 & 0.29 & 0.52 & 1.26\tablenotemark{a} \\
       & 2530 & 0.023 & 2.324 & 0.981 & 0.28 & 1.02 & 1.01\tablenotemark{a} \\
       & 2531 & 0.011 & 2.983 & 0.984 & 0.30 & 1.09 & 1.01\tablenotemark{a} \\
       & 2532 & 0.027 & 1.987 & 0.833 & 0.25 & 0.51 & 0.93\tablenotemark{a} \\
       & 2533 & 0.085 & 1.293 & 0.288 & 0.26 & 0.23 & 1.23\tablenotemark{a} \\
       & 2534 & 0.007 & 1.690 & 0.834 & 0.30 & 0.43 & 1.09\tablenotemark{a} \\
\tableline
\enddata
\tablenotetext{a}{Spectral fits on these observations were carried out
without letting all parameters float in the final fitting step.}
\tablenotetext{b}{Spectral fits on these observations were conducted
by letting all parameters (except $N_{H}$, if known) float in the
final fitting step.}
\end{deluxetable}

\clearpage

\begin{deluxetable}{llrrrrrrr}
\tablewidth{0pt}
\tablecaption{Hardness ratios and normalized quantiles for the stars
in X-Atlas, derived from the predicted ACIS-I aimpoint spectra.
Hardness ratios and quantiles calculated from the predicted ACIS-S
aimpoint spectra are available on the X-Atlas website. \label{tblqh}}
\tablehead{
\colhead{Star}           & \colhead{ObsID}      &
\colhead{$HR_{1}$}          & \colhead{$HR_{2}$}  &
\colhead{$Q_{25}$}          & \colhead{$Q_{50}$}  &
\colhead{$Q_{75}$} & \colhead{$log_{10}(\frac{m}{1-m})$} & \colhead{$3 \frac{Q_{25}}{Q_{75}}$}
\tabletypesize{5pt}
\tablecolumns{9}
}
\startdata
\tableline
HD 93129AB & 5397 & -0.150 & 0.304 & 0.081 & 0.119 & 0.180 & -0.868 & 1.357 \\
           & 7201 & -0.369 & 0.521 & 0.085 & 0.114 & 0.159 & -0.892 & 1.606 \\
           & 7202 & -0.221 & 0.373 & 0.083 & 0.117 & 0.172 & -0.876 & 1.441 \\
           & 7203 & -0.282 & 0.448 & 0.085 & 0.121 & 0.171 & -0.860 & 1.499 \\
           & 7204 & -0.244 & 0.503 & 0.089 & 0.125 & 0.182 & -0.845 & 1.468 \\
           & 7228 & -0.269 & 0.566 & 0.091 & 0.127 & 0.172 & -0.837 & 1.582 \\
           & 7229 & -0.271 & 0.355 & 0.081 & 0.116 & 0.163 & -0.884 & 1.499 \\
           & Combined & -0.291 & 0.528 & 0.089 & 0.123 & 0.169 & -0.852 & 1.584 \\
HD 93250 & 5399 & -0.178 & 0.434 & 0.087 & 0.121 & 0.184 & -0.860 & 1.422 \\
         & 5400 & -0.151 & 0.421 & 0.087 & 0.125 & 0.188 & -0.845 & 1.393 \\
         & 7189 & -0.262 & 0.514 & 0.089 & 0.123 & 0.174 & -0.852 & 1.532 \\
         & 7341 & -0.180 & 0.407 & 0.085 & 0.121 & 0.182 & -0.860 & 1.405 \\
         & 7342 & -0.185 & 0.489 & 0.089 & 0.123 & 0.191 & -0.852 & 1.395 \\
         & Combined & -0.205 & 0.466 & 0.087 & 0.123 & 0.180 & -0.852 & 1.452 \\
HD 150136  & 2569 & -0.421 & 0.321 & 0.078 & 0.104 & 0.146 & -0.934 & 1.597 \\
9 Sgr & 5398 & -0.591 & 0.038 & 0.064 & 0.085 & 0.119 & -1.031 & 1.618 \\
      & 6285 & -0.591 & 0.008 & 0.064 & 0.085 & 0.119 & -1.031 & 1.618 \\
      & Combined & -0.613 & 0.077 & 0.066 & 0.087 & 0.119 & -1.020 & 1.666 \\
$\zeta$ Pup & 640 & -0.602 & 0.155 & 0.070 & 0.091 & 0.127 & -1.000 & 1.656 \\
$\theta^{1}$ Ori C & 3 & 0.056 & 0.646 & 0.106 & 0.152 & 0.226 & -0.748 & 1.411 \\
                   & 4 & 0.064 & 0.631 & 0.104 & 0.153 & 0.226 & -0.742 & 1.386 \\
                   & 2567 & 0.009 & 0.659 & 0.104 & 0.148 & 0.216 & -0.761 & 1.447 \\
                   & 2568 & 0.068 & 0.614 & 0.104 & 0.152 & 0.231 & -0.748 & 1.352 \\
Cyg OB2 8A & 2572 & 0.121 & 0.836 & 0.127 & 0.163 & 0.222 & -0.711 & 1.717 \\
HD 206267 & 1888 & -0.709 & -0.054 & 0.061 & 0.080 & 0.112 & -1.063 & 1.626 \\
          & 1889 & -0.695 & -0.040 & 0.064 & 0.080 & 0.114 & -1.063 & 1.699 \\
15 Mon & 5401 & -0.767 & -0.472 & 0.047 & 0.064 & 0.081 & -1.162 & 1.743 \\
       & 6247 & -0.816 & -0.456 & 0.044 & 0.062 & 0.083 & -1.176 & 1.567 \\
       & 6248 & -0.630 & -0.342 & 0.045 & 0.066 & 0.093 & -1.149 & 1.468 \\
       & Combined & -0.726 & -0.338 & 0.055 & 0.070 & 0.089 & -1.123 & 1.850 \\
$\xi$ Per & 4512 & -0.698 & -0.075 & 0.062 & 0.080 & 0.106 & -1.063 & 1.767 \\
$\zeta$ Oph & 2571 & -0.706 & -0.067 & 0.062 & 0.080 & 0.102 & -1.063 & 1.832 \\
            & 4367 & -0.715 & -0.037 & 0.062 & 0.080 & 0.106 & -1.063 & 1.767 \\
$\iota$ Ori & 599  & -0.797 & -0.283 & 0.053 & 0.070 & 0.089 & -1.123 & 1.786 \\
            & 2420 & -0.759 & -0.334 & 0.051 & 0.068 & 0.089 & -1.136 & 1.722 \\
$\tau$ CMa & 2525 & -0.753 & -0.285 & 0.044 & 0.059 & 0.080 & -1.110 & 1.686 \\
           & 2526 & -0.771 & -0.333 & 0.040 & 0.055 & 0.077 & -1.136 & 1.561 \\
$\theta^{2}$ Ori A & 4473 & -0.690 & 0.114 & 0.068 & 0.085 & 0.114 & -1.031 & 1.799 \\
                   & 4474 & 0.115 & 0.577 & 0.108 & 0.155 & 0.241 & -0.735 & 1.346 \\
$\sigma$ Ori & 3738 & -0.808 & -0.329 & 0.053 & 0.070 & 0.087 & -1.123 & 1.825 \\
$\delta$ Ori & 639 & -0.723 & -0.163 & 0.059 & 0.076 & 0.099 & -1.086 & 1.787 \\
$\zeta$ Ori & 610  & -0.785 & -0.225 & 0.055 & 0.072 & 0.095 & -1.110 & 1.739 \\
            & 1524 & -0.726 & -0.211 & 0.055 & 0.074 & 0.097 & -1.098 & 1.705 \\
$\gamma$ Cas & 1895 & 0.320 & 0.594 & 0.123 & 0.186 & 0.362 & -0.642 & 1.020 \\
$\epsilon$ Ori & 3753 & -0.797 & -0.254 & 0.055 & 0.072 & 0.091 & -1.110 & 1.811 \\
$\tau$ Sco & 638  & -0.642 & 0.079 & 0.066 & 0.085 & 0.116 & -1.031 & 1.720 \\
           & 2305 & -0.656 & 0.073 & 0.066 & 0.085 & 0.114 & -1.031 & 1.749 \\
$\beta$ Cru & 2575 & -0.865 & -0.525 & 0.044 & 0.061 & 0.080 & -1.191 & 1.641 \\
$\gamma$ Vel & 629 & 0.510 & 0.834 & 0.153 & 0.203 & 0.294 & -0.595 & 1.567 \\
WR 140 & 2337 & 0.411 & 0.796 & 0.142 & 0.203 & 0.354 & -0.595 & 1.203 \\
       & 2338 & 0.946 & 0.420 & 0.281 & 0.394 & 0.518 & -0.186 & 1.626 \\
       & 5419 & 0.094 & 0.745 & 0.114 & 0.159 & 0.243 & -0.723  & 1.406 \\
       & 6286 & 0.124 & 0.751 & 0.114 & 0.163 & 0.258 & -0.711 & 1.323 \\
       & 6287 & 0.112 & 0.742 & 0.114 & 0.161 & 0.254 & -0.717 & 1.343 \\
$\eta$ Car & 632  & 0.967 & 0.652 & 0.336 & 0.453 & 0.590 & -0.082 & 1.707 \\
           & 3745 & 0.983 & 0.521 & 0.362 & 0.480 & 0.622 & -0.035 & 1.747 \\
           & 3746 & 0.751 & 0.218 & 0.193 & 0.368 & 0.557 & -0.235 & 1.041 \\
           & 3747 & 0.984 & 0.027 & 0.493 & 0.618 & 0.770 & 0.209 & 1.921 \\
           & 3748 & 0.981 & 0.349 & 0.389 & 0.508 & 0.658 & 0.014 & 1.772 \\
           & 3749 & 0.967 & 0.607 & 0.320 & 0.440 & 0.578 & -0.105 & 1.662 \\
Algol & 604 & -0.271 & 0.395 & 0.083 & 0.112 & 0.167 & -0.900 & 1.499 \\
Canopus & 636 & -0.686 & 0.105 & 0.068 & 0.085 & 0.110 & -1.031 & 1.861 \\
HD 220657 & 3731 & -0.738 & 0.074 & 0.068 & 0.083 & 0.106 & -1.041 & 1.928 \\
TZ CrB & 15 & -0.506 & 0.232 & 0.074 & 0.093 & 0.133 & -0.990 & 1.671 \\
44 Boo & 14 & -0.555 & 0.157 & 0.070 & 0.089 & 0.127 & -1.010 & 1.656 \\
ER Vul & 1887 & -0.568 & 0.259 & 0.074 & 0.093 & 0.129 & -0.990 & 1.720 \\
XI Uma & 1894 & -0.673 & 0.003 & 0.064 & 0.081 & 0.110 & -1.052 & 1.758 \\
HD 111812 & 1891 & -0.625 & 0.336 & 0.076 & 0.093 & 0.125 & -0.990 & 1.817 \\
HD 223460 & 1892 & -0.060 & 0.470 & 0.093 & 0.133 & 0.205 & -0.816 & 1.360 \\
AR Lac & 6 & -0.481 & 0.328 & 0.076 & 0.099 & 0.140 & -0.961 & 1.621 \\
       & 7 & -0.203 & 0.413 & 0.085 & 0.117 & 0.178 & -0.876 & 1.435 \\
       & 8 & -0.282 & 0.365 & 0.081 & 0.110 & 0.163 & -0.909 & 1.499 \\
       & 9 & -0.457 & 0.310 & 0.076 & 0.099 & 0.142 & -0.961 & 1.599 \\
       & 10 & -0.455 & 0.293 & 0.076 & 0.097 & 0.140 & -0.971 & 1.621 \\
       & 11 & -0.422 & 0.305 & 0.076 & 0.099 & 0.146 & -0.961 & 1.558 \\
V987 Tau & 3756 & -0.371 & 0.388 & 0.080 & 0.104 & 0.153 & -0.934 & 1.555 \\
SU Aur & 3755 & 0.101 & 0.675 & 0.108 & 0.157 & 0.258 & -0.729 & 1.257 \\
24 Ursae Majoris & 2564 & -0.701 & -0.002 & 0.066 & 0.081 & 0.108 & -1.052 & 1.841 \\
                 & 3471 & -0.726 & -0.029 & 0.064 & 0.080 & 0.104 & -1.063 & 1.854 \\
TY Pyx & 601 & -0.478 & 0.304 & 0.076 & 0.097 & 0.140 & -0.971 & 1.621 \\
UX Ari & 605 & -0.390 & 0.358 & 0.080 & 0.102 & 0.152 & -0.943 & 1.574 \\
Capella & 57 & -0.749 & -0.059 & 0.064 & 0.080 & 0.102 & -1.063 & 1.888 \\
        & 1010 & -0.759 & -0.076 & 0.064 & 0.078 & 0.100 & -1.075 & 1.924 \\
	& 1099 & -0.722 & -0.077 & 0.064 & 0.078 & 0.102 & -1.075 & 1.888 \\
	& 1100 & -0.750 & -0.053 & 0.064 & 0.080 & 0.102 & -1.063 & 1.888 \\
	& 1101 & -0.741 & -0.059 & 0.064 & 0.080 & 0.102 & -1.063 & 1.888 \\
	& 1103 & -0.744 & -0.080 & 0.064 & 0.078 & 0.102 & -1.075 & 1.888 \\
        & 1199 &  &  &  &  &  &  & \\
	& 1235 & -0.750 & -0.077 & 0.064 & 0.078 & 0.102 & -1.075 & 1.888 \\
	& 1236 & -0.753 & -0.074 & 0.064 & 0.078 & 0.102 & -1.075 & 1.888 \\
	& 1237 & -0.755 & -0.063 & 0.064 & 0.080 & 0.102 & -1.063 & 1.888 \\
	& 1318 & -0.759 & -0.091 & 0.064 & 0.078 & 0.100 & -1.075 & 1.924 \\
	& 2583 & -0.725 & -0.038 & 0.064 & 0.080 & 0.104 & -1.063 & 1.854 \\
	& 3674 & -0.746 & -0.072 & 0.064 & 0.078 & 0.102 & -1.075 & 1.888 \\
	& 5040 & -0.743 & -0.023 & 0.066 & 0.080 & 0.104 & -1.063 & 1.908 \\
	& 5955 & -0.757 & -0.022 & 0.066 & 0.080 & 0.104 & -1.063 & 1.908 \\
	& 6471 & -0.747 & -0.009 & 0.066 & 0.080 & 0.104 & -1.063 & 1.908 \\
HD 93497 & 1890 & -0.738 & -0.013 & 0.066 & 0.080 & 0.104 & -1.063 & 1.908 \\
         & 3410 & -0.710 &  0.003 & 0.066 & 0.081 & 0.106 & -1.052 & 1.874 \\
FK Com & 614 & -0.031 & 0.418 & 0.091 & 0.135 & 0.207 & -0.808 & 1.320 \\
$\lambda$ And & 609 & -0.485 & 0.328 & 0.076 & 0.099 & 0.142 & -0.961 & 1.599 \\
DoAr 21 & 3761 & 0.567 & 0.908 & 0.161 & 0.231 & 0.402 & -0.522 & 1.202 \\
Speedy Mic & 2536 & -0.459 & 0.249 & 0.074 & 0.097 & 0.138 & -0.971 & 1.602 \\
           & 3491 & -0.441 & 0.193 & 0.070 & 0.093 & 0.138 & -0.990 & 1.520 \\
HIP 92680 & 3729 & -0.431 & 0.304 & 0.076 & 0.099 & 0.144 & -0.961 & 1.578 \\
VW Cep & 3766 & -0.595 & 0.178 & 0.070 & 0.089 & 0.123 & -1.010 & 1.707 \\
$\beta$ Ceti & 974 & -0.689 & 0.081 & 0.068 & 0.083 & 0.112 & -1.041 & 1.830 \\
HDE 245059 & 5420 & -0.469 & 0.230 & 0.072 & 0.095 & 0.136 & -0.980 & 1.582 \\
           & 6241 & -0.316 & 0.229 & 0.074 & 0.100 & 0.153 & -0.952 & 1.444 \\
           & 7253 & -0.346 & 0.265 & 0.076 & 0.100 & 0.152 & -0.952 & 1.499 \\
V824 Ara & 2538 & -0.506 & 0.257 & 0.074 & 0.095 & 0.135 & -0.980 & 1.647 \\
AB Dor & 16 & -0.464 & 0.232 & 0.072 & 0.095 & 0.138 & -0.980 & 1.561 \\
YY Men & 2557 & -0.069 & 0.462 & 0.093 & 0.133 & 0.199 & -0.816 & 1.399 \\
$\sigma$ Gem & 5422 & -0.225 & 0.391 & 0.083 & 0.116 & 0.174 & -0.884 & 1.434 \\
             & 6282 & -0.285 & 0.386 & 0.081 & 0.110 & 0.165 & -0.909 & 1.482 \\
IM Peg & 2527 & -0.250 & 0.392 & 0.083 & 0.112 & 0.169 & -0.900 & 1.482 \\
       & 2528 & -0.355 & 0.387 & 0.081 & 0.106 & 0.155 & -0.926 & 1.572 \\
       & 2529 & -0.363 & 0.408 & 0.081 & 0.108 & 0.155 & -0.917 & 1.572 \\
       & 2530 & -0.361 & 0.408 & 0.081 & 0.106 & 0.155 & -0.926 & 1.572 \\
       & 2531 & -0.330 & 0.392 & 0.081 & 0.108 & 0.159 & -0.917 & 1.535 \\
       & 2532 & -0.300 & 0.405 & 0.083 & 0.110 & 0.163 & -0.909 & 1.534 \\
       & 2533 & -0.315 & 0.416 & 0.083 & 0.110 & 0.161 & -0.909 & 1.552 \\
       & 2534 & -0.348 & 0.392 & 0.081 & 0.108 & 0.157 & -0.917 & 1.553 \\
II Peg & 1451 & -0.165 & 0.423 & 0.087 & 0.121 & 0.184 & -0.860 & 1.422 \\
TV Crt & 3728 & -0.448 & 0.192 & 0.072 & 0.093 & 0.135 & -0.990 & 1.605 \\
V4046 Sgr & 5423 & -0.535 & 0.129 & 0.070 & 0.087 & 0.123 & -1.020 & 1.707 \\
          & 6265 & -0.493 & 0.134 & 0.068 & 0.089 & 0.125 & -1.010 & 1.635 \\
TW Hya & 5 & -0.501 & 0.139 & 0.070 & 0.087 & 0.123 & -1.020 & 1.707 \\
CC Eri & 4513 & -0.617 & 0.171 & 0.070 & 0.089 & 0.117 & -1.010 & 1.789 \\
       & 6132 & -0.285 & 0.369 & 0.081 & 0.108 & 0.165 & -0.917 & 1.482 \\
AU Mic & 17 & -0.530 & 0.179 & 0.070 & 0.091 & 0.127 & -1.000 & 1.656 \\
Hen 3-600 & 4502 & -0.458 & 0.150 & 0.070 & 0.093 & 0.133 & -0.990 & 1.585 \\
AD Leo & 2570 & -0.619 & 0.073 & 0.066 & 0.085 & 0.112 & -1.031 & 1.779 \\
EV Lac & 1885 & -0.504 & 0.101 & 0.066 & 0.089 & 0.127 & -1.010 & 1.566 \\
Proxima Centauri & 2388 & -0.678 & -0.031 & 0.062 & 0.081 & 0.106 & -1.052 & 1.767 \\
\tableline
\enddata
\end{deluxetable}

\clearpage

\begin{deluxetable}{lcccc}
\tablewidth{0pt}
\tablecaption{Overview of the Gregory-Loredo variability analysis
results. Stars were graded by their variability index.  Those with
indices of 7-10 have a greater than 99\% probability of being variable.  Those with
indices of 6 were probably variable (greater than 90\% probability)
and those scoring below 6 (less than 90\% probability) were graded not
variable \citep{rots06}.  For stars for which multiple observations
have been made, the star's variability was judged by the observation
with the highest variability index. \label{tblvar}}
\tablehead{
\colhead{ }  & \colhead{Number of Stars}      & \colhead{Variable}      &
\colhead{Probably Variable}          & \colhead{Not Variable}
\tabletypesize{5pt}
\tablecolumns{5}
}
\startdata
\tableline
High-Mass Stars & 24 & 10 & 1 & 13\\
Low-Mass Stars & 40 & 34 & 1 & 5\\
\tableline
Total & 64 & 44 & 2 & 18\\
\tableline
\enddata
\end{deluxetable}


\clearpage

\begin{thebibliography}{}
\bibitem[Audard et al.(2001)]{2001A&A...365L.318A} Audard, M., G{\"u}del,                                                                                 
M., \& Mewe, R.\ 2001, \aap, 365, L318                                               
\bibitem[Brickhouse et al. (2000)]{bri00} Brickhouse, N. S., Dupree,
A. K., Edgar, R. J., Liedahl, D. A., Drake, S. A., White, N. E., \&
Singh, K. P., 2000, \apj, 530, 387
\bibitem[Canizares et al. (2000)]{can00} Canizares et al., 2000, \apj,
539, L41
\bibitem[Canizares et al. (2005)]{can05} Canizares, C. R., Davis,
J. E., Dewey, D., Flanagan, K. A., Galton, E. B., Huenemoerder, D. P.,
Ishibashi, K., Markert, T. H., Marshall, H. L., McGuirk, M.,
Schattenburg, M. L., Schulz, N. S., Smith, H. I., Wise, M., 2005,
\pasp, 117, 1144
\bibitem[CIAO (2006)]{ciao06} Chandra Interactive Analysis of Observations (CIAO), http://cxc.harvard.edu/ciao/
\bibitem[Proposers' Observatory Guide (2005)]{cxc05} Chandra X-ray Center, 'The
Chandra Proposers' Observatory Guide,' Vers. 8.0, Cambridge, MA:
Dec. 2005 
\bibitem[Corcoran et al. (2005)]{cor05} Corcoran, M. F., Hamaguchi,
K., Henley, D. B., Ishibashi, K., Gull, T., Nielsen, K., Pittard,
J. M., 2005, \baas, 37, 1347
\bibitem[De Becker et al. (2004)]{deb04} De Becker, M., Rauw, G., \&
Swings, J. P., 2004, \apss, 297, 291
\bibitem[De Marco \& Schmutz (1999)]{dem99} De Marco, O., \& Schmutz,
W., 1999, \aap, 345, 163
\bibitem[Donati, et al. (2006)]{don06}Donati, J.-F., Howarth, I. D.,
Jardine, M. M., Petit, P., Catala, C., Landstreet, J. D., Bouret, J.-C.,
Alecian, E., Barnes, J. R., Forveille, T., Paletou, F., and Manset, N.
2006, \mnras, 370, 629  
\bibitem[Drake et al. (2006)]{dra06} Drake, J. J., Ratzlaff, P.,
Kashyap, V., Edgar, R., Izem, R., Jerius, D., Siemiginowska, A.,
Vikhlinin, A., 2006, \procspie, 6270, 62701I
\bibitem[Dupree \& Brickhouse (1996)]{dup96} Dupree, A.K., \&
Brickhouse, N.S., 1996, in Poster Proc., IAU Symp.\ 176: Stellar
Surface Structure (Wien: Institut f\"{u}r Astronomie), 184 
\bibitem[Evans et al. (2003)]{eva03} Evans, N. R., Seward, F. D.,
Krauss, M. I., Isobe, T., Nichols, J., Schlegel, E. M., Wolk, S. J.,
2003, \apj, 589, 509
\bibitem[Favata et al. (2005)]{fav05} Favata, F., Flaccomio, E.,
Reale, F., Micela, G., Sciortino, S., Shang, H., Stassum, K. G.,
Feigelson, E. D., 2005, \apjs, 160, 469
\bibitem[Fruscione et al. (2006)]{fru06} Fruscione, A., 2006,
\procspie, 6270, 62701V, D. R. Silvia \& R. E. Doxsey, eds.
\bibitem[Gagn\'{e} et al. (2005)]{gag05} Gagn\'{e}, M., Oksala, M. E.,
Cohen, D. H., Tonnesen, S. K., ud-Doula, A., Owocki, S. P., Townsend,
R. H. D., \& MacFarlane, J. J., 2005, \apj, 628, 986
\bibitem[Getman et al. (2002)]{get02} Getman, K. V., Feigelson, E. D.,
Townsley, L., Bally, J., Lada, C. J., Reipurth, B., 2002, \apj, 575, 354
\bibitem[Getman et al. (2005)]{get05} Getman, K. V., et al., 2005,
\apjs, 160, 319
\bibitem[Gies et al. (1993)]{gie93} Gies, D. R., et al., 1993, \aj,
106, 2072
\bibitem[Gonz\'{a}lez-Riestra \& Rodriguez-Pascual (2007)]{gon07}
Gonz\'{a}lez-Riestra, R. \& Rodriguez-Pascual, P. M., 2007, ``BiRD: A
Browsing Utility for RGS Spectra,'' X-ray Spectroscopy Workshop, July 11-13
\bibitem[Gregory \& Loredo (1992)]{gl92} Gregory, P. C. \& Loredo,
T. J., 1992, \apj, 398, 146
\bibitem[G{\"u}del et al.(2001)]{2001A&A...365L.336G} G{\"u}del, M., et              
al.\ 2001, \aap, 365, L336                                                           
\bibitem[Hamaguchi et al. (2007)]{ham07} Hamaguchi, K., et al., 2007,
\apj, 663, 522
\bibitem[Harnden et al. (1979)]{har79} Harnden, F. R., Jr.,
Branduardi, G., Gorenstein, P., Grindlay, J., Rosner, R., Topka, K.,
Elvis, M., Pye, J. P., \& Vaiana, G. S., 1979, \apjl, 234, L51
\bibitem[Hong et al. (2004)]{hon04} Hong, J., Schlegel, E. M. \&
Grindlay, J. E., 2004, \apj, 614, 508
\bibitem[Houck \& Denicola (2000)]{hou00} Houck, J. C., \& Denicola,
L. A., 2000, in ASP Conf. Ser. 216: Astronomical Data Analysis
Software and Systems IX, Vol. 9, 591
\bibitem[Huenemoerder et al. (1993)]{hue93} Huenemoerder, D. P.,
Ramsey, L. W., Buzasi, D. L., Nations, H. L., 1993, \apj, 404, 316
\bibitem[Huenemoerder et al. (2001)]{hue01} Huenemoerder, D. P.,
Canizares, C. R., \& Schulz, N. S., 2001, \apj, 559, 1135
\bibitem[Ishibashi et al. (2006)]{ish06} Ishibashi, K., Dewey, D.,
Huenemoerder, D. P., \& Testa, P., 2006, \apj, 644, 117
\bibitem[Johnson et al. (2002)]{joh02} Johnson, O., Drake, J. J.,
Kashyap, V., Brickhouse, N. S., Dupree, A. K., Freeman, P., Young, P. R.,
\& Kriss, G.A., 2002, \apj, 565, 97
\bibitem[Kaastra (1992)]{kaa92} Kaastra, J. S., 1992, 'An X-Ray
Spectral Code for Optically Thin Plasmas, Internal SRON-Leiden Report,
version 2.0
\bibitem[Kashyap \& Drake(2000)]{kas00} Kashyap, V. L. \& Drake,
    J. J., 2000, BASI, 28, 475
\bibitem[Liedahl et al. (1995)]{lie95} Liedahl, D. A., Osterheld,
A. L., \& Goldstein, W. H., 1995, \apjl, 438, 115
\bibitem[Linsky et al. (1998)]{lin98} Linsky, J. L., Wood, B. E.,
Brown, A., \& Osten, R. A., 1998, \apj, 492, 767
\bibitem[Lopes de Oliveira et al. (2006)]{lop06} Lopes de Oliveira, R.,
Motch, C., Haberl, F., Negueruela, I., \& Janot-Pacheco, E., 2006,
\aap, 454, 265
\bibitem[MacFarlane et al. (1991)]{mac91} MacFarlane, J. J.,
Cassinelli, J. P., Welsh, B. Y., Vedder, P. W., Vallerga, J. V., \&
Waldron, W. L., 1991, \apj, 380, 564 
\bibitem[Ma\'{i}z-Apell\'{a}niz et al. (2004)]{mai04}
Ma\'{i}z-Apell\'{a}niz, J., Walborn, N. R., Galu\'{e}, H. \'{A}., Wei, L. H., 2004, \apjs, 151, 103
\bibitem[Mewe et al. (1985)]{mew85} Mewe, R., Gronenschild,
E. H. B. M., \& van den Oord, G. H. J., 1985, \aap, 62, 197
\bibitem[Mewe et al. (1986)]{mew86} Mewe, R., Lemen, J. R., \& van den
Oord, G. H. J., 1986, \aap, 65, 511
\bibitem[Millour et al. (2007)]{mil07} Millour, F., et al., 2007, \aap, 464, 107
\bibitem[Morrison \& McCammon (1983)]{mor83} Morrison, R. \& McCammon,
D., \apj, 270, 119
\bibitem[Mullan \& Waldron (2006)]{mul06} Mullan, D. J. \& Waldron,
W. L., \apj, 637, 506
\bibitem[Park et al. (2006)]{par06} Park, T., Kashyap, V. L.,
Siemiginowska, A., van Dyk, D. A., Zezas, A., Heinke, C., Wargelin,
B. J., 2006, \apj, 652, 610
\bibitem[Preibisch (1999)]{pre99} Preibisch, T., 1999, \aap, 345, 583
\bibitem[Raassen \& Kaastra (2007)]{raa07} Raassen, A. J. J., Kaastra,
J. S., \aap, 461, 679
\bibitem[Rakowski et al. (2006)]{rak06} Rakowski, C. E., Schulz,
N. S., Wolk, S. J., \& Testa, P., \apj, 649, L111
\bibitem[Raymond \& Smith (1977)]{ray77} Raymond, J. C. \& Smith, B. W.,
1977, \apjs, 35, 419
\bibitem[Reale (2002)]{rea02} Reale, F., 2002, in ASP Conf. Ser.,
Vol. 277, 103
\bibitem[Rots (2006)]{rots06} Rots, A. H., 2006, in ASP
Conf. Ser. 351: Astronomical Data Analysis Software and Systems XV,
Vol. 351, 73
\bibitem[Schulz et al. (2000)]{sch00} Schulz, N. S., Canizares, C. R.,
Huenemoerder, D., \& Lee, J. C., 2000, \apjl, 545, 135 
\bibitem[Schulz et al. (2001)]{sch01} Schulz, N. S., Canizares, C.,
Huenemoerder, D., Kastner, J. H., Taylor, S. C., \& Bergstrom, E. J.,
2001, \apj, 549, 441
\bibitem[Schulz et al. (2003)]{sch03} Schulz, N. S., Canizares, C.,
Huenemoerder, D., Tibbets, K., 2003, \apj, 595, 365
\bibitem[Schulz et al. (2006)]{sch06} Schulz, N. S., Testa, P.,
Huenemoerder, D. P., Ishibashi, K., \& Canizares, C. R., 2006, \apj, 653,
636
\bibitem[Scelsi et al. (2007)]{sce07} Scelsi, L., Maggio, A., Micela,
G., Briggs, K., \& Gudel, M., preprint (astro-ph 0707.2857)
\bibitem[Schmidt-Kaler (1982)]{sk82} Schmidt-Kaler, T., 1982,
Landolt-B\"ornstein, Group VI, Vol 2b, eds. K. Schaifers and
H. H. Voigt (New York: Springer Verlag), 14
\bibitem[Seward et al. (1979)]{sew79} Seward, F. D., Forman, W. R.,
Giacconi, R., Griffiths, R. E., Harnden, F. R., Jr., Jones, C., \& Pye,
J. P., 1979, \apjl, 234, L55
\bibitem[Seward \& Chlebowski (1982)]{sew82} Seward, F. D. \&
Chlebowski, T., 1982, \apj, 256, 530
\bibitem[SIMBAD (2006)]{sim06} SIMBAD Astronomical Database, Strausborg, France
\bibitem[Smith et al.(2001)]{2001AJ.556..91L} Smith, R.~K.,                          
Brickhouse, N.~S., Liedahl, D.~A., Raymond, J~ C. 2001, \apj, 556, L91               
\bibitem[Smith et al. (2004)]{smi04} Smith, M. A., Cohen, D. H.,
Gu. M. F., Robinson, R. D., Evans, N. R., Schran, P. G., 2004, \apj,
600, 972
\bibitem[Testa et al. (2004)]{tes04} Testa, P., Drake, J. J., \& Peres, G.,
2004, \apj, 617, 508 
\bibitem[ud-Doula \& Owocki (2002)]{udd02} ud-Doula, A. \& Owocki,
S. P., 2002, \apj, 576, 413
\bibitem[van der Hucht (2001)]{huc01} van der Hucht, K. A., 2001,
\nar, 45, 135
\bibitem[Walborn (2006)]{walb06} Walborn, N. R., {\it The UV Universe:
Stars from Birth to Death}, 2006, $26^{th}$ Meeting of the IAU, Joint
Discussion 4, Prague, Czech Republic, 1
\bibitem[Waldron (1994)]{wal94} Waldron, W. L., 1994, AIP
Conf. Proceed., 313, 279
\bibitem[Weisskopf et al. (2002)]{wei02} Weisskopf, M. C., Brinkman,
B., Canizares, C., Garmire, G., Murray, S., Van Speybroeck, L. P., 2002
\pasp, 114, 1
\bibitem[White et al. (1980)]{whi80} White, N. E., Holt, S. S., Boldt,
E. A., \& Serlemitsos, P. J., 1980, \apjl, 239, 69
\bibitem[Wolk et al. (2006)]{wol06} Wolk, S. J., Spitzbart, B. D.,
Bourke, T. L., \& Alves, J., 2006, \apj, 132, 1100
\bibitem[Young et al. (2001)]{you01} Young, P. R., Dupree, A. K.,
Wood, B. E., Redfield, S., Linsky, J. L., Ake, T. B., \& Moos, H. W., 2001, \apj, 555, L121
\end{thebibliography}
\end{document}